\documentclass[fleqn,usenatbib]{mnras}

\usepackage{newtxtext,newtxmath}
\usepackage{tablefootnote}
\usepackage{subcaption}
\usepackage{booktabs}
\usepackage{threeparttable}
\usepackage{ulem}
\usepackage[T1]{fontenc}
\usepackage{verbatim}
\usepackage{fancyvrb}
\DeclareRobustCommand{\VAN}[3]{#2}
\let\VANthebibliography\thebibliography
\def\thebibliography{\DeclareRobustCommand{\VAN}[3]{##3}\VANthebibliography}

\usepackage{graphicx}	% Including figure files
\usepackage{amsmath}	% Advanced maths commands
\usepackage{bm}
\newcommand\degs{\ifmmode ^{\circ}\else$^{\circ}$\fi}

\newcommand{\verbat}[1]{%
  \begin{Verbatim}
#1
  \end{Verbatim}
}

\newcommand{\insertinlineverbatim}[1]{\verb|#1|}

\newcommand{\Rstar}{\ensuremath{R_\star}}

\newcommand{\Msun}{\ensuremath{M_{\sun}}}

\newcommand{\Rp}{\ensuremath{R_{\text{pl}}}}
\newcommand{\Bp}{\ensuremath{B_{\text{pl}}}}
\newcommand{\Reff}{\ensuremath{R_{\text{eff}}}}
\newcommand{\Rmp}{\ensuremath{R_{\text{mp}}}}
\newcommand{\kmp}{\ensuremath{k_{\text{mp}}}}
\newcommand{\vrel}{\ensuremath{v_{\text{rel}}}}
\newcommand{\vorb}{\ensuremath{v_{\text{orb}}}}
\newcommand{\vw}{\ensuremath{v_{\text{sw}}}}
\newcommand{\vA}{\ensuremath{v_{\text{A}}}}
\newcommand{\nw}{\ensuremath{n_{\text{sw}}}}
\newcommand{\rhow}{\ensuremath{\rho_{\text{w}}}}

\newcommand{\Pmagn}{\ensuremath{P_{\text{B}}}}
\newcommand{\Pdyn}{\ensuremath{P_{\text{dyn}}}}

\newcommand{\Tw}{\ensuremath{T_{\text{sw}}}}

\newcommand{\Bsw}{\ensuremath{B_{\text{sw}}}}
\newcommand{\Mdotstar}{\ensuremath{\dot{M}_{\star}}}
\newcommand{\Mdotsun}{\ensuremath{\dot{M}_{\sun}}}
\newcommand{\MA}{\ensuremath{M_{\rm A}}}
\newcommand{\Rss}{\ensuremath{R_{\rm SS}}}

\newcommand{\vsw}{\ensuremath{v_{\rm sw}}}

\newcommand{\proxb}{\mbox{Proxima b}}
\newcommand{\proxd}{\mbox{Proxima d}}

\newcommand{\kB}{\ensuremath{k_{\text{B}}}}

\title[Modelling SPI in three iconic M dwarf systems]{Modelling magnetic star-planet interaction in the iconic M dwarfs Proxima Centauri, YZ Ceti and GJ 1151}

\author[Pe\~na-Mo\~nino L. \&  P\'erez-Torres M.]{
Luis Pe\~na-Mo\~nino$^{1}$\thanks{Corresponding authors: lpm@iaa.es, torres@iaa.es}
\& Miguel Pérez-Torres$^{1}$
\\
$^{1}$IAA-CSIC, Instituto de Astrof\'isica de Andaluc\'ia, 
Glorieta de la Astronom\'ia s/n, 18008, Granada, Spain 
}

\date{Accepted XXX. Received YYY; in original form ZZZ}

\pubyear{\the\year{}}

\begin{document}
\label{firstpage}

\maketitle

\begin{abstract}
The unambiguous detection of magnetic star-planet interaction (SPI) via radio observations would provide a novel method for detecting exoplanets and probing their magnetic fields. Although direct radio detection of sub-Jovian planets is hindered by the low frequencies involved, models of sub-Alfvénic SPI predict that Earth-like planets in close-in orbits around M dwarfs may induce detectable emission.
Here, we revisit the modelling of the expected radio emission from magnetic star-planet interaction in the iconic M-dwarf systems Proxima Centauri, YZ Ceti, and GJ 1151, where claims of SPI-related radio detections have been made. For this, we use \texttt{\textbf{SIRIO}} (\textbf{S}tar-planet \textbf{I}nteraction and \textbf{R}adio \textbf{I}nduced \textbf{O}bservations), a public Python code that models radio emission from sub-Alfvénic SPI. We benchmark \texttt{SIRIO} results against those paradigmatic systems, whose SPI modeling has been previously discussed in the literature.
Our results support previous findings that Proxima b, YZ Cet b, and the putative planet GJ~1151 b are most likely in the sub-Alvénic regime (assuming a hybrid PFSS geometry), so SPI should be at work in all of them. We find that the Alfvén wing model generally predicts a very low level of radio emission, while if magnetic reconnection takes place, prospects for detection are significantly better.
We also find that free-free absorption may play a relevant role, in particular in YZ Ceti. 
Our \texttt{SIRIO} code can also be used to evaluate the feasibility of radio proposals aimed at detecting SPI, and to constrain the stellar wind mass-loss rate or planetary magnetic field.

\end{abstract}

\begin{keywords}
planet-star interactions -- radio continuum: stars -- stars: late-type -- stars: low-mass -- (stars:) planetary systems -- software: public release 
\end{keywords}

%%%%%%%%%%%%%%%%%%%%%%%%%%%%%%%%%%%%%%%%%%%%%%%%%%

%%%%%%%%%%%%%%%%% BODY OF PAPER %%%%%%%%%%%%%%%%%%

%%%%%%%%%%%%%%%%%%%%%%%%%%%%%%%%%%%%%%%%%%%%%%%%%%%%%%%%%%%%
%%%%%%%%%%%%%%%%%%% Intro        %%%%%%%%%%%%%%%%%%%%%%%%%%%
%%%%%%%%%%%%%%%%%%%%%%%%%%%%%%%%%%%%%%%%%%%%%%%%%%%%%%%%%%%%
\section{Introduction}

Over the past three decades, thousands of exoplanets have been discovered using techniques such as transit photometry, radial velocity measurements, direct imaging, and gravitational microlensing. These methods primarily rely on observations at higher frequencies, such as infrared or optical wavelengths. However, despite its potential as a powerful detection method, radio observations have yet to yield definitive results. Radio observations are uniquely suited to probe the magnetic properties of exoplanets, which are crucial for understanding their potential habitability and internal structure.

However, detecting auroral radio emission directly from planetary magnetospheres remains challenging, due to their relatively weak magnetic fields. The resulting emission would typically fall below the ionospheric cutoff frequency of Earth, making it undetectable from the ground, except in the case of giant, Jupiter-like planets. 
Hence,  searching for radio emission arising from magnetic star-planet interactions (SPI) offers a more promising approach, and in particular in the case of M-dwarf stars, which possess strong magnetic fields, often reaching several hundred Gauss, or even kGauss. This makes them promising candidates for detection of SPI-induced emission, which would fall within the frequency range of several hundred MHz to a few GHz, potentially observable with modern radio telescopes like LOFAR, uGMRT, and the VLA \citep{Zarka1997,Farrell1999, Zarka2004a}.

SPI-induced radio signals are expected to exhibit a high degree of circular polarization and a frequency close to the local cyclotron frequency (e.g., \citealt{Melrose1982}). 
To confirm an SPI origin, the signal must also exhibit periodicity, approximately correlated with the planet's orbital or synodic period \citep{Zarka2004a,Fischer2019}. However, a  variety of system characteristics can affect this periodicity, including among others the geometric configuration of the planet and the star, the topology of the stellar magnetic field, and the relative motions of the system's bodies \citep{Strugarek2019,Strugarek2022}.

Tentative detections of SPI emission have been reported in systems such as GJ 1151 \citep{Vedantham2020}, Proxima Centauri \citep{PerezTorres2021}, and YZ Cet \citep{Pineda2023,Trigilio2023}. However, definitive detection of SPI emission remains elusive, highlighting the need for robust theoretical models and targeted observational campaigns.
Theoretical modeling of SPI faces significant challenges, including uncertainties in key parameters such as the efficiency of the conversion of Poynting flux into radio power, and the angle of the emission cone. While magneto-hydrodynamic (MHD) simulations provide the most accurate predictions, they are computationally very expensive. Thus, there is a pressing need for computationally efficient yet reliable methods to model SPI and guide observational efforts.

Here, we revisit the tentative detections of SPI-induced radio emission in the M dwarf systems Proxima Centauri, YZ Ceti, and GJ~1151, using \texttt{SIRIO} (\textbf{S}tar-planet \textbf{I}nteraction and  \textbf{R}adio \textbf{I}nduced \textbf{O}bservations), a public Python code designed to predict the radio emission arising from SPI. We benchmark \texttt{SIRIO} against the above systems, which host close-in exoplanets and have been the subject of intense radio campaigns and radio modelling of their magnetic star-planet interaction by different groups. The paper is organized as follows: we discuss the radio emission from sub-Alfvénic star-planet interaction in Sect.~\ref{sec:sub-alfven}; we present the code in Sect.~\ref{sec:SIRIO}, and apply it to the cases of Proxima Centauri, YZ Ceti and GJ~1151, in Sect.~\ref{sec:case-studies}, where we compare our results against those obtained by other groups. Finally, we summarize our main results and conclusions in Sect.~\ref{sec:summary}.

%%%%%%%%%%%%%%%%%%%%%%%%%%%%%%%%%%%%%%%%%%%%%%%%%%%%%%%%%%%%
%%%%%%%%%%%%%%%%%%% SPI          %%%%%%%%%%%%%%%%%%%%%%%%%%%
%%%%%%%%%%%%%%%%%%%%%%%%%%%%%%%%%%%%%%%%%%%%%%%%%%%%%%%%%%%%
\section{Sub-Alfvénic star-planet interaction}
\label{sec:sub-alfven}

The stellar wind surrounding an exoplanet is a magnetized plasma. As the exoplanet moves through this medium, it acts as an obstacle to the plasma flow, generating a Poynting flux that can transfer energy back to the host star (e.g., \citealt{Handbook_Saur2018,Zarka2024,Handbook_Strugarek2018}).

However, this energy transfer requires two specific conditions to be met. First, the relative speed of the stellar wind with respect to the planet, $\vec{v}_{\rm rel} = \vec{v}_{\rm w}  - \vec{v}_{\rm orb} $,  must be smaller than the Alfvén speed, \vA = \Bsw /$(4\,\pi\, \rhow)^{1/2}$, where \Bsw\ and \rhow\ are the stellar wind magnetic field intensity and the mass density, respectively, at the orbit of the planet. This condition defines the sub-Alfvénic regime, where the Alfvén Mach number 
$M_A = \vrel / v_A < 1$. Second, the radial stellar wind velocity must be less than the radial Alfvén speed, allowing the Alfvén wing to travel back to the star and transfer energy. 
This (large-scale) Alfvén wave is thought to cascade to smaller spatial scales, where kinetic Alfvén waves develop parallel electric fields.
At this point, the energy of the kinetic (non-linear) Alfvén wave can accelerate  electrons, leading to cyclotron radio emission (e.g., \citealt{Melrose1982}). In fact, kinetic simulations and Juno observations show that electrons are indeed accelerated by dispersive/kA waves in Jupiter's auroral zones \citep{Saur2018}.
The mechanism responsible for this emission is the cyclotron maser instability (CMI), which produces a population of unstable electrons. As these electrons precipitate along magnetic field lines, they generate non-thermal radio emission. The CMI can be triggered by several processes, including the interaction between the stellar wind and a planet’s magnetic field—both on the dayside and in the magnetotail (analogous to the Dungey cycle)—as well as the breakdown of plasma corotation and interactions with major satellites.

A classic example of sub-Alfvénic interaction is observed between Jupiter and its Galilean moons, which produces intense radio emissions and auroral footprints on Jupiter's atmosphere (e.g., \citealt{Clarke1996,Zarka1998,Saur2004}). In the solar system, such interactions do not occur between the Sun and its planets, as the orbital separation are large enough that the solar wind remains super-Alfvénic in the planets' reference frames. However, for exoplanets in close orbits around their host stars, the interaction can be sub-Alfvénic, making them promising candidates for detecting analogous radio emissions.

In this study, we investigate the detectability of radio emission generated by sub-Alfvénic interactions between exoplanets and their host stars. We focus on three iconic M-dwarf systems: Proxima Centauri, YZ Ceti and GJ 1151.
We consider both magnetized and unmagnetized (yet conductive) exoplanets. The former case resembles the interaction between Ganymede and Jupiter, while the latter is similar to the interaction between Io (or Europa) and Jupiter. Both types of interactions can produce highly circularly polarized auroral radio emissions.
The radio emission resulting from interactions like those between Io and Jupiter serves as the foundation to search for similar emissions arising from the magnetic interaction between exoplanets and their host stars. In this analogy, the exoplanet plays the role of Io, while the host star plays that of Jupiter. 

%%%%%%%%%%%%%%%%%%%%%%%%%%%%%%%%%%%%%%%%%%%%%%%%%%%%%%%%%%%%
%%%%%%%%%%%%%%%%%%% SIRIO       %%%%%%%%%%%%%%%%%%%%%%%%%%%
%%%%%%%%%%%%%%%%%%%%%%%%%%%%%%%%%%%%%%%%%%%%%%%%%%%%%%%%%%%%
\section{SIRIO}
\label{sec:SIRIO}

\texttt{SIRIO}\footnote{https://github.com/mapereztorres/sirio} (\textbf{S}tar-planet \textbf{I}nteraction and  \textbf{R}adio \textbf{I}nduced \textbf{O}bservations) is a public Python code that predicts the expected radio emission arising from sub-Alfvénic SPI. The code uses a one-dimensional isothermal stellar wind model and incorporates three different stellar wind magnetic geometries: closed dipole, open Parker spiral, and Potential Field Source Surface (PFSS). \texttt{SIRIO} also accounts for free-free absorption of the radio emission within the stellar wind. 

\texttt{SIRIO} can predict the radio emission from magnetic star-planet interaction for any number of exoplanetary systems as a function of stellar wind mass loss and/or exoplanetary magnetic field. \texttt{SIRIO} can also compute the expected radio emission as a function of the exoplanet's separation from its host star, which is useful to study the potential radio emission from systems with no confirmed exoplanet, or whose orbit is poorly known. In this paper, we apply \texttt{SIRIO} to various M dwarfs, and benchmark our results against published literature.
 \texttt{SIRIO} is designed to evaluate the feasibility of radio observations aimed at detecting SPI and to constrain stellar wind mass-loss rates and exoplanetary magnetic fields.
We note that, while 3D magnetohydrodynamic (MHD) codes like PLUTO and BATS-R-US provide more detailed simulations, they are computationally very expensive. In contrast, \texttt{SIRIO}'s 1D approach is computationally very efficient, making it ideal for guiding observational proposals. 
We also note that a key addition of \texttt{SIRIO} with respect to other codes is its incorporation of free-free absorption effects, which may explain non-detections in previous campaigns, such as the GJ 486 observations with the uGMRT \citep{penamonino2025}.

%%%%%%%%%%%%%%%%%%%%%%%%%%%%%%%%%%%%%%%%%%%%%%%%%%%%%%%%%%%%
%%%%%%%%%%%%%%%%% Stellar wind model %%%%%%%%%%%%%%%%%%%%%%%
%%%%%%%%%%%%%%%%%%%%%%%%%%%%%%%%%%%%%%%%%%%%%%%%%%%%%%%%%%%%
\subsection{Stellar wind model and magnetic field geometry}
\label{sec:sw-model}

The current version of \texttt{SIRIO} considers an isothermal stellar wind in which the (ideal) gas is subject to only two forces: an inward force due to the gravity of the star and an outward force due to the gradient gas pressure of the wind \citep{Parker1958}. 
We provide a concise description of the equation of motion for this well-known model in Appendix \ref{app:sw_model}. We only recall here 
that in an isothermal stellar wind,  the stellar wind velocity profile, $\vw (r)$, depends only on the stellar wind temperature, \Tw, and the stellar mass, $M_\star$.
We obtain the stellar wind velocity profile, $\vw (r)$, by solving the equation of motion (\citealt{Cranmer2004}; see also Appendix \ref{app:sw_model}):

\begin{equation}
\vw^2 = 
\begin{cases} 
-a^2 W_0[-D(r)], & \text{if } r \leq r_c \\ 
-a^2 W_{-1}[-D(r)],  & \text{if } r \geq r_c 
\end{cases}
\label{eq:v_sw_profile}
\end{equation}

where $W$ is the Lambert function. Once the stellar wind velocity profile, $\vw (r)$ is known, we can determine the density profile of the stellar wind for a given stellar mass-loss rate, \Mdotstar:

\begin{equation}
    \nw = \frac{\dot{M}_\star}{4\pi\,r^2\,\vw\, \mu m_{\rm p}},
\label{eq:rho-wind}
\end{equation}

%%%%%%%%%%%%%%%%%%%%%%%%%%%%%%%%%%%%%%%%%%%%%%%%%%%%%%%%%%%%
%%%%%%%%% Stellar wind magnetic field geometry   %%%%%%%%%%%
%%%%%%%%%%%%%%%%%%%%%%%%%%%%%%%%%%%%%%%%%%%%%%%%%%%%%%%%%%%%

\texttt{SIRIO} considers three different geometries of the stellar wind magnetic field: a classical pure dipole, a Parker spiral and hybrid model with a transition at the Potential Field Source Surface (PFSS). The pure dipolar geometry considers that all magnetic field lines are closed up to an infinity distance, while the Parker spiral assumes that all magnetic field lines are open and radial, starting at the stellar surface and extending up to infinity. The PFSS model assumes all magnetic field lines are closed up to a source surface radius $R_{\rm SS}$, beyond which they become open and radial. For the Sun, the estimated value of $R_{\rm SS}$ is $2.5 R_{\sun}$ \citep{Altschuler-Newkirk1969}. MHD simulations for early M-dwarfs by \citet{Vidotto2014} show that $R_{\rm SS}$ varies from $\sim$2.8 \Rstar (similar to the solar case) to $\sim$4.6 \Rstar.
We note that \citet{Reville2015} offers an analytical way to estimate the value of \Rss. which, for slowly rotating stars, can be approximated to a pressure balance between the magnetic  pressure of the stellar wind ($P_{\rm B}$) and the sum of the dynamic ($P_{\rm dyn}$) and thermal pressures ($P_{\rm th}$). This value is similar to the Alfvén surface radius on the dipolar geometry, where $P_{\rm B} = P_{\rm dyn}$, just differing by the effect of $P_{\rm th}$, which is usually rather small. 
In this work, we adopt $R_{\rm SS} = 4.5 R_*$, the same value used by \citet{Pineda2023}, thus easily allowing comparisons.
We note that lower (higher) values of \Rss\ would result in significantly higher (lower) values of the radio emission.

%%%%%%%%%%%%%%%%%%%%%%%%%%%%%%%%%%%%%%%%%%%%%%%%%%%%%%%%%%%%
%%%%%%%%% Figure: Sketch of the Alfvén wing      %%%%%%%%%%%
%%%%%%%%%%%%%%%%%%%%%%%%%%%%%%%%%%%%%%%%%%%%%%%%%%%%%%%%%%%%
\begin{figure}%[htb!]
\centering
 \includegraphics[width=0.5\textwidth]{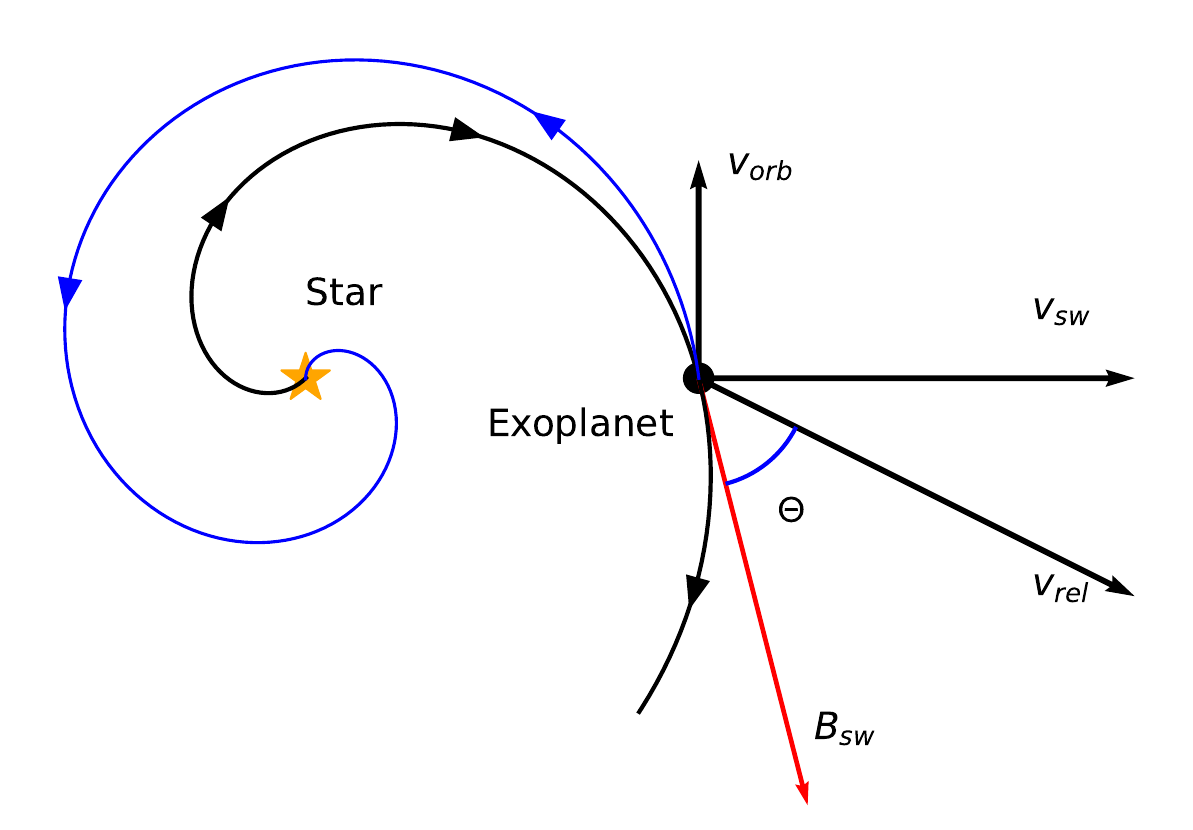}
\caption{Sketch depicting magnetic star planet interaction, adapted from \citet{Turnpenney2018}. The black spiral that originates at the star represents the magnetic field of the Parker spiral, while the blue one, which originates at the planet, represents the Alfvén wing that travels back to the star from the planet. The red colored vector is the magnetic field of the stellar wind at the position of the planet ($\Bsw$), while the three black-colored vectors correspond to the stellar wind velocity ($\vsw$), the orbital velocity of the planet ($\vorb$) and the velocity of the stellar wind with respect to the planet ($\vrel$). The angle between $\Bsw$ and $\vrel$ is defined as $\Theta$. 
Note that in \citet{Saur2013}, $\Theta$ is measured from the normal to the stellar wind magnetic field, making it the complementary angle to the one used in this work.}
{\small 
}
\label{fig:sketch}
\end{figure}%

%%%%%%%%%%%%%%%%%%%%%%%%%%%%%%%%%%%%%%%%%%%%%%%%%%%%%%%%%%%%
%%%%%%%%% Section: Radio emission from SPI       %%%%%%%%%%%
%%%%%%%%%%%%%%%%%%%%%%%%%%%%%%%%%%%%%%%%%%%%%%%%%%%%%%%%%%%%
\subsection{Radio emision from sub-Alfvénic star-planet interaction}   
\label{sec:radio_spi}

The energy that powers the observed ECM radio emission due to star-planet interaction comes from the Poynting flux,
$S_{\rm Poynt}$, generated at the orbit of the planet and transferred back to the star, provided the planet is in the sub-Alfvénic region. The current version of \texttt{SIRIO} implements three of the most popular models of sub-Alfvénic star-planet interaction: the Alfvén wing model (see e.g. \citealt{Zarka2007, Saur2013}), the reconnection model \citep{Lanza2009} and the stretch-and-break model \citep{Lanza2013,Strugarek2022}.
In the Alfvén wing model, the Poynting flux released by the star-planet interaction propagates back to the host star via one of the Alfvén wings (see figure \ref{fig:sketch} for an idealized depiction), 
whereas in the reconnection model, 
the Poynting flux is released through magnetic reconnection
at the boundary of the planetary magnetosphere.
In the stretch-and-break model, reconnection is induced by the orbital motion of the planet, which twists the magnetic field lines and triggers stronger reconnection events.
In all three scenarios, the final result is a release of magnetic energy, close to the surface of the host star. Motivated by the interaction of Jupiter with its natural satellites, we assume that a part of this energy is converted into kinetic energy of accelerated electrons and, if the conditions are adequate for the temporal formation of an electron cyclotron maser, coherent radio emission is generated. 

The Poynting flux released by sub-Alfvénic interaction is $S_{\rm Poynt}
\propto \, B_\perp^2\, v_{\rm rel} \, R_{\rm eff}^2$, where 
 $v_{\rm rel}$ is the
relative velocity between that of the stellar wind flow, $\vw$, and that of the planet 
(Fig.~\ref{fig:sketch}); $B_\perp = B_{\rm sw} \, {\rm sin}\Theta$, is the
component of the stellar wind magnetic field perpendicular to the plasma velocity at
the location of the planet; and  
$R_{\rm eff}$ is the effective radius of the exoplanet.  
In the Alfvén wing model, the planet may be magnetized, or unmagnetized, as long as the (unmagnetized) planet is an electric conductor. Therefore, $\Reff \geq \Rp$. In the reconnection scenario, not only must the planet be magnetized for the model to work, but the intensity of the magnetic field needs to be large enough, so that a planetary magnetosphere is formed. If those conditions are not met, the reconnection model cannot work. The same requirements apply to the stretch-and-break model.

The Poynting flux for the Alfvén wing model is given by (e.g \citealt{Saur2013}):

\begin{equation}
    \label{eq:Saur_eq55}
    S_{\rm Alf} 
    = \frac{1}{2}(\Reff\, \overline{\alpha}\,M_{\rm A}\,B_{\rm sw} \ \rm {sin\Theta})^2\,v_{\rm A}, 
\end{equation}

where $R_{\rm eff}$ is the effective radius of the planet, $M_{\rm A}$ is the Alfvén Mach number, $B_{\rm sw} $ is the magnetic field of the stellar wind and $v_{\rm A}$ is the Alfvén speed; $\Theta$ is the angle between the relative wind speed and the stellar wind magnetic field, as defined in figure \ref{fig:sketch}; and  $\overline{\alpha}$ is a measure of how effective the interaction is. For the reconnection model, the expression is given by \citep{Lanza2009}:

\begin{equation}
\label{eq:Lanza_eq8}
S_{\rm rec} = \frac{\gamma}{4\epsilon\ }  (B_{\rm sw} \Rmp \rm {sin\Theta})^2 v_{\rm rel}, 
\end{equation}
where   0 < $\gamma$ < 1, and its exact value depends on the angle between the interacting magnetic field lines, reaching its maximum when the magnetic fields of the planet and the star are anti-aligned.
We note here that the expression in \citet{Lanza2009} is for the dissipated power, which is related to the Poynting flux as follows: $P_{\rm d} = \epsilon \, S_{\rm Poynt}$, where the efficiency $\epsilon = 0.2 \pm 0.1$, based 
on measurements for the Earth’s magnetosphere and satellite-Jupiter interactions (see \citealt{Zarka2024} and references therein).

Finally, the Poynting flux in the stretch-and-break model is defined as (e.g. \citet{Strugarek2022}):

\begin{equation}
\label{eq:stretch_and_break}
    S_{\text{sb}} = \frac{1}{2} R_{\text{obs}}^2 \xi^{-2} f_{\text{ap}}\ \rm {sin^2\Theta}\  B_{\text{sw}}^2 v_{\text{rel}}
\end{equation}

where $\xi = B_{\text{sw}}/B_{\text{pl}}$ is the ratio between the magnetic field of the stellar wind and the planet, and $f_{\text{ap}}$ is the fractional area of the planetary disc where magnetic field lines are connected to the stellar wind, which can be obtained from the following expression \citep{Adams2011}:

\begin{equation}
\label{eq:f_ap}
    f_{\text{ap}} = 1 - \sqrt{1 - \left(\frac{3 \, \xi^{1/3}}{2 + \xi}\right)}
\end{equation}

The ratio of the Poynting flux in the reconnection model to that in the Alfvén wing model is

\begin{equation}
        \label{eq:flux_ratio}
    \frac{S_{\rm rec}}{S_{\rm Alf}} = \frac{\gamma}{2\,\epsilon\,\overline{\alpha}^2\, M_{\rm A}\,}
\end{equation}

Similarly, the ratio of the Poynting flux in the stretch-and-break model to that in the reconnection model is

\begin{equation}
        \label{eq:flux_ratio_REC_vs_SB}
    \frac{S_{\rm rec}}{S_{\rm sb}} = \frac{\gamma \xi^2}{2\epsilon f_{\text{ap}}}
\end{equation}

Since essentially in all realistic cases $\xi$ is much less than unity, the value of this ratio is mostly driven by the factor $\xi^{-2}$.
In this paper, we fix $\gamma = 1$, its maximum value, 
and adopt $\epsilon = 0.2$, motivated by the value obtained for the Io-Jupiter interaction, which is in the range from 0.15 to 0.30 \citep{Zarka2007}.
Finally, we use $\overline\alpha = 1$, as the exact value for each of the three cases is indistinguishable from unity (see Appendix \ref{app:alpha_interaction}).
Thus, $S_{\rm rec}/S_{\rm Alf} =  2.5/\MA$ and 
$S_{\rm rec} / S_{\rm sb} = 2.5\, \xi^2/f_{\rm ap}$.
This has important implications in the predicted fluxes, as we discuss in the next section.

The total radio power received at the Earth can is $P_R = \beta\,S_{\rm Poynt}$, where $\beta$ is the efficiency factor in converting Poynting
flux to ECM radio emission. Although the value of $\beta$ is poorly known, it is expected to be 
in the range from $10^{-4}$ to $10^{-2}$, with a nominal value of 
$\beta = 10^{-3}$ \citep{Zarka2024}, which we assume throughout the paper.
Finally, the radio flux density is

 \begin{equation}
   \label{eq:flux_density}
   F_R = \frac{P_R}{\Omega\,D^{2} \,\Delta\nu}, 
 \end{equation}
 
\noindent
where $\Omega$ is the solid angle into which the ECM radio emission is beamed, $D$ is the distance to the star from Earth, and $\Delta\nu$ is the total bandwidth of the ECM emission. Observations of the Io-DAM emission indicate that the beaming solid angle of a single flux tube is of about 0.16 sr \citep{Kaiser2000,Queinnec2001}. However, various characteristics can lead to a larger solid beam of about a few times that value, including a thick emission cone wall, a planetary plasma wake, or a distorted stellar magnetic field. On the other hand, there is no reason to consider a full auroral oval around the star, which in the case of Jupiter yields a solid beam angle of 1.6 sr (see, e.g., \citealt{Zarka2004}). We therefore used $\Omega$ values in the range from 0.16 str (the size of a single magnetic flux tube for the Jupiter-Io system) to 3 times that value, 0.48 str.
We make the standard assumption that ECM radio emission has $\Delta\nu = \nu_{\rm g}$, where $\nu_g \approx 2.8\, B_\star$ MHz is the cyclotron frequency, and $B_\star$ is the average surface magnetic field strength of the star.

%%%%%%%%%%%%%%%%%%%%%%%%%%%%%%%%%%%%%%%%%%%%%%%%%%%%%%%%%%%%
%%%%%%%%% Table - Sample of stellar systems      %%%%%%%%%%%
%%%%%%%%%%%%%%%%%%%%%%%%%%%%%%%%%%%%%%%%%%%%%%%%%%%%%%%%%%%%

\begin{table*}%[htb!]
\centering
\caption{Stellar and exoplanetary parameters used in our simulations}
\label{tab:sample}  
\begin{threeparttable}
\begin{tabular}{|l|l|l|l|l|l|l|l|l|l|l|l|l|l|l|}
\hline
Star & Spectral & D & $R_\star$ & $M_\star$ & $P_{\rm rot}$ & $B_*$ & Planet & $M_{\rm p}$ & $R_{\rm p}$ & $P_{\rm orb}$ & $a$ & $B_{\rm pl}$ & $T_c$ & $\dot{M}$ \\
name & type & pc & [$R_{\odot}$] & [$M_{\odot}$] & [days] & [G] & name & [$M_{\oplus}$] & [$R_{\oplus}$] & [days] & [au] & [G] & [$10^6$\,K] & [\Mdotsun]\tnote{$\dagger$}  \\
\hline
Proxima \tnote{a}  & M5.5 & 1.3 & 0.15 & 0.12 & 82.6 & 600 & Proxima b & 1.3 & 1.1 & 11.19 & 0.05 & 0.05  & 2.0 & 1.5 \\ %turnpenney
                   &      &     &      &      &      &     &           &     &     &       &      &       &     & 0.20 \tnote{b} \\ %reville
                   &      &     &      &      &      & 200\tnote{c} &           &     &     &       &      &       &     & 0.25 \tnote{c} \\ %kavanagh
YZCet  \tnote{d}  & M4.5 & 3.7 & 0.16 & 0.14 & 68.5 & 220 & YZCet b & 0.7 & 0.89 & 2.02 & 0.02 & 0.20  & 1.5 & 5\tnote{e} \\
 & & &  & & &  & &  &  &  &  &  &  & 0.25\tnote{e}\\
GJ1151 \tnote{f} & M4.5 & 8.04 & 0.2 & 0.17 & 125 & 50\tnote{g}  & GJ 1151 b  \tnote{h} & 0.73\tnote{i}  &  & 1 &  &0.35  & 1.6 & 1 \\
 &  &  &  &  & & &  & 1.25\tnote{i} &  & 5 &  & 0.11 &  & \\

\hline
\end{tabular}
\begin{tablenotes}
\footnotesize
\item[$\dagger$] We assume throughout the paper  mass-loss rate value for the Sun \Mdotsun = $2\times 10^{-14} \Msun$ yr$^{-1}$.
\item[a] Stellar and planetary parameters from \citet{Turnpenney2018}.
\item[b] Stellar mass-loss rate estimated by \citet{Reville2024}.
\item[c] Values taken from \citet{Kavanagh2021}.
\item[d] Stellar and planetary parameters from \citet{Pineda2023}. %The two $\Mdotstar$ values correspond to the models A and B.
\item[e] Mass loss rate values corresponding to models A and B from \citet{Pineda2023}. 
\item[f] GJ1151 stellar parameters from \citet{Vedantham2020}.
\item[g] Stellar magnetic field for which the detection reported by \citet{Vedantham2020} would be caused by electron-cyclotron radiation.
\item[h] Non-confirmed planet.
\item[i] GJ1151 planetary constraints from \citet{Blanco-Pozo2023}.
\end{tablenotes}
\end{threeparttable}

\end{table*}

%%%%%%%%%%%%%%%%%%%%%%%%%%%%%%%%%%%%%%%%%%%%%%%%%%%%%%%%%%%%
%%%%%%%%% Exoplanetary magnetic field and R_eff  %%%%%%%%%%%
%%%%%%%%%%%%%%%%%%%%%%%%%%%%%%%%%%%%%%%%%%%%%%%%%%%%%%%%%%%%
\subsection{Exoplanetary magnetic field and effective radius}
\label{sec:Bplanet}

The effective radius of the exoplanet, \Reff \,, in Eqs. \ref{eq:Saur_eq55} and \ref{eq:Lanza_eq8}, depends on the nature of the planet. If the planet is unmagnetized, then \Reff = \Rp \, for the Alfvén model, and \Reff \, = 0 for the reconnection model, as the stellar wind magnetic field lines cannot reconnect. However, if the planet is magnetized, and its intensity is high enough that it forms a magnetosphere above the surface of the planet,  \Reff  \,is then equal to the magnetopause radius, \Rmp \, .
We determine \Rmp \, by balancing the exoplanetary magnetic pressure with the sum of the dynamic (ram), thermal, and magnetic stellar wind pressures: 

\begin{equation}
    \label{eq:Rmp}
    \Rmp = \kmp^{1/3} \Bigg[\frac{(B_{\rm pl}/2)^2/8\pi}{ \,n_{\rm sw}\,(\mu\,m_{\rm p}\,v_{\rm rel}^2 + k_B\,T_{\rm sw}) + \Bsw^2/8\pi} \Bigg]^{1/6} \Rp.    
\end{equation}

We estimate the (unknown) exoplanetary magnetic fields, $\Bp = \mathfrak{M}/  \Rp^3$, using the empirical scaling law  of \citet{Sano1993}, where the planetary dipole magnetic moment, $\mathfrak{M}$ is 

\begin{equation}
   \mathfrak{M}  \propto \Omega_{\rm pl} \rho_{\rm core}^{1/2} r_{\rm core}^{7/2}, 
   \label{eq:Sano}
\end{equation}

where 
$\rho_{\rm core}$ is the outer core density of the planet, defined as $M_{\rm pl} / R_{\rm pl}^3$, and $r_{\rm core}$ is the radius of the planet core, which can be determined from the exoplanetary mass by using the scaling law of \citet{CurtisNess1986} ($r_{\rm core} \propto M_{\rm pl}^{0.44}$). 
 We assume that the planets are tidally locked, so that the rotational speed of each planet is equal to its orbital speed. 
Hence, our values of $\Bp$ in Table~\ref{tab:sample} are lower limits to their true exoplanetary magnetic field values.

\subsection{Free-free absorption within the stellar wind}
\label{app:free-free}

The thermal electrons in the stellar wind will partially absorb the outgoing radio emission via the free-free mechanism, thus attenuating the radio signal by a factor of $e^{-\tau_{\nu}}$, where the optical depth, $\tau_{\nu}$, is defined as \citep{Cox2000}

\begin{equation}
    \label{eq:ff_tau}
    \tau_\nu =  \int_{R_\star}^{ \infty } \kappa_\nu \,dz ,
\end{equation}
\noindent

The integration is performed along the line of sight, starting from the location where the emission originates—conservatively assumed to be the stellar surface—up to the observer at infinity. In practice, we integrate up to 10$^4$ \Rstar, beyond which the effect of free-free absorption becomes negligible in all cases.
The absorption coefficient $\kappa_\nu$ in Eq.~\ref{eq:ff_tau} is defined as (e.g., \citealt{Cox2000}):

\begin{equation}
    \label{eq:ff}
    \kappa_\nu = 3.692 \times 10^{8} \left(1 - e^{-h \nu / k_B T}  \right) Z^{2}\, g T^{-1/2}\, \nu ^{-3}\, n_e\,n_p \, 
%\bigr[\mathrm{cm^{-1}}\bigr],
\end{equation}
\noindent

where $\nu$ is the emission frequency, $h$ is Planck's constant, $k_B$ is Boltzmann's constant, $Z$ is the ionization state,  $T$ is the temperature of the medium (assumed to be $T_c$ for an isothermal wind); and 
$g$ is the Gaunt factor, which in the radio regime is 
$g = 10.6 + 1.9  \log _{10} (T) - 1.26 \log _{10} (Z \nu )$ \citep{Cox2000}.
The parameters $n_e$ and $n_i$ represent the number densities of electrons and ions, respectively. 
Since we assume a fully ionized hydrogen plasma, $Z = 1$ and $n_i = n_p = n_e$, where $n_p$ is the proton number density. Hence $\mu=1/2$. 
All quantities in the above equations are in cgs units.

%%%%%%%%%%%%%%%%%%%%%%%%%%%%%%%%%%%%%%%%%%%%%%%%%%%%%%%%%%%%
%%%%%%%%% Case studies                           %%%%%%%%%%%
%%%%%%%%%%%%%%%%%%%%%%%%%%%%%%%%%%%%%%%%%%%%%%%%%%%%%%%%%%%%
\section{Case studies}
\label{sec:case-studies}

In this section, we apply \texttt{SIRIO} to several well-known star-planet systems to demonstrate its capabilities and validate its predictions against existing models in the literature. We focus on three iconic M-dwarf systems--Proxima Centauri, YZ Ceti, and GJ~1151--which have been the subject of intense observational campaigns searching for radio signals of star-planet interaction, and benchmark our results against those obtained by other teams, e.g.
\citet{Turnpenney2018} (Proxima Cen), \citet{Pineda2023} (YZ Cet), or
\citet{Vedantham2020} (GJ~1151). Since we consider some of the effects that were not taken into account in several, or all of those papers (e.g. free-free absorption in the wind, or a PFSS geometry), we highlight not only the agreements, but also the differences in the obtained results, which in some cases lead to significantly different conclusions. For the sake of simplicity, we assume in all cases an efficiency factor, $\beta$, in converting Poynting flux to radio emission, to 10$^{-3}$ and, as mentioned earlier, we use a range of values for the solid angle covered by the ECM emission, $\Omega$, from 0.16 sr (as in the case of the flux tube for the Jupiter-Io interaction; e.g. \citealt{Zarka2004}) up to three times this value.

%%%%%%%%%%%%%%%%%%%%%%%%%%%%%%%%%%%%%%%%%%%%%%%%%%%%%%%%%%%%
%%%%%%%%%%%%%%%%%%% Proxima Cen  %%%%%%%%%%%%%%%%%%%%%%%%%%%
%%%%%%%%%%%%%%%%%%%%%%%%%%%%%%%%%%%%%%%%%%%%%%%%%%%%%%%%%%%%
\subsection{Proxima b}
 \label{sec:Proxima}

%%%%%%%%%%%%%%%%%%%%%%%%%%%%%%%%%%%%%%%%%%%%%%%%%%%%%%%%%%%%%%%%%%%%%%
%%%%%%%%% Fig. stellar wind parameters -  Proxima  - Parker geometry
%%%%%%%%%%%%%%%%%%%%%%%%%%%%%%%%%%%%%%%%%%%%%%%%%%%%%%%%%%%%%%%%%%%%%%
\begin{figure*}%[htbp!]
\centering
   \includegraphics[width=\textwidth]{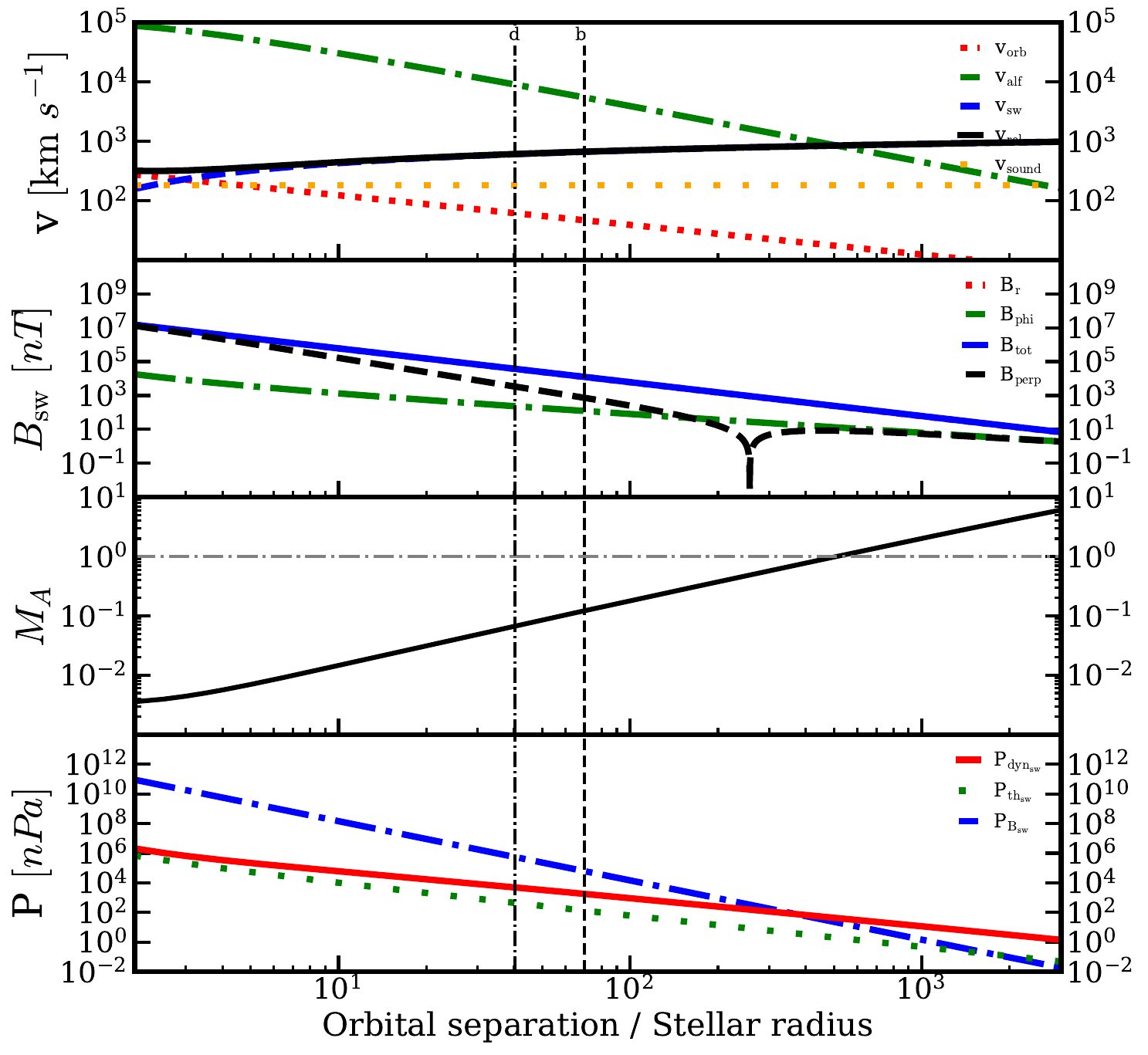}
\caption{
{\small Stellar wind parameters for Proxima Centauri vs. orbital separation, for an open Parker spiral geometry. \textit{Top panel:}  Keplerian orbital speed \vorb\ (red close-dotted line),  Alfvén speed \vA\ (green dash-dotted line),  stellar wind speed \vsw\ (dashed blue line), relative speed between the stellar wind and the exoplanet \vrel\ (black solid line), and the (constant) sound speed $a$ (orange sparse-dotted line). 
\textit{Top-middle:} Stellar wind magnetic field components: radial (red dotted line), azimuthal (green dash-dotted line), total (blue solid line) and perpendicular (black dashed line). 
\textit{Bottom-middle:} Alfvén Mach number \MA, where the horizontal dashed line drawn at $\MA = 1$ signaling the transition from the sub-Alfvénic to the super-Alfvénic regime.
\textit{Bottom:}
 Dynamic pressure (red solid line), thermal pressure (green dotted line) and magnetic pressure (blue dash-dotted line) of the stellar wind.
The vertical lines across all four panels correspond to the orbits of the known planets around Proxima Centauri, Proxima b (dashed line) and Proxima d (dash-dotted line).
}}
 \label{fig:proxima_b_parker_diagnostic}
 \end{figure*}
%%%%%%%%%%%%%%%%%%%%%%%%%%%%%%%%%%%%%%%%%%%%%%%%%%%%%%%%%%%%%%%%%%%%%%

%%%%%%%%%%%%%%%%%%%%%%%%%%%%%%%%%%%%%%%%%%%%%%%%%%%%%%%%%%%%%%%%%%%%%%
%%%%%%%%% Fig. diagnostic plots Proxima  - pure dipolar and PFSS %%%
%%%%%%%%%%%%%%%%%%%%%%%%%%%%%%%%%%%%%%%%%%%%%%%%%%%%%%%%%%%%%%%%%%%%%%
\begin{figure*}%[htbp!]
\centering
 \includegraphics[width=8.8cm]{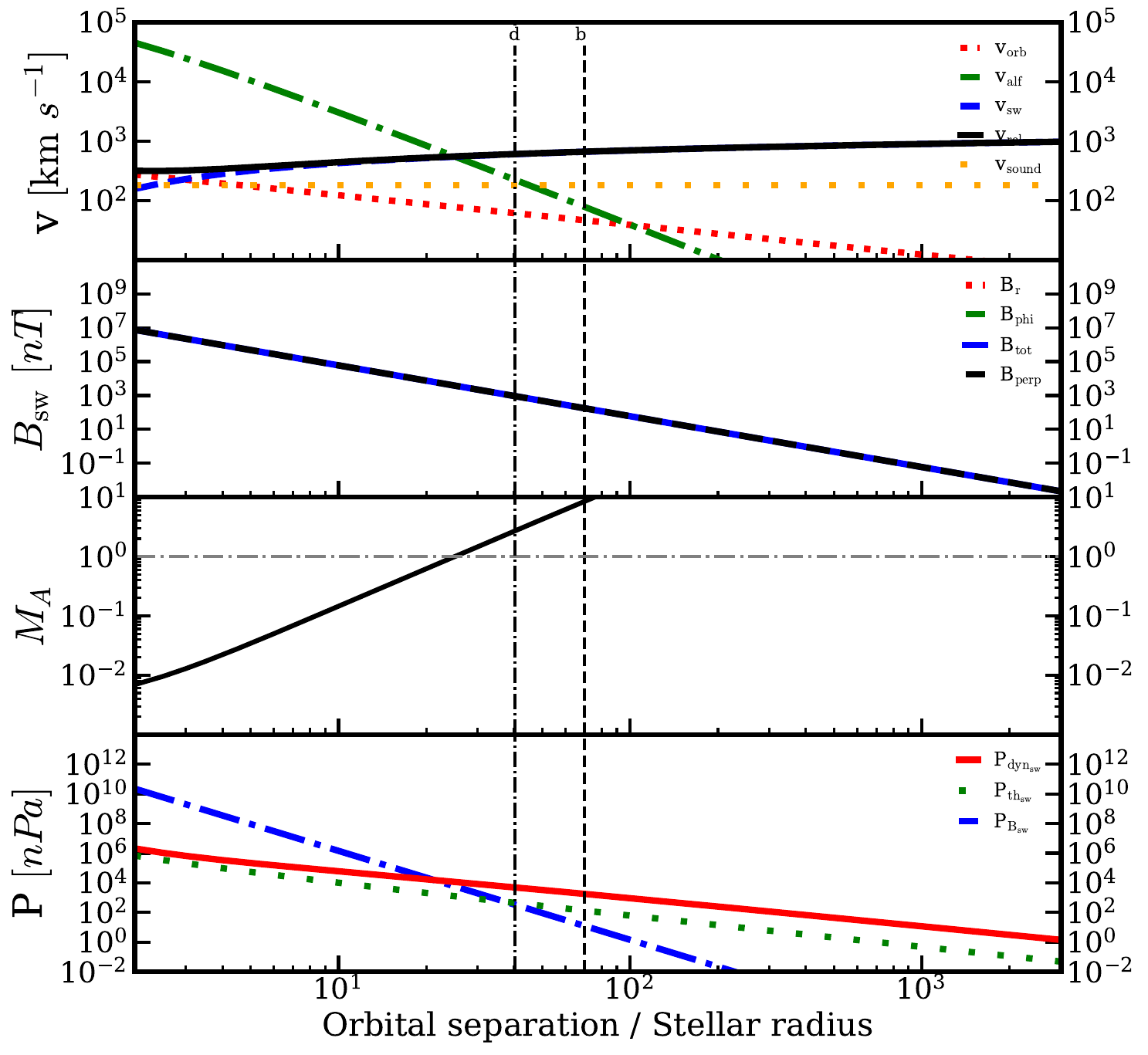}
 \includegraphics[width=8.8cm]{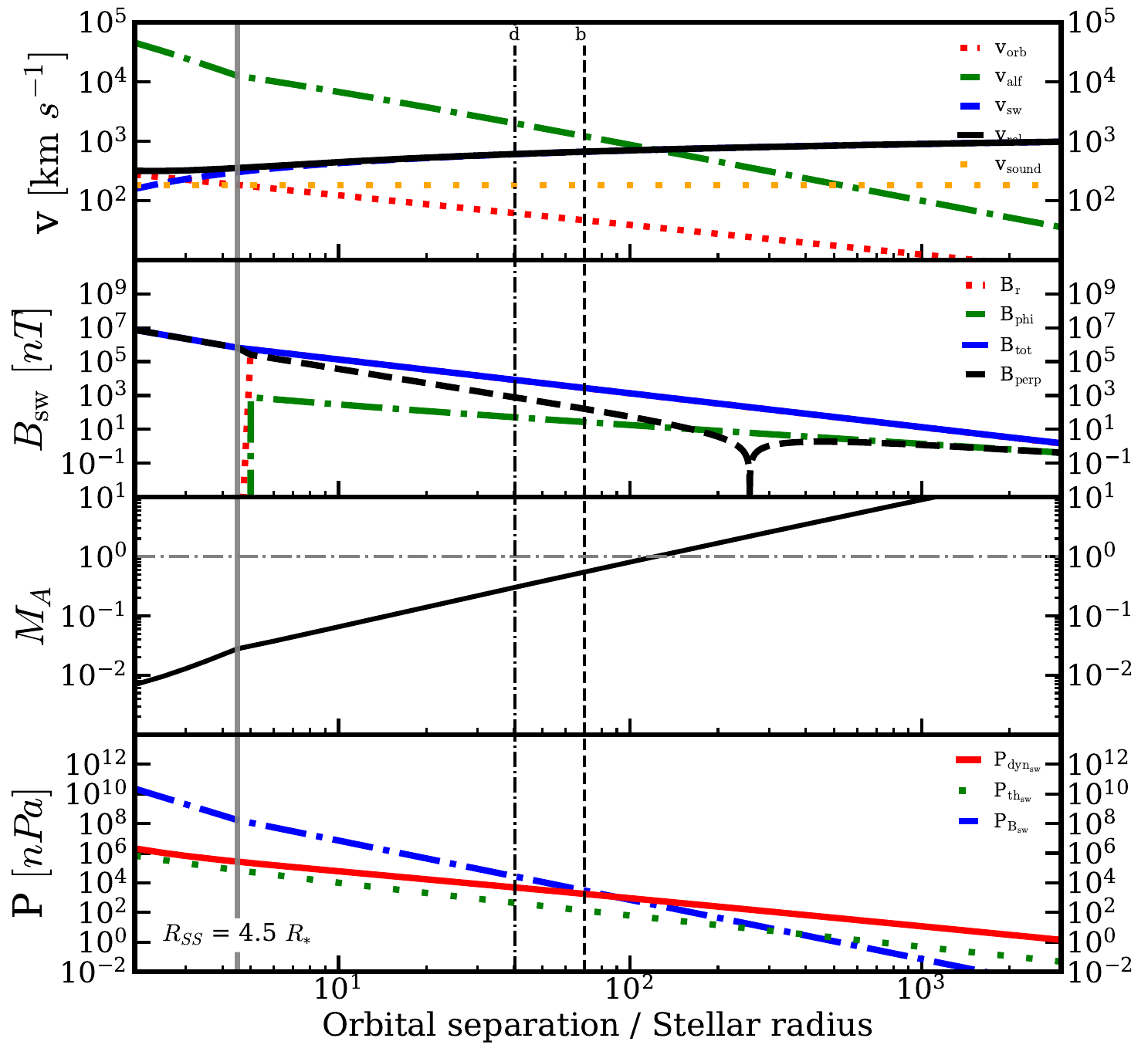}
\caption{
{\small Same as in Figure \ref{fig:proxima_b_dipole_pfss_diagnostic}, but for a purely dipolar stellar wind geometry (left), and a PFSS geometry (right). The solid grey line crossing all four PFSS plots are drawn at an orbital separation equal to $\Rss$. For values smaller than $\Rss$, a purely dipolar regime applies.
}
\label{fig:proxima_b_dipole_pfss_diagnostic}
}
\end{figure*}

Proxima Centauri, the closest star to the Sun, hosts at least two Earth-like exoplanets: Proxima b \citep{Anglada-Escude2016} and Proxima d \citep{suarezmascareno2020}, making it a key target for star-planet interaction studies. 
In fact, \citet{PerezTorres2021} detected highly circularly polarized emission occurring at the expected cyclotron frequency  that happened approximately at about half the orbital period of Proxima b, which they interpreted as evidence 
of sub-Alfvénic star-planet interaction, as the observing frequency of the signal ($\nu \sim 1.7$ GHz) agreed very well with the expected cyclotron frequency for a stellar magnetic field of $\sim 600$ G \citep{Reiners2008}.
Here, we apply \texttt{SIRIO} to Proxima b and compare our results with those published in \citet{Turnpenney2018}, \citet{Kavanagh2021} and \citet{Reville2024}. 

We first address the comparison of the results from \texttt{SIRIO} with those published in \citet{Turnpenney2018}, who assumed a Parker spiral geometry for the stellar wind magnetic field.  To this end, we generated the same set of diagnostic plots as in their paper. 
Fig.~\ref{fig:proxima_b_parker_diagnostic} shows the velocity components, stellar wind magnetic field, Alfvén Mach number, and pressure, drawn versus orbital separation, using the same units as in \citet{Turnpenney2018} to facilitate comparisons. Our results are essentially identical to those in Fig.~9 of their work.

Figure~\ref{fig:proxima_b_dipole_pfss_diagnostic} shows the 
diagnostic plots for the two other magnetic field geometries considered: a pure dipole (left panel) and a PFSS (right panel).
In the case of a pure dipole, $\Bsw$ decreases much faster than in the case of the open Parker spiral, as the azimuthal component of the dipole varies with $r^{-3}$ compared to the $r^{-2}$ dependence of the radial component in a Parker spiral. This implies that the Alfvén speed, $\vA$,  decreases much faster. Since the relative speed of the planet with respect to that of the stellar wind, $\vrel$, is the same in both geometries, this implies that the super-Alfvénic regime, when $\MA > 1$, is reached much earlier, at an orbital separation of about 25\,\Rstar. 
We can also see this reflected in the pressure diagnostic plot, since the magnetic pressure of the stellar wind decreases far more sharply than in the open geometry, and $\Pdyn = \Pmagn$ happens at a lower orbital separation. In this case, both Proxima b and d would be in the Super-Alfvénic regime, thus precluding the possibility of transporting energy and momentum back to the star.

The PFSS geometry shows, as expected, an intermediate behavior between that of an open Parker spiral and a closed dipolar geometry. Note that for orbital separations smaller than $\Rss$, the diagnostic plots are identical to those of a pure dipolar geometry. Beyond \Rss, the behavior follows that of an open Parker spiral, albeit with a significant difference: the magnetic field of the stellar wind is lower due to the effect of the PFSS. This implies Alfvén speeds  considerably smaller than in the open Parker spiral case, and therefore the super-Alfvén regime starts closer to the star. 
While \proxb\ and \proxd\ are well within the sub-Alfvénic regime, the more realistic PFSS geometry places both planets close to the Alfvén surface.

\begin{figure*}
\centering

   \includegraphics[width=5.7cm]{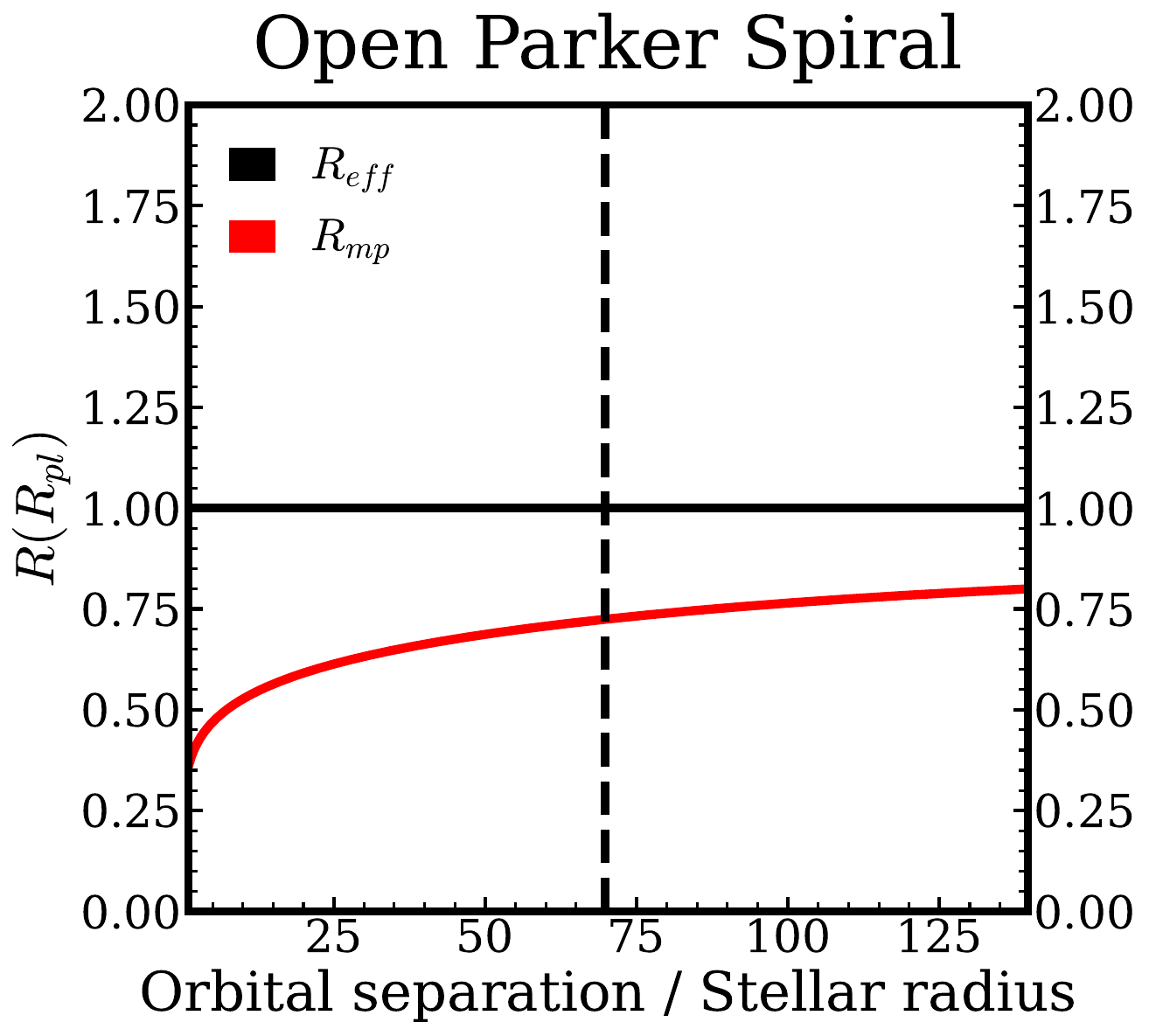}
   \includegraphics[width=5.7cm]{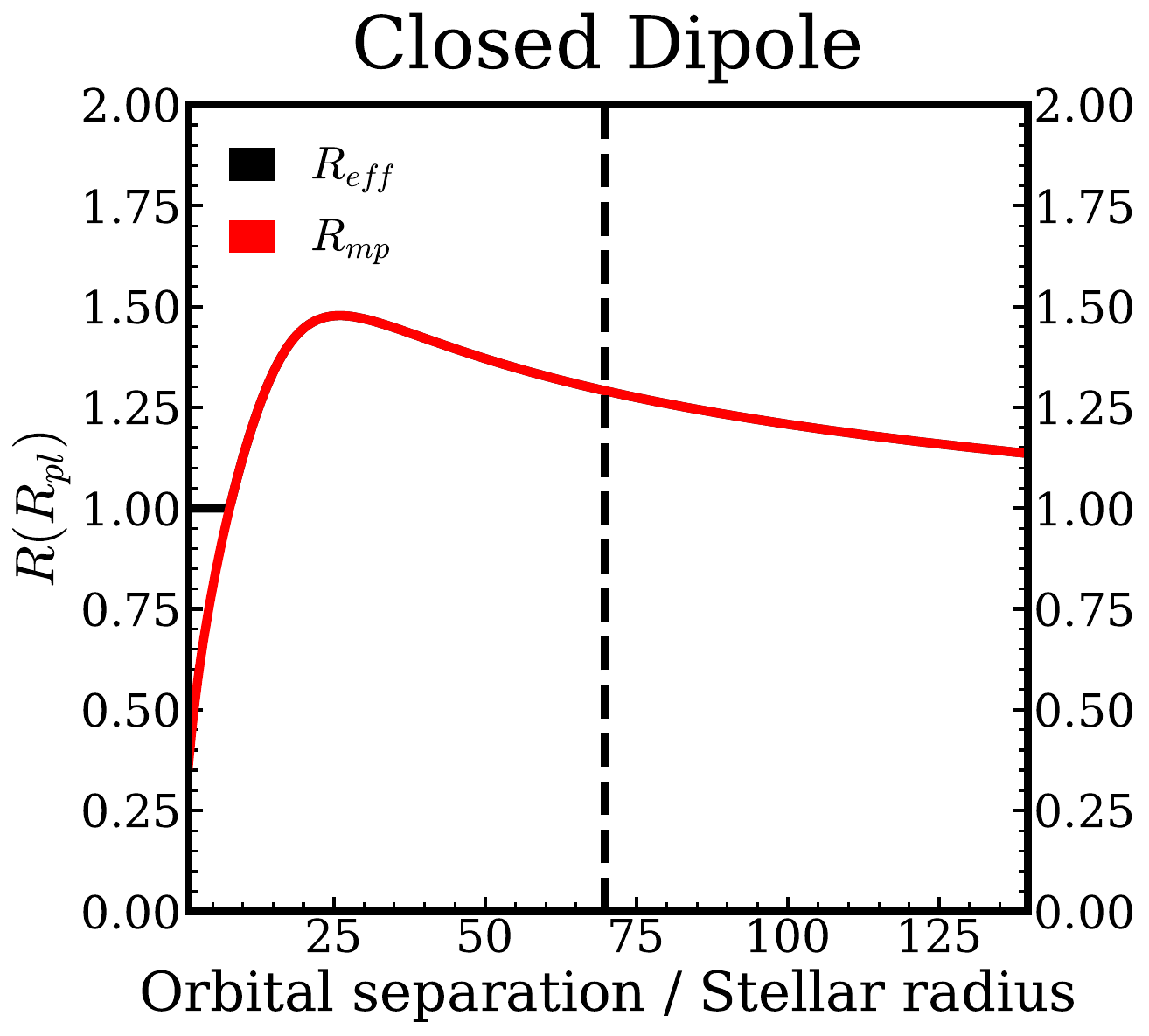}
   \includegraphics[width=5.7cm]{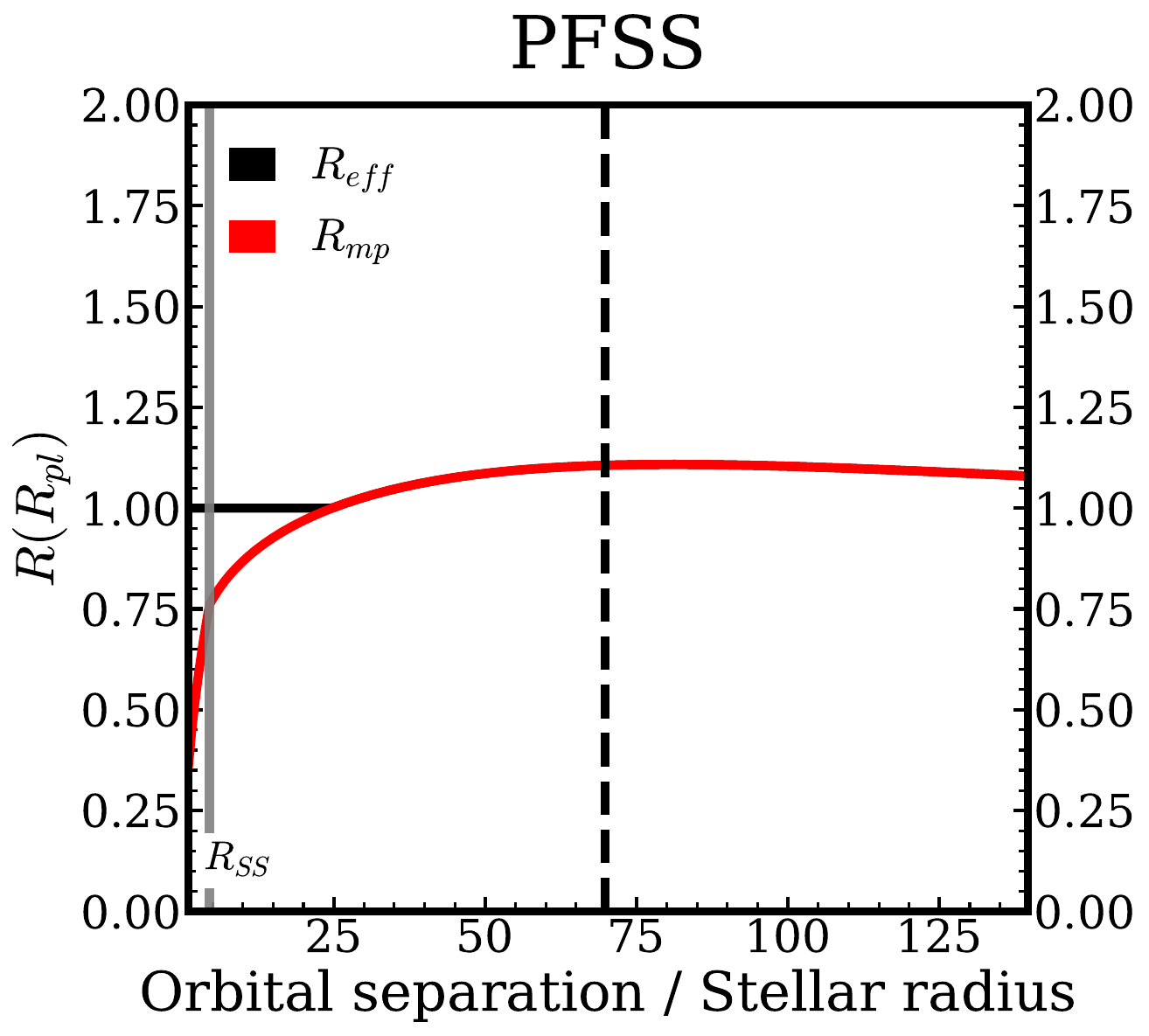}

   \includegraphics[width=5.8cm]{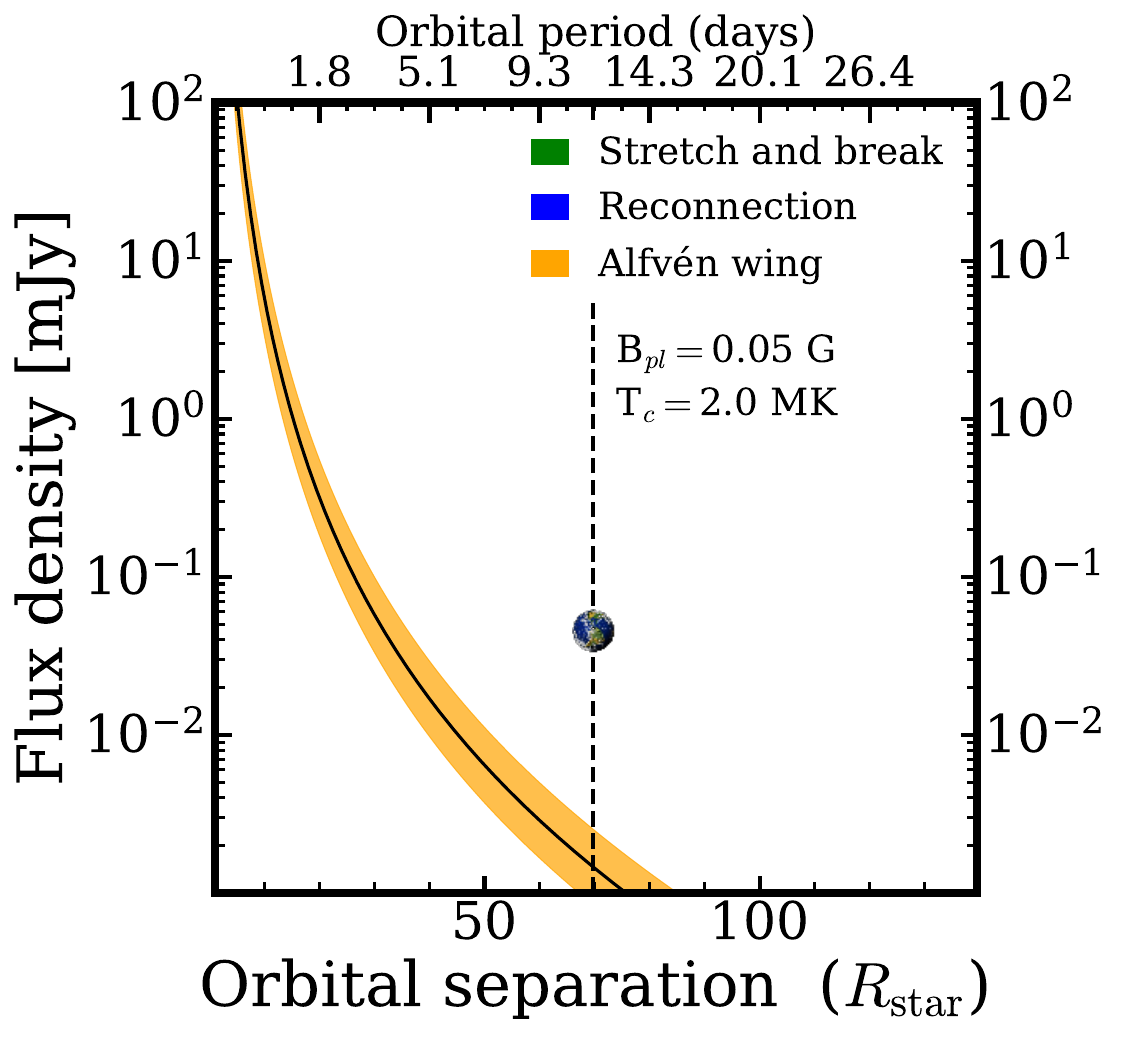}
   \includegraphics[width=5.8cm]{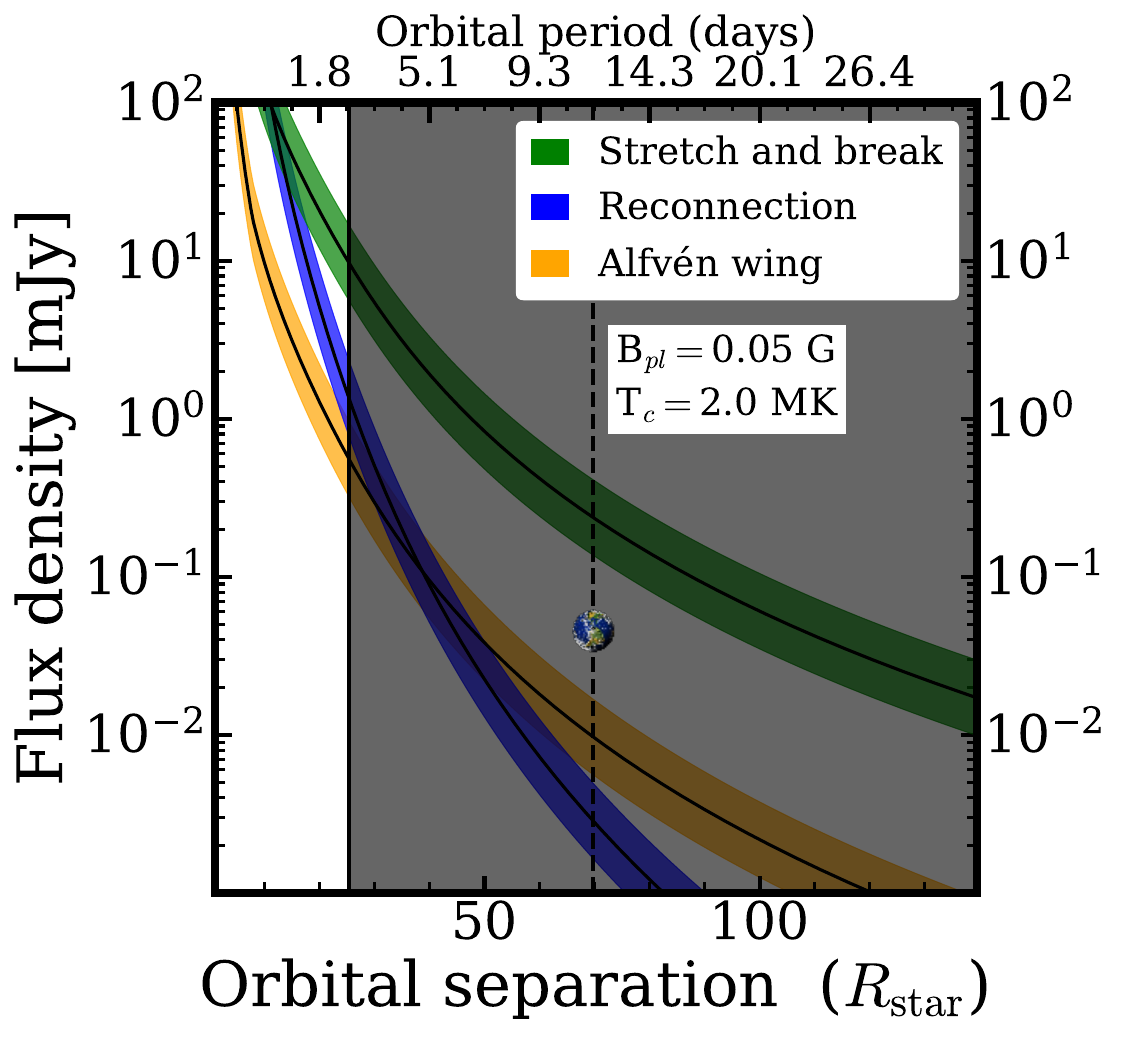}
   \includegraphics[width=5.8cm]{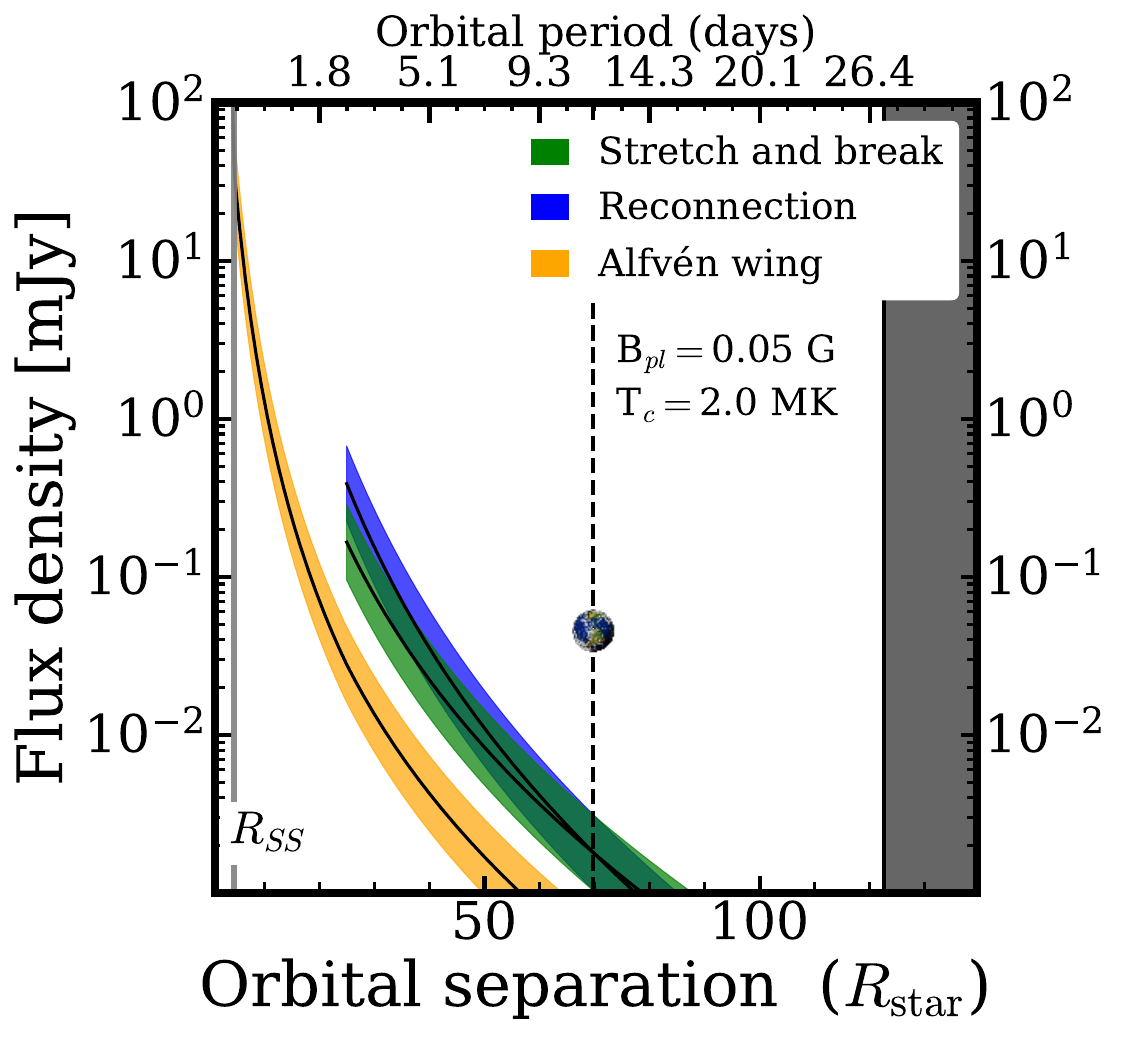}

   \includegraphics[width=5.8cm]{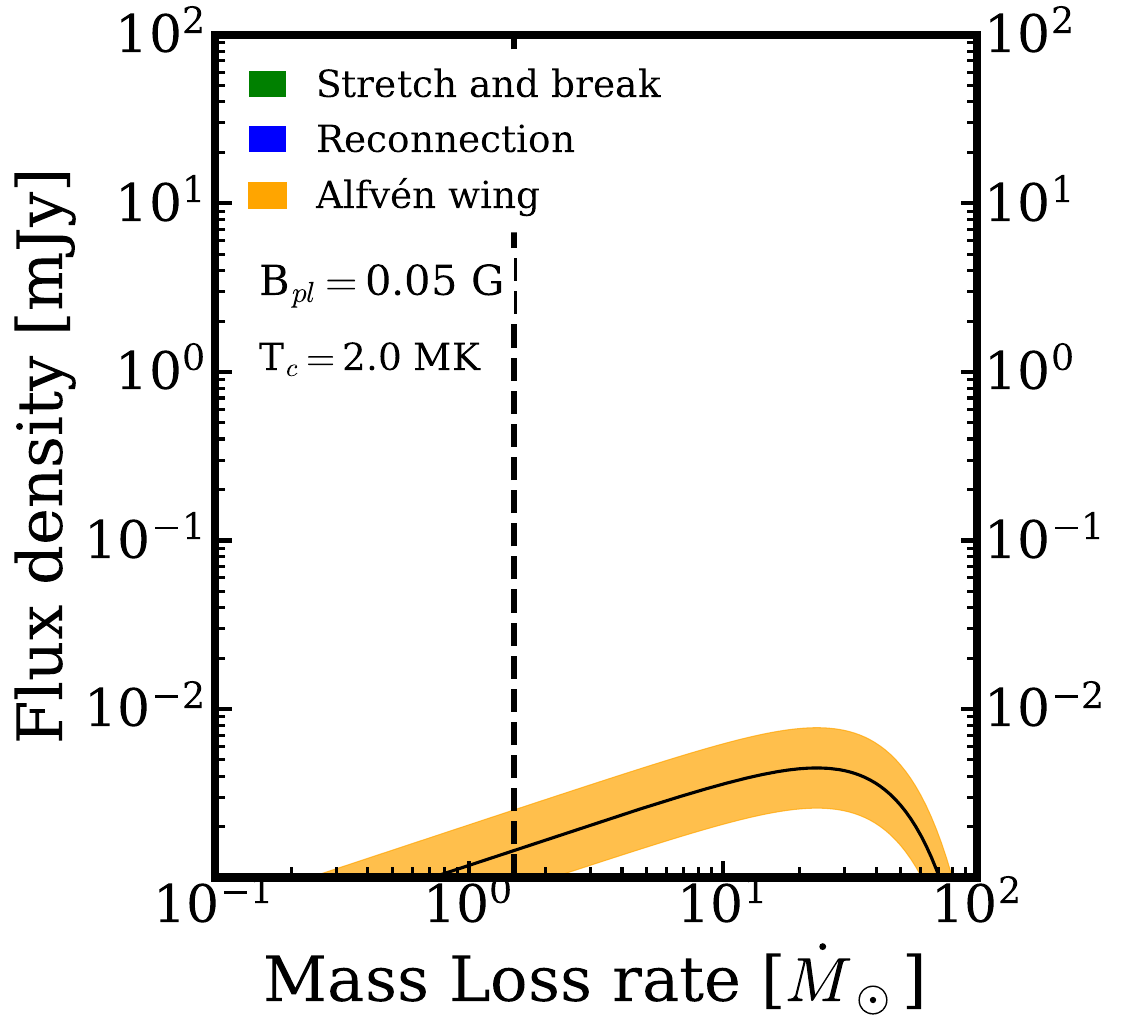}
   \includegraphics[width=5.8cm]{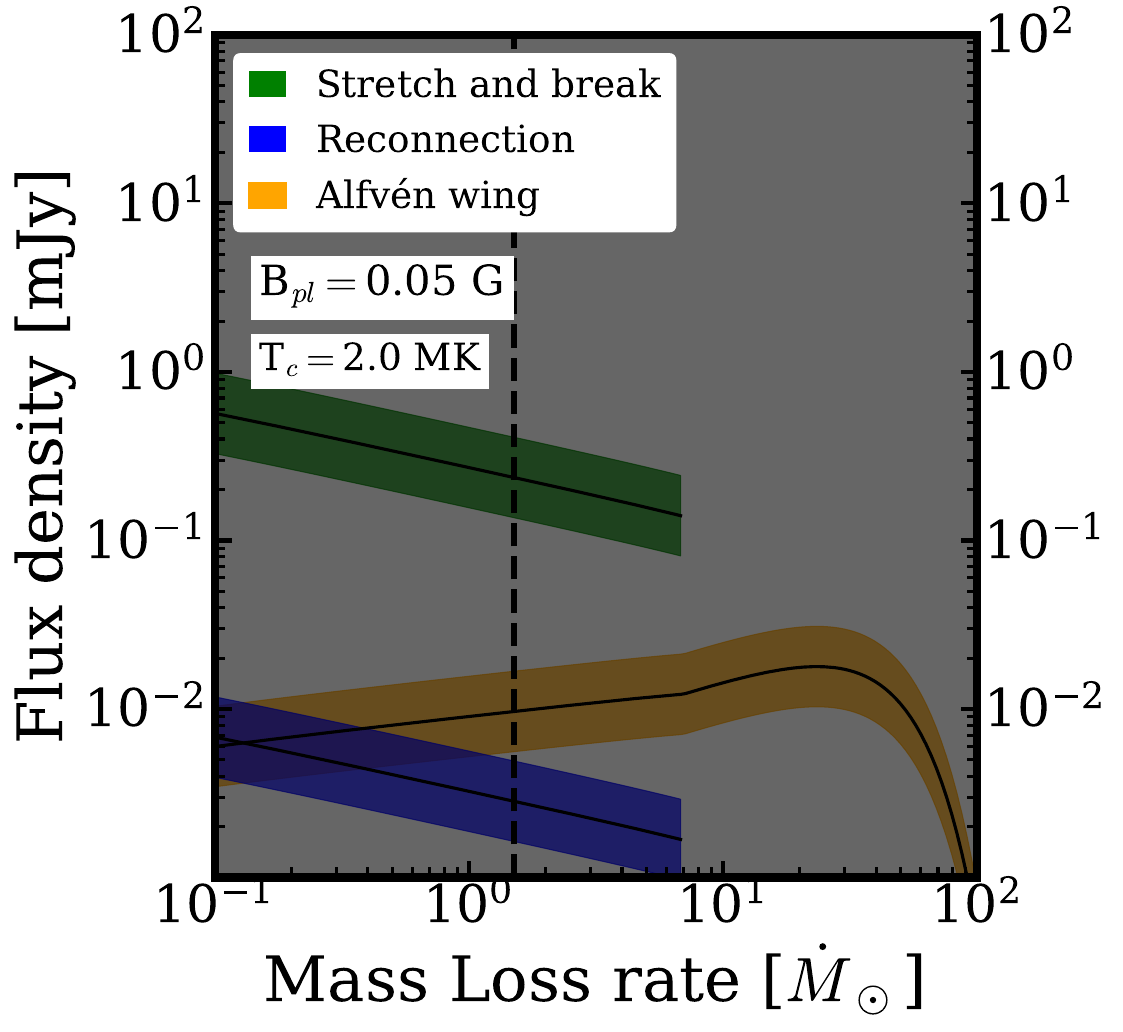}
   \includegraphics[width=5.8cm]{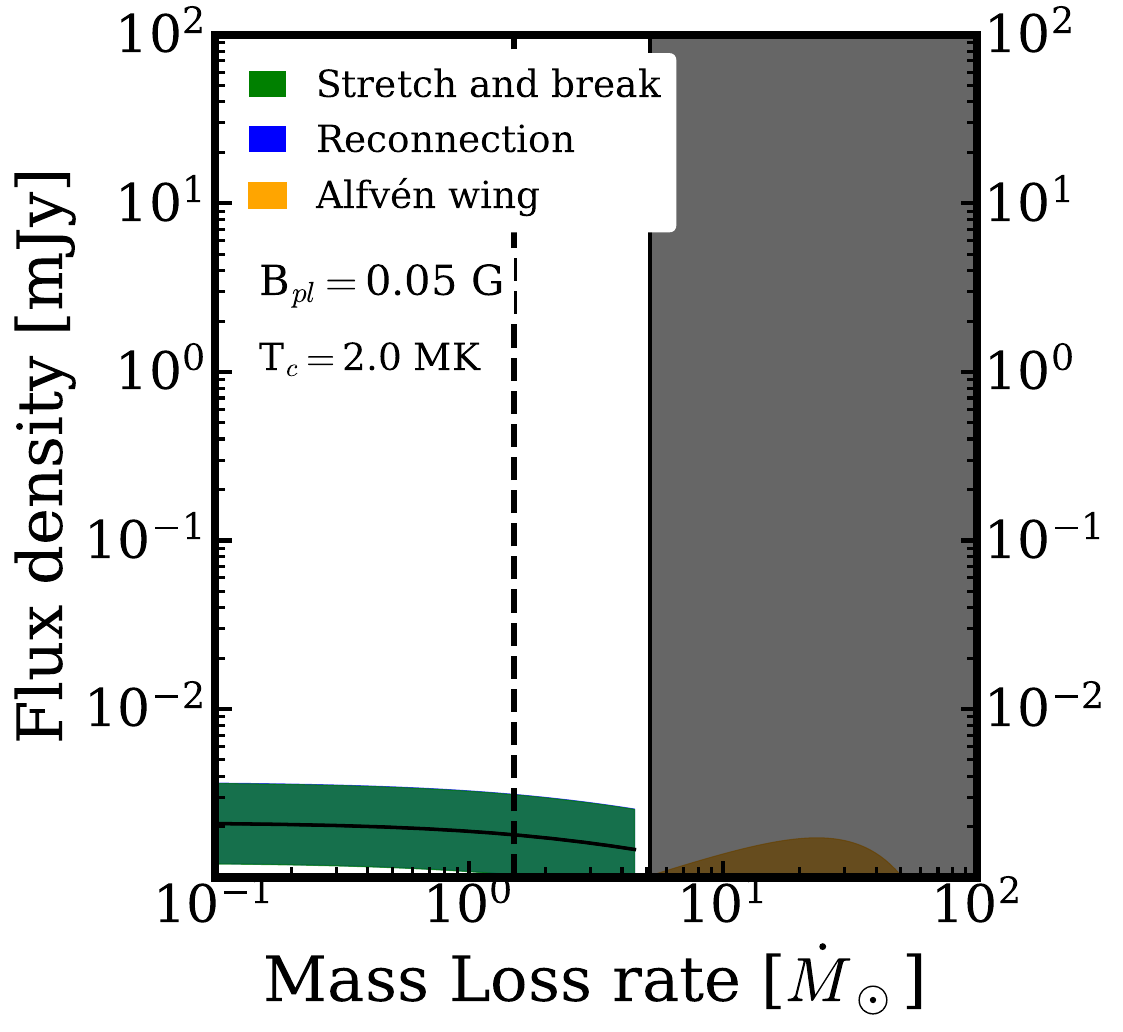}

   \includegraphics[width=5.8cm]{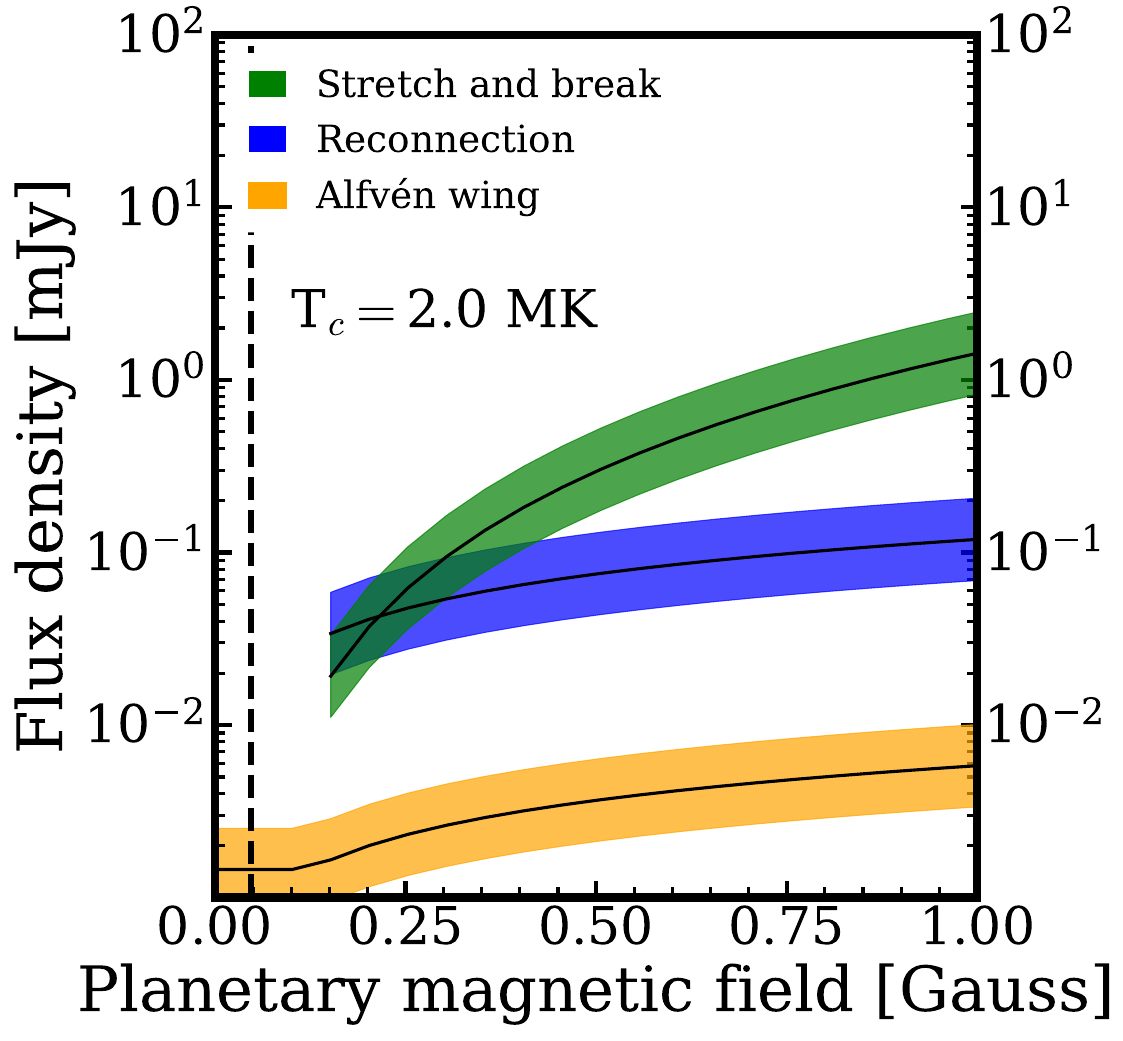}
   \includegraphics[width=5.8cm]{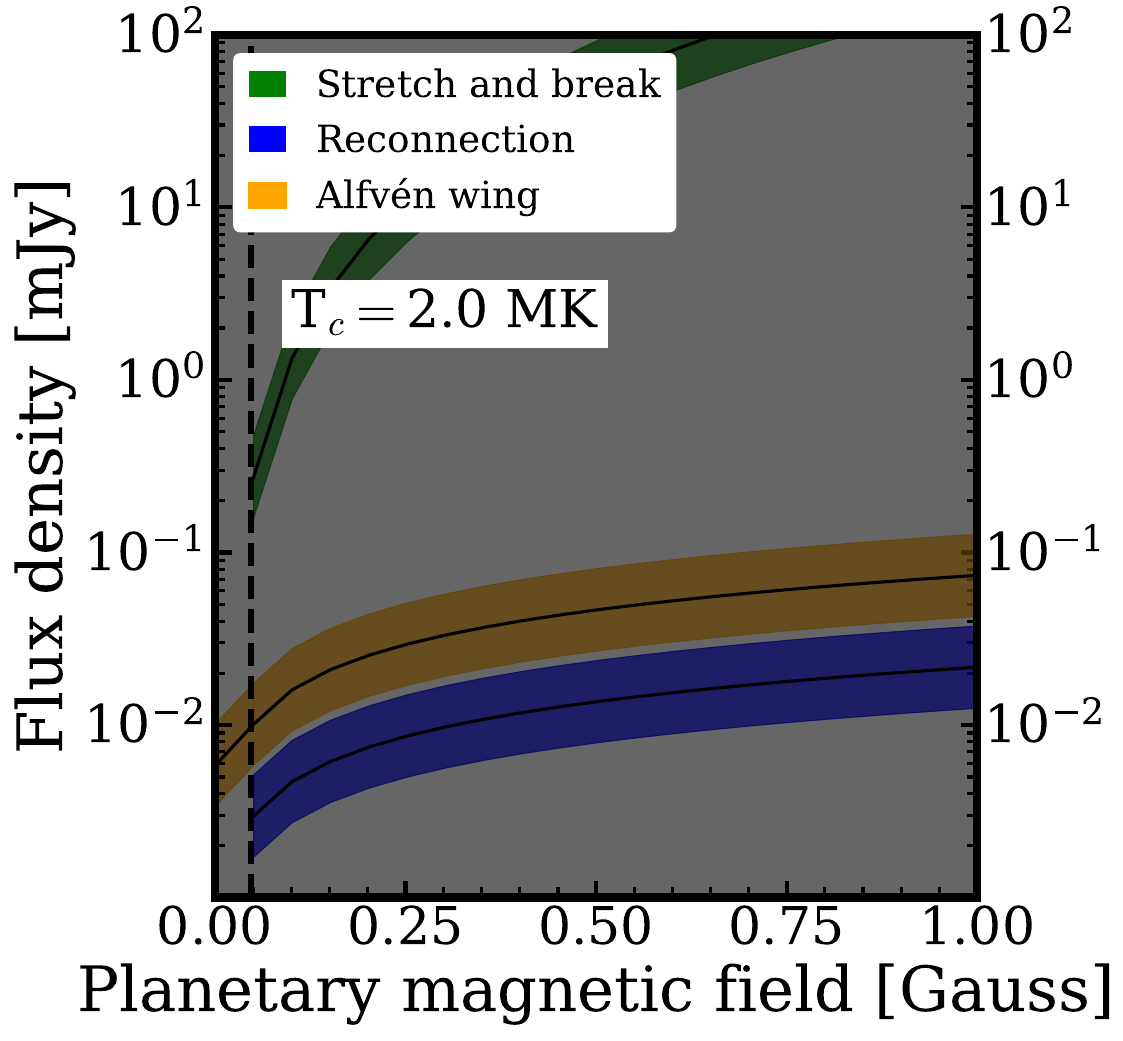}
   \includegraphics[width=5.8cm]{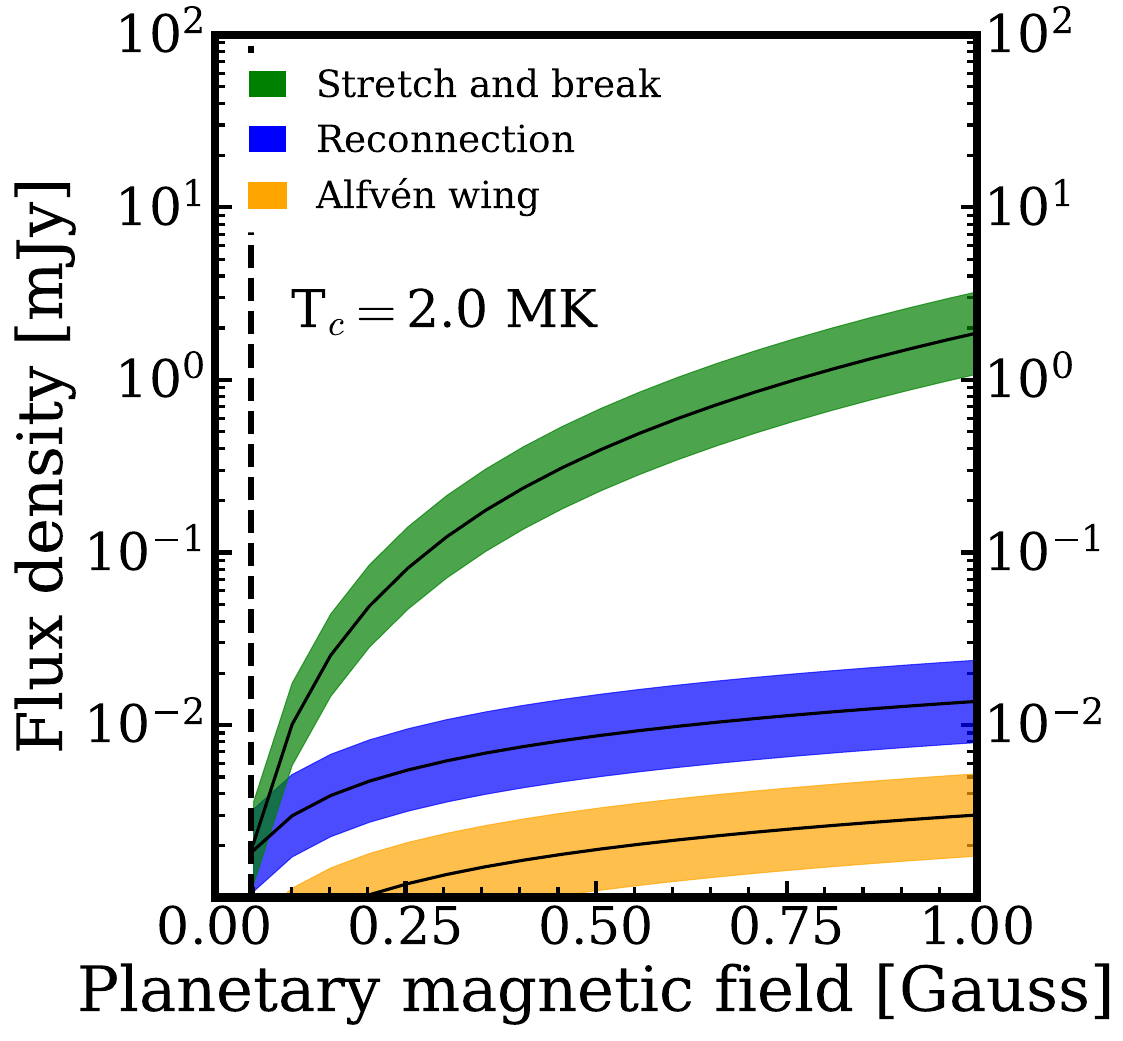}

\caption{Effective radius of Proxima Centauri (top panels) and predicted flux density from SPI as a function of orbital separation, mass-loss rate and exoplanetary magnetic field, for three different geometries of the stellar wind magnetic field: Parker spiral (left), dipolar (middle), and PFSS (right).  The black lines in the top panels correspond to the computed magnetosphere radii using Eq.~\ref{eq:Rmp}, while the red lines correspond to  the effective radii of the interaction in the Alfvén wing model. $\Reff$ coincides with \Rmp\  if  $\Rmp \geq R_{\rm pl}$, otherwise $\Reff = \Rp$. The predicted radio emission for the stretch-and-break, reconnection, and Alfvén wing models are shown in green, blue, and orange, respectively. Each colored area corresponds to the emission for an opening cone with an aperture ranging from that of the Jupiter-Io interaction (0.16 strrad) to three times that value. The vertical dashed lines shown in each figure correspond to the nominal values of orbital separation (first and second rows),  and the reference values of stellar mass-loss rate (third row) and exoplanetary magnetic field (fourth row). Dark-grey shaded regions indicate that the planet is in the super-Alfvénic regime, while the light-grey vertical line in the PFSS geometry corresponds to \Rss =4.5 \Rstar.}
\label{fig:Proxima_b_flux}
\end{figure*}

Fig.~\ref{fig:Proxima_b_flux} illustrates the capabilities of \texttt{SIRIO}, for each 
of the three magnetic field geometries considered, for the case of Proxima - Proxima b: Parker spiral (left panels), pure dipolar (middle panels), and PFSS (right panels).
In the first row, we show plots of the effective and magnetopause radii of the exoplanet, as a function of its orbital separation.  In the second row, we show 
the predicted flux density arising from magnetic star-planet interaction between Proxima b and its host star, also as a function of the orbital separation, for the reference values of stellar wind mass-loss rate and exoplanetary magnetic field (see Table~\ref{tab:sample}). We notice that, although the orbital separation of Proxima b is well-known (see Table~\ref{tab:sample}), predicting radio emission as a function of orbital separation is a particularly useful tool when dealing with cases of unconfirmed, or unknown, planets, as the case of GJ~1151 b that we discuss in Sect.~\ref{sec:gj1151}.
Finally, we show in the last two rows the dependence of the predicted flux density as a function of the stellar wind mass loss rate, \Mdotstar,  and exoplanetary magnetic field, \Bp, respectively. We show the flux density predictions 
for  the Alfvén wing model (yellow shaded areas), the reconnection model (blue shaded areas) and stretch-and-break model (green shaded areas), and take into account free-free absorption effects in all simulations. 

We note that the magnetopause radius, \Rmp\ (red line) of \proxb, as determined by 
Eq.~\ref{eq:Rmp}, is always smaller than \Rp\ (black solid line) for the Parker spiral (top left panel), independently of the planet's orbital separation. This is due to the low value of the reference magnetic field (\Bp $\approx$ 0.05 G; see Table \ref{tab:sample}). As a consequence, only the Alfvén wing model is able to yield radio emission from SPI, since in this scenario  the mere existence of an obstacle (the planet) causes a perturbation---independently of whether the planet is magnetized or not---which in turn generates an Alfvén wave.  On the contrary, 
the reconnection and stretch-and-break models require that
\Rmp\ (black line) must be equal or larger than that of the exoplanet, \Rp,  so that reconnection can happen. If this condition is not fulfilled, the effective radius of the planet is zero, and no radio emission arises in this case. 

Hence, in the Parker spiral geometry, \Rmp\ is always less than \Rp\ for the reference value of the magnetic field, and therefore only the Alfvén wing model is viable. 
For the pure dipolar and PFSS geometries, \Rmp\ is larger than \Rp\ at the orbital separation of Proxima, hence all three models (Alfvén wing, reconnection and stretch-and-break) are viable. However, the planet would be in the super-Alfvénic regime in the dipolar case, so no radio emission from SPI is expected in that geometry.  We also note that the predicted radio emission is very low. Indeed, the Alfvén wing model predicts values around the $\mu$Jy level in a Parker spiral geometry, and well below that value for a PFSS geometry. The reconnection and stretch-and-break model predict very similar values, around a few $\mu$Jy (PFSS geometry). 

For our study of the SPI radio signal as a function of the stellar wind mass loss rate, we used values of $\Mdotstar$  from 0.1 up to 100 \Mdotsun, which encompass the whole range of values inferred for M dwarfs \citep{Wood2021}. 
The predicted flux density shows a similar behavior to that shown in the predictions as a function of the orbital separation: if the Alfvén wing model is operating, chances of detecting \proxb\ in a Parker spiral geometry, or in the more realistic PFSS geometry are meager, given the expected faintness of the radio signal. 
We also emphasize  that the predicted flux density increases with mass-loss rate in the Alfvén wing model, while it decreases both for the reconnection and stretch-and-break models. In fact, 
$S_{\rm rec} \propto {\rm R}^2_{\rm mp}$ and 
$S_{\rm Alf} \propto {\rm R}^2_{\rm eff}\,\MA^2\,\vA \propto {\rm R}^2_{\rm eff}\, \rho^{1/2}_{\rm sw}$. Since ${\rm R}^2_{\rm mp}$  decreases (very slowly) as the  ram pressure increases (Eq.~\ref{eq:Rmp}), the reconnection model predicts smaller flux densities as the stellar wind mass loss rate increases, while the  Alfvén wing model predicts increasingly higher values up to the point where free-free absorption starts to become relevant. The stretch-and-break model shows a similar dependence to the reconnection model, so the same comments apply to that model. In fact, as Eq.~\ref{eq:flux_ratio_REC_vs_SB} shows, for the given values of orbital separation and planetary magnetic field, the only difference between the predictions of the stretch-and-break and the reconnection models is a constant factor.
All of the above highlights the importance of the availability of stellar wind mass-loss rates for assessing the feasibility of detecting radio emission from sub-Alfvénic star-planet interaction.

We also note that the predicted flux density in the Alfvén wing model increases with stellar wind mass-loss rate approximately as $\dot{M}_\star^{1/2}$, due to the scaling $S_{\rm Poynt} \propto R_{\rm eff}^2 \rho^{1/2}$, which follows from the isothermal Parker wind assumption. In more general models, such as polytropic winds or full MHD simulations, the density profile typically falls off more steeply with distance (e.g., \citealt{Cranmer2004, Suzuki2013}), reducing both the local density and the Alfvén speed at the planet’s location. This would modify the scaling, potentially yielding a weaker dependence on density, i.e., $S_{\rm Poynt} \propto R_{\rm eff}^2 \rho^\alpha$ with $\alpha < 1/2$.

For our study of the predicted flux as a function of the planetary magnetic field (bottom panels in Fig.~\ref{fig:Proxima_b_flux}), we used  the range from 0 to 4 G for all planets in Table~\ref{tab:sample}, which allows tackling exoplanetary cases ranging from non-magnetized planets to Earth-like planets and up to super-Earths. Those plots illustrate clearly two aspects of the radio emission. First, as soon as the exoplanetary magnetic field is strong enough to form a magnetopause with \Rmp$\geq$\Rp, the reconnection and stretch-and-break models are at work. 
Second, the three models yield significantly different predictions. 
The reconnection and Alfvén wing models differ by a constant factor, as seen in Eq.~\ref{eq:flux_ratio}. For the open magnetic field geometry, this factor is larger, while in the dipolar is smaller. This discrepancy is explained by the value of \MA\ differing from one geometry to another. For the Parker spiral case (Fig.~\ref{fig:proxima_b_parker_diagnostic}), \MA\ reaches lower values, resulting in the largest ratio $S_{\rm rec}/{S_{\rm Alf}}$. Conversely, in the dipolar case, \MA\ is highest, making the difference between the reconnection and Alfvén wing model predictions smaller (Fig.\ref{fig:proxima_b_dipole_pfss_diagnostic}, left panel). The PFSS hybrid geometry (Fig.~\ref{fig:proxima_b_dipole_pfss_diagnostic} right) shows an intermediate behavior between the two other geometries. The stretch-and-break model also predicts substantially higher fluxes, and steeper dependence on the exoplanetary magnetic field strength (as it scales with $B_{\rm pl}^2$), as Eq.~\ref{eq:flux_ratio_REC_vs_SB} shows.

We notice here that 
the stretch-and-break model is the only one predicting values in agreement with with the tentative detection of radio emission from SPI in Proxima - Proxima b, as claimed in \citet{PerezTorres2021}, if the planet has a magnetic field similar to that of the Earth, or larger (bottom panels in Fig.~\ref{fig:Proxima_b_flux}). On the contrary, 
the Alfvén wing model predicts negligible radio emission from SPI for any realistic exoplanetary magnetic field or and stellar wind mass loss rate, and the reconnection model fells short by a factor of a few for a Parker spiral geometry, and is negligible in the case of the PFSS geometry.

%%%%%%%%%%%%%%%%%%%%%%%%%%%%%%%%%%%%%%%%%%%%%%%%%%%%%%%%%%%%%%%%%%
%%%%%%%%%%% Plots of R_eff/R_mp vs. d_orb
%%%%%%%%%%%   and of flux density vs. d_orb, M_dot and B_planet
%%%%%%%%%%%%%%%%%%%%%%%%%%%%%%%%%%%%%%%%%%%%%%%%%%%%%%%%%%%%%%%%%%

%%%%%%%%%%%%%% Fig. M_A - Proxima
%%%% BEGIN
\begin{figure*}
\centering
\includegraphics[width=0.68\columnwidth]{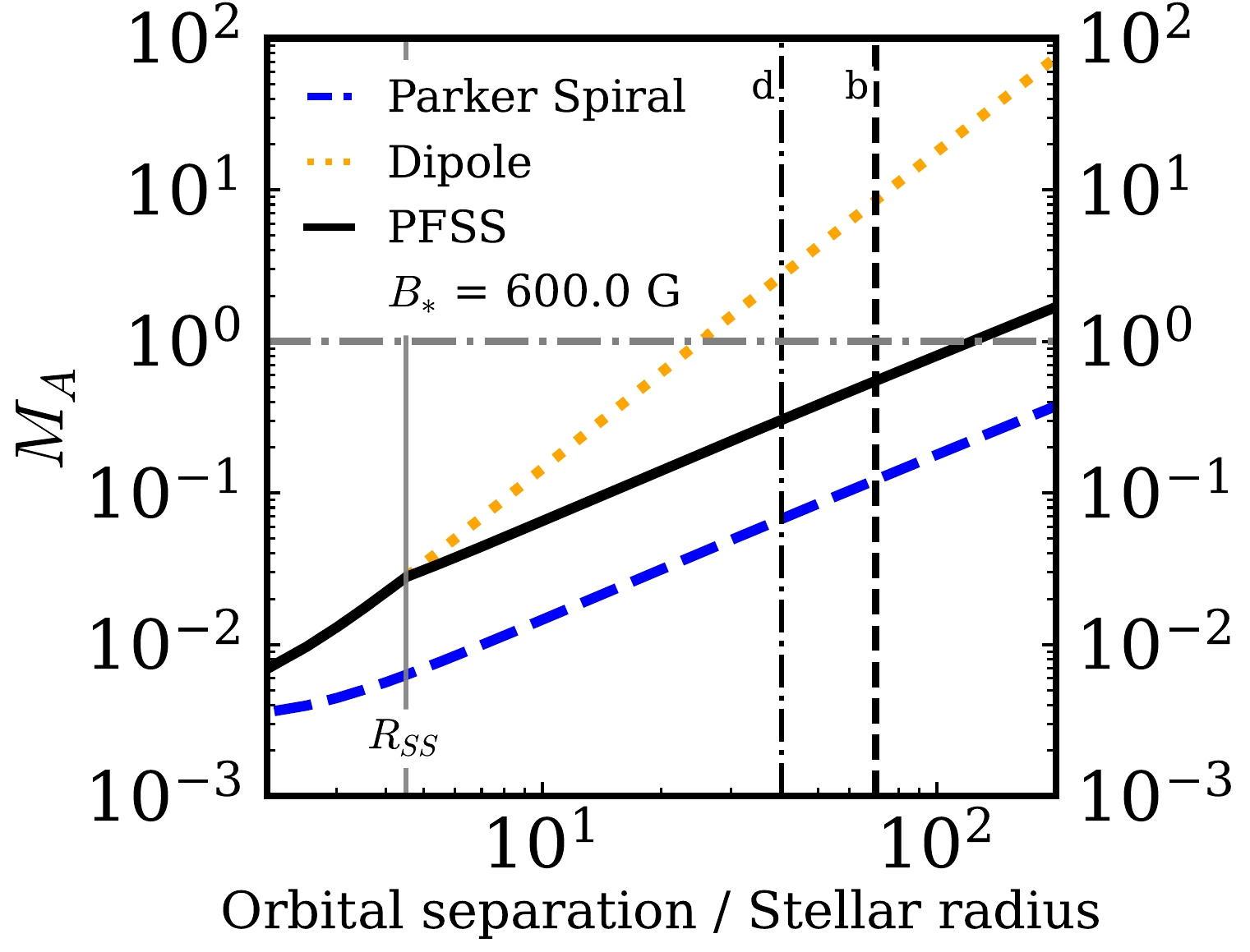}
\includegraphics[width=0.68\columnwidth]{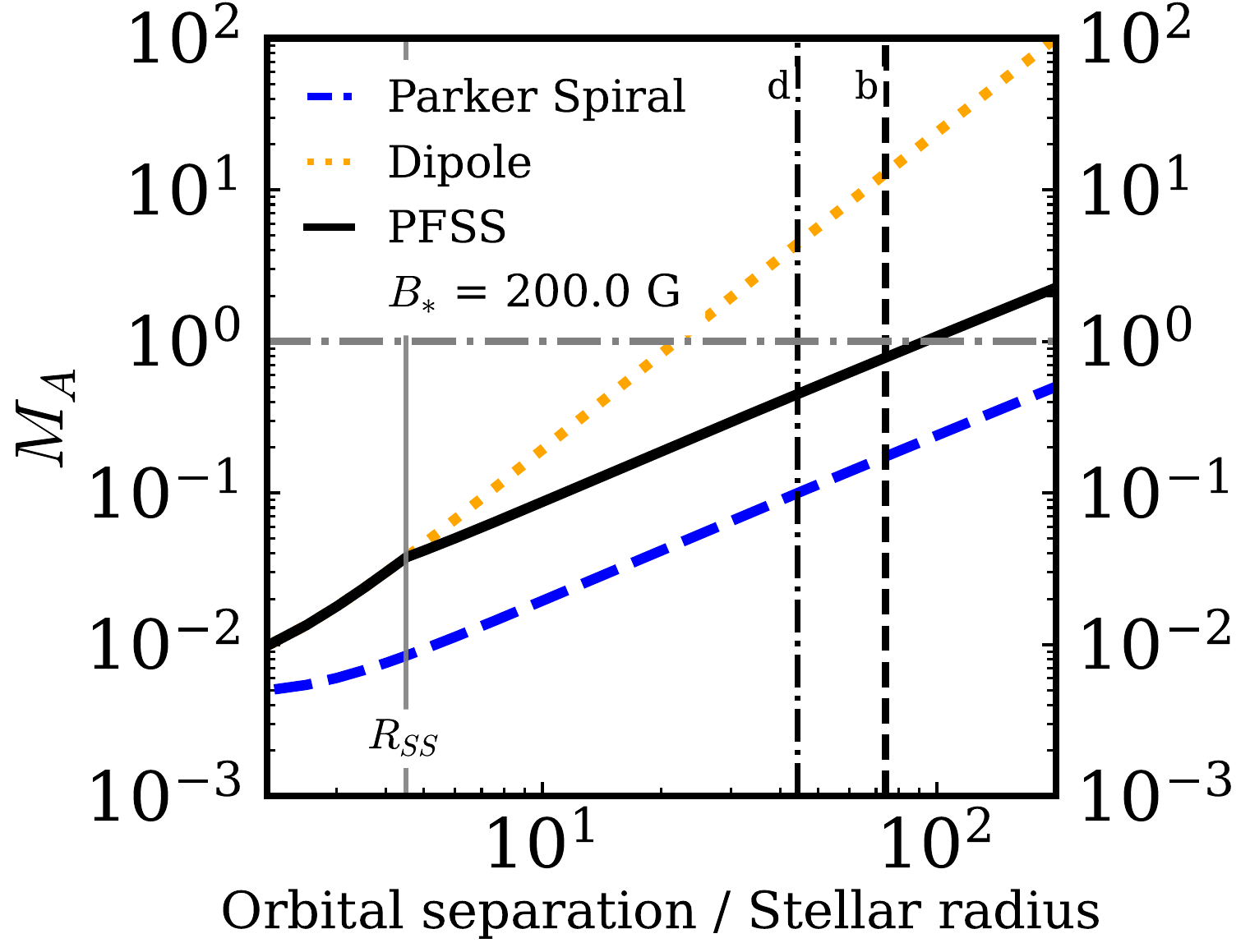}
\includegraphics[width=0.68\columnwidth]{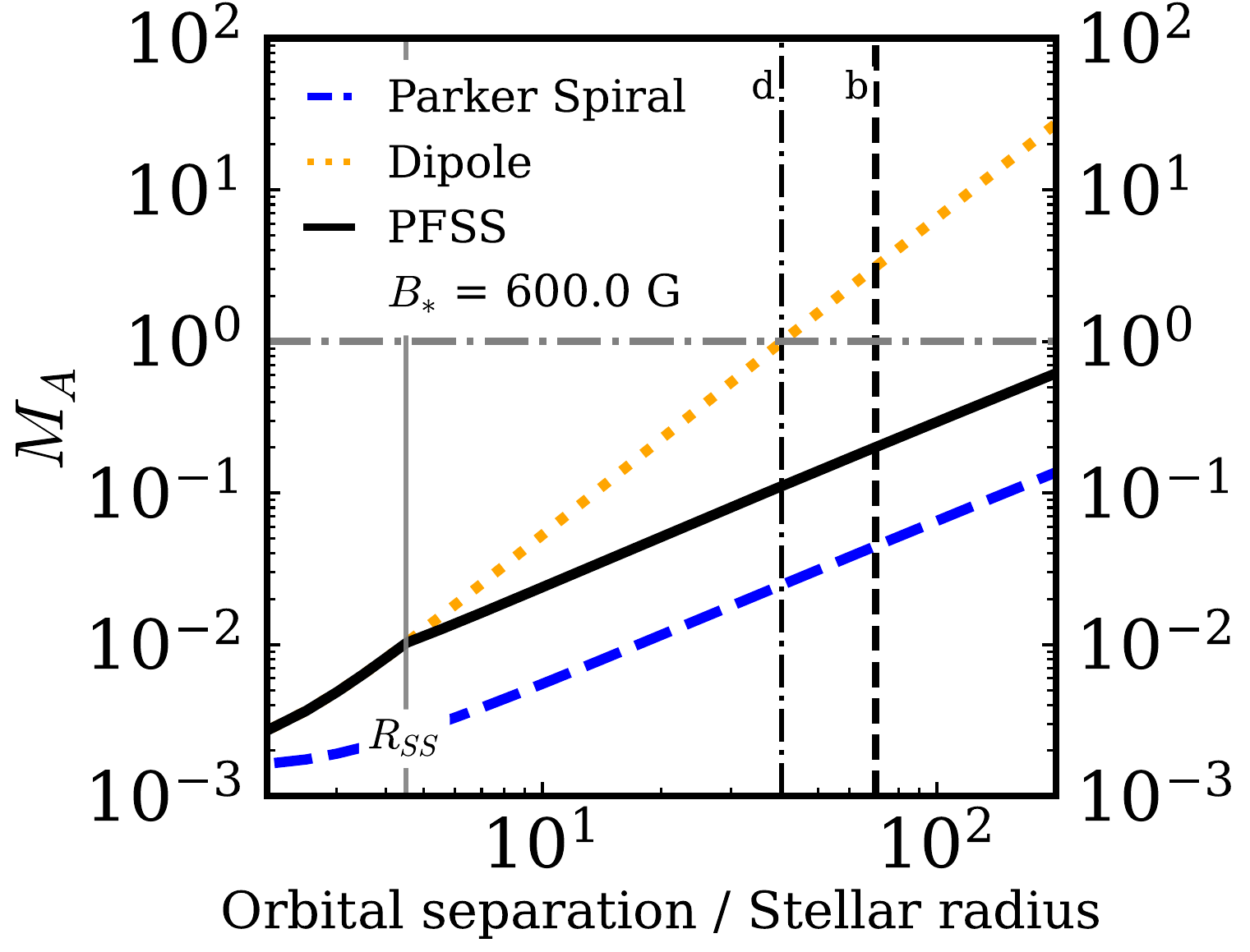}
\\
\includegraphics[width=0.68\columnwidth]{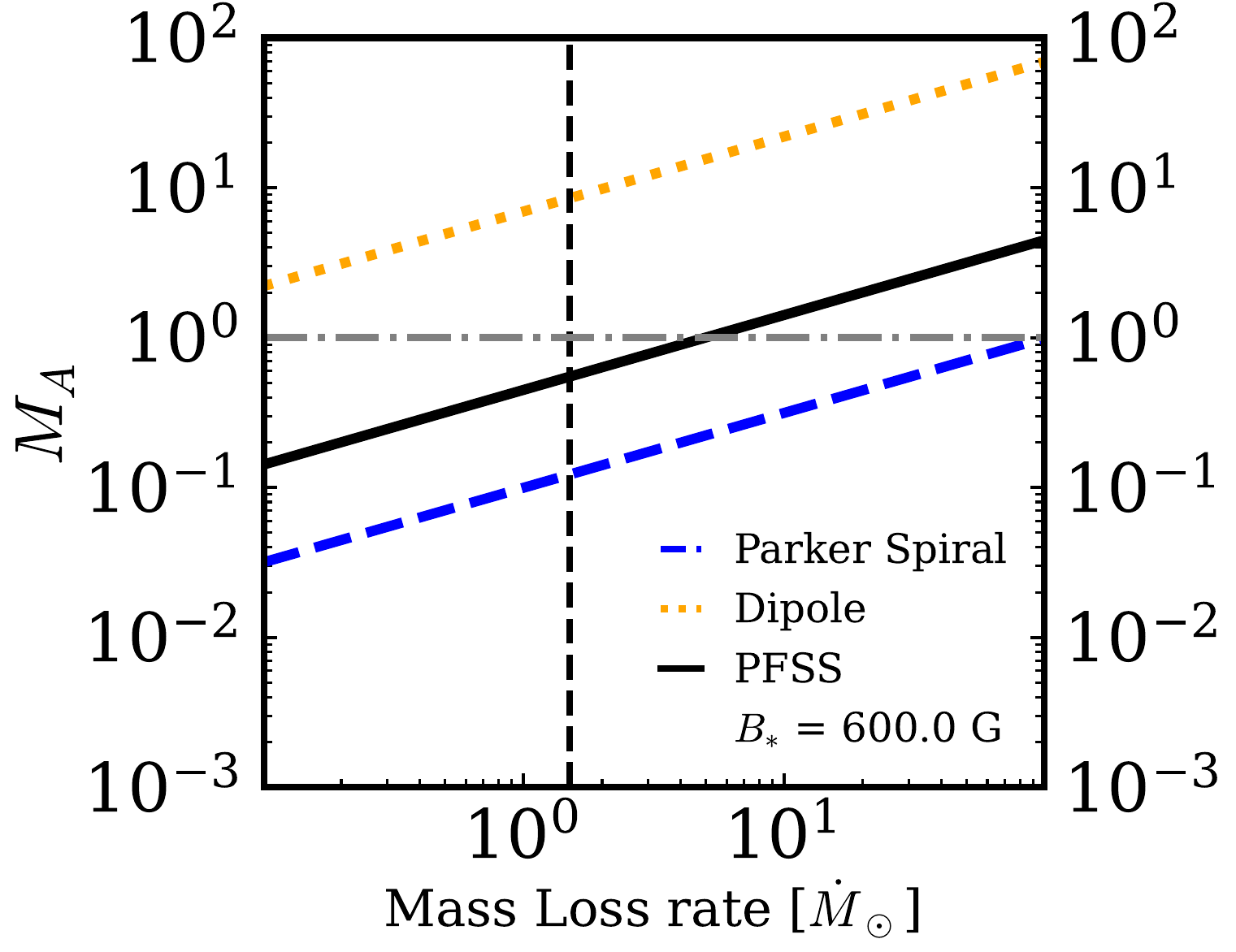} 
\includegraphics[width=0.68\columnwidth]{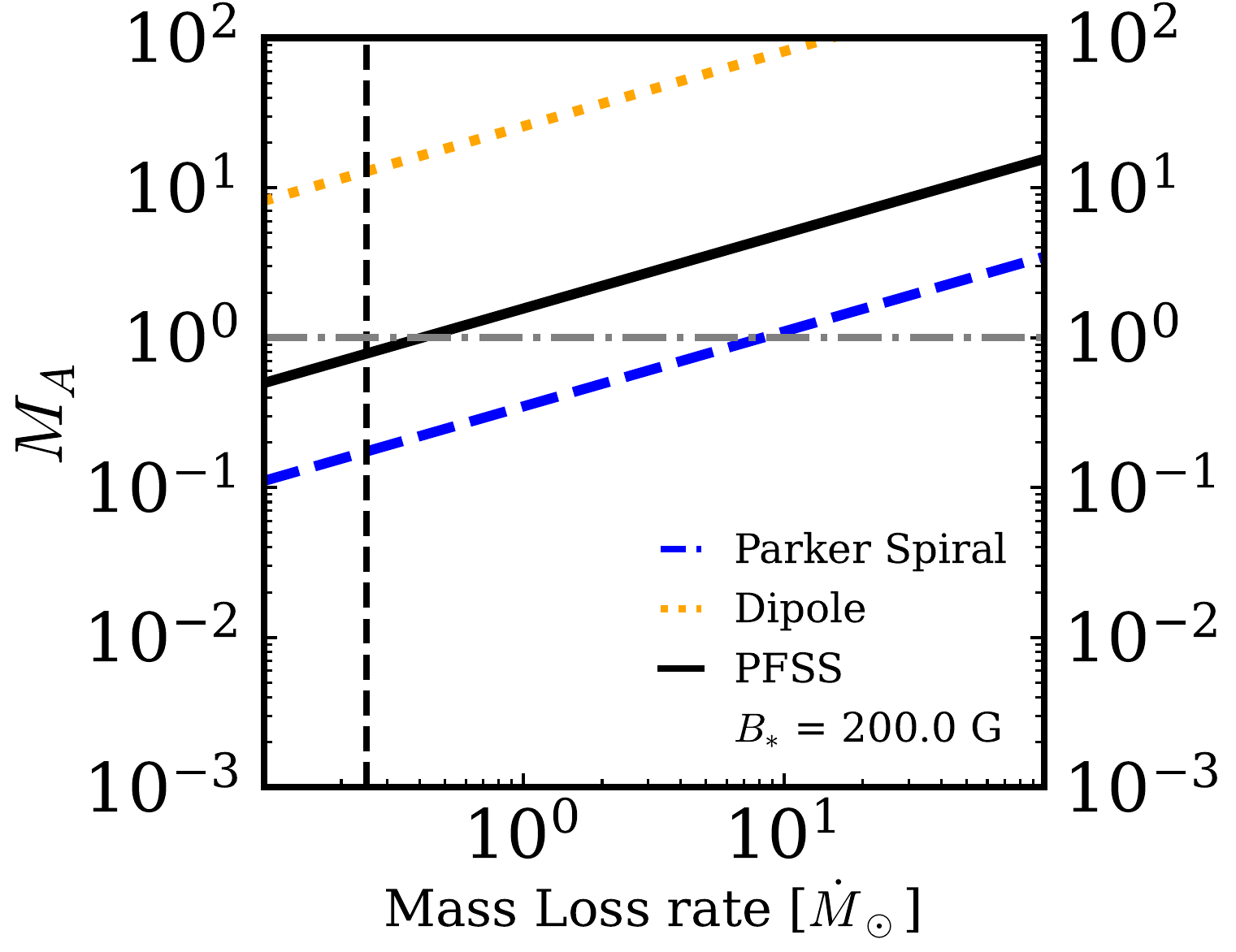}
\includegraphics[width=0.68\columnwidth]{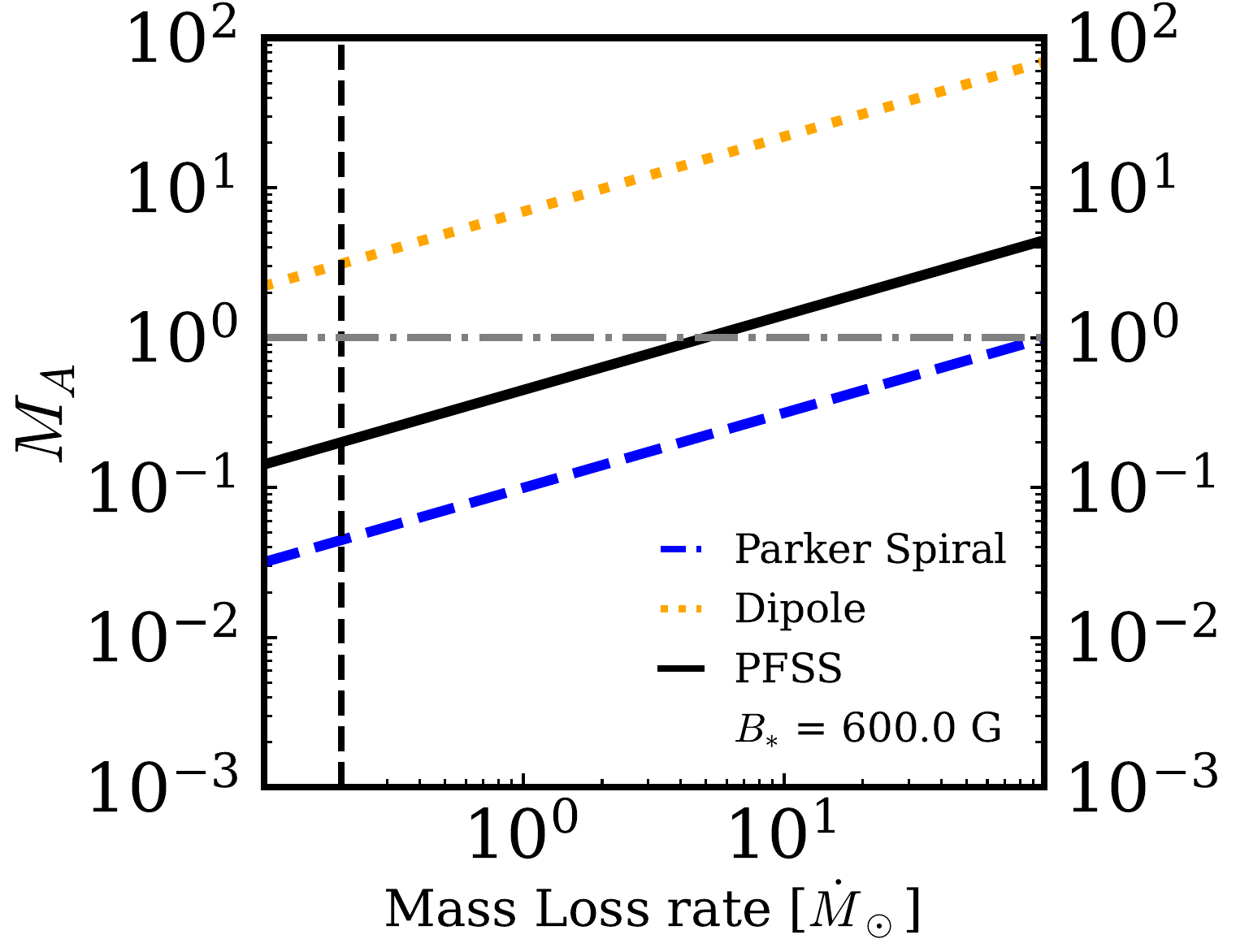}   
\caption{
{\small Comparison of the Alfvén Mach number in the Proxima system for the parameters used in \citet{Turnpenney2018}  (left), \citet{Kavanagh2021}  (middle) and \citet{Reville2024} (right), as a function of the orbital separation  and stellar mass-loss rate, for the three  geometries considered in this paper: open Parker spiral (blue dashed line), pure dipole (orange dotted line) and hybrid PFSS (solid black line).
}}
\label{fig:Proxima_M_A_all_geometries}
\end{figure*}%
%%%%% END 

We now compare the results obtained with \texttt{SIRIO} for Proxima with those in \citet{Kavanagh2021} and \citet{Reville2024}, using the same parameters published in those works. We show in Fig,~\ref{fig:Proxima_M_A_all_geometries} the Mach Alfvén number, $\MA$, as function of the orbital separation (top panels) and of the stellar wind mass loss rate (bottom panels), for all three stellar wind magnetic field geometries. We also used
two different values of the intensity of the magnetic field at the surface of Proxima: 600 G (left and right, for different stellar mass-loss rates), and 200 G (middle), to take into account the two significantly different estimates of the stellar surface magnetic field intensity of Proxima b discussed in the literature ($\sim 600$ G, \citealt{Reiners2008}; $\sim$200 G, \citealt{Klein2021}), as well as to facilitate comparisons with the results published in \citet{Turnpenney2018}, \citet{Kavanagh2021} and \citet{Reville2024}.
Note also that the top panels correspond to different values of \Mdotstar: $1.5\, ,\Mdotsun$ (left) $0.25\,\Mdotsun$ (middle) and $0.20\,\Mdotsun$ (right), which correspond to the values used in \citet{Turnpenney2018}, \citet{Kavanagh2021} and \citet{Reville2024}, respectively (see also Table~\ref{tab:sample}).

As previously discussed, $\MA$ increases most rapidly for a dipolar geometry. In the case of Proxima, this implies that both planets would be in the super-Alfvénic regime for the reference values of \Mdotstar\ considered. Therefore,  no star-planet interaction should be expected in the pure dipole scenario, whether the stellar magnetic field is 200 G or 600 G.  This result agrees with the much more detailed and complex 3D MHD simulations by \citet{Kavanagh2021}. If a PFSS geometry is in place, Proxima b is formally in the sub-Alfvénic regime for the reference value of \Mdotstar, although $\MA$ is so close to unity that star-planet interaction could be very inefficient. However, for the low-$\Mdotstar$, 600 G case the value would be significantly lower ($\MA \sim$ 0.2, roughly by a factor of 5), consistent with the MHD simulations results from \citet{Reville2024}. The lower panels show \MA\ as a function of \Mdotstar\ at the orbital separation of Proxima b. As mentioned earlier, the less realistic scenario of a pure Parker spiral predicts that Proxima b is in the sub-Alfvénic regime, unless \Mdotstar $\gtrsim 7$ (100) \Mdotsun\ for a stellar surface magnetic field of 200 (600) G. On the contrary, if the stellar wind magnetic field is mainly dipolar up to the orbital separation of Proxima b,  then the planet would always be in the super-Alfvénic regime, even if \Mdotstar\ is much smaller than $0.1\,\Mdotsun$.

The hybrid, PFSS geometry shows that if the magnetic field of the stellar surface is around 200 G (600 G), Proxima b would be in the super-Alfvénic regime for a value $\Mdotstar \gtrsim 0.4\,\Mdotsun$ ($\gtrsim 4.5\,\Mdotsun$).
We also note that for a stellar magnetic field of 200 G, Proxima b is always in the super-Alfvénic regime if $\Rss$=5.5 \Rstar. For 600 G, that value of $\Rss$ increases to 8 (22) \Rstar, for $\Mdotstar$= 1.5 (0.2) $\Mdotsun$.

\begin{figure}
\centering
\includegraphics[width=\linewidth]{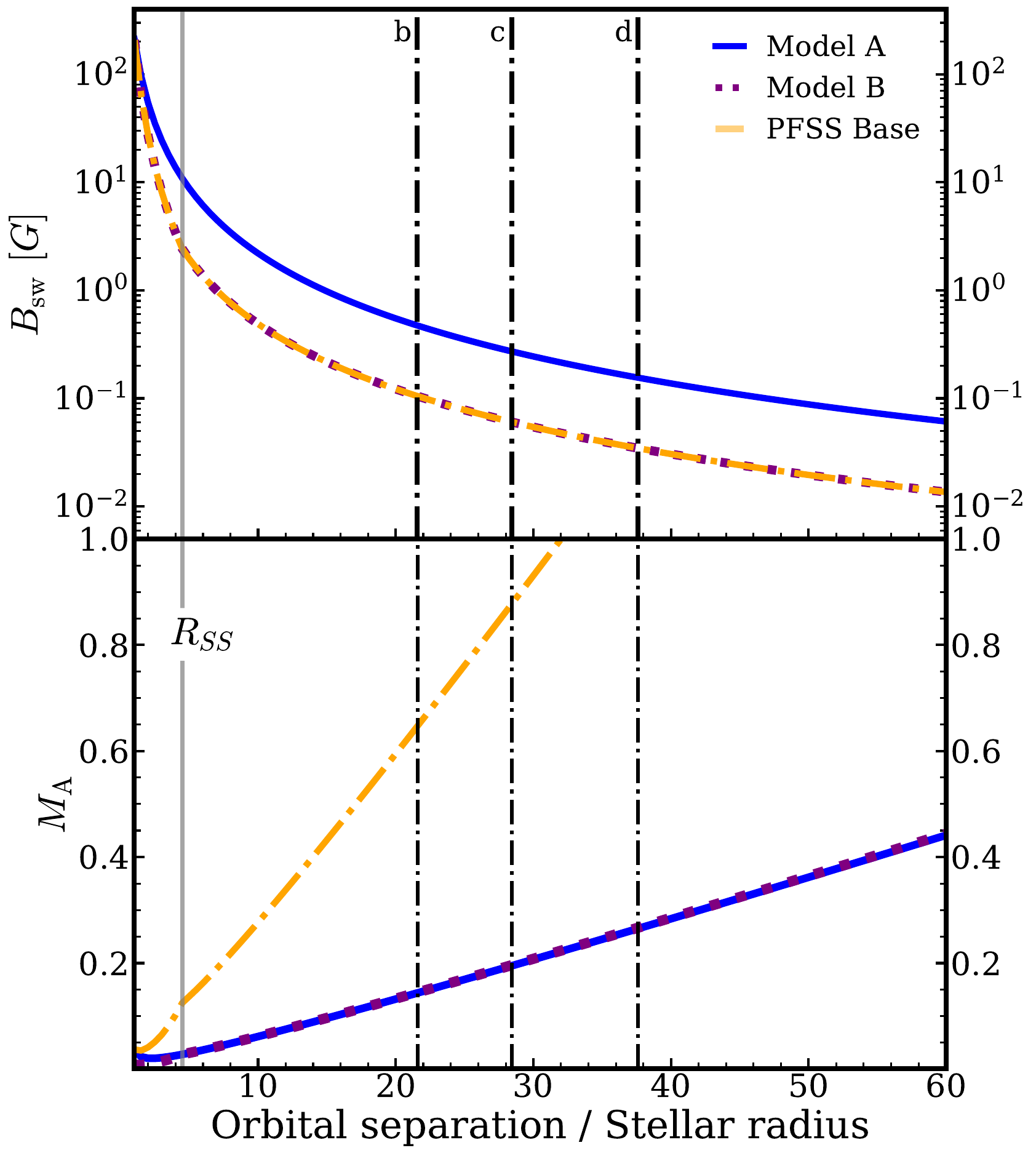}
\caption{
{\small Stellar wind magnetic field, $\Bsw$ (top panel) and Alfvén Mach number, \MA, (bottom panel) for YZ Ceti as a function of the orbital separation, for models A (blue solid line), B (magenta dotted line), and PFSS base (dash-dotted orange line), as described in \citet{Pineda2023} (see also Table~\ref{tab:sample}). The  vertical, solid grey line corresponds to the value of the \Rss, while vertical dash-dotted lines show the positions of the three known exoplanets.
Note that the slope of $\Bsw$ for model B is steeper than for model A, below $\Rss$. For orbital separations larger than $\Rss$, both models have the same slope, as the magnetic field geometry of the stellar wind is open in both cases. 
The lower panel shows that $\MA$ for models A and B is below unity up to and beyond the orbital separation of YZ Cet d, so all three planets are in the sub-Alfvénic regime. Note also that both models A and B predict the same values for \MA\ (see main text). 
}}
\label{fig:Bsw_MA_pineda}
\end{figure}

\begin{figure}
\centering
\includegraphics[width=\linewidth]{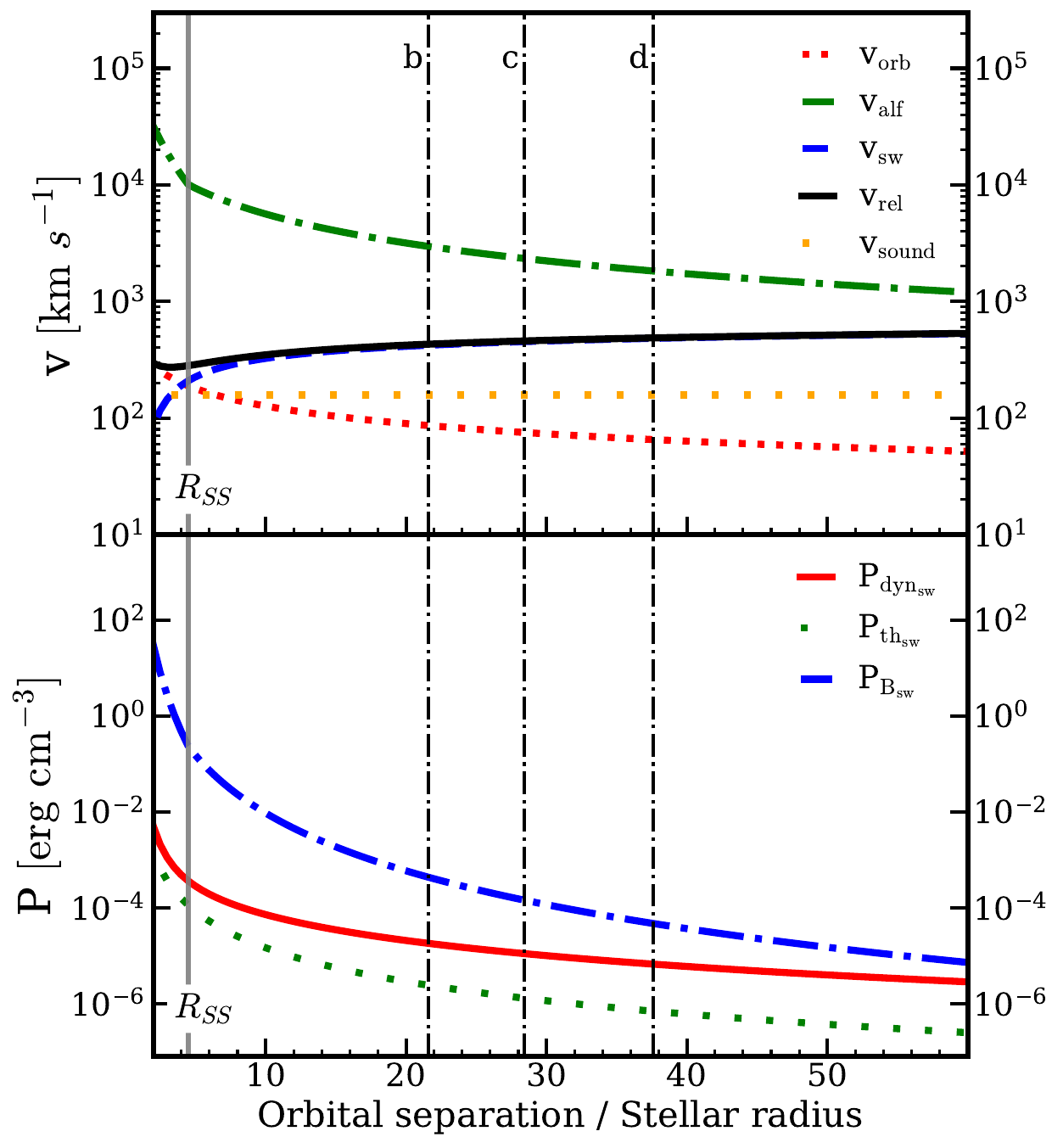}
\caption{
{\small  Stellar wind velocity (top) and pressure (bottom) parameters for YZ Cet as a function of orbital separation, for Model B.
\textit{Top:} Keplerian orbital speed (red close-dotted line),  Alfvén speed (green dash-dotted line),  stellar wind speed  (dashed blue line), relative speed between the stellar wind and the exoplanet (black solid line), and sound speed (orange sparse-dotted line). 
\textit{Bottom:}
 Dynamic pressure (red solid line), thermal pressure (green dotted line) and magnetic pressure (blue dash-dotted line) of the stellar wind.
}}
\label{fig:yzcet-vw-pressure}
\end{figure}

\begin{figure*}
\centering
\includegraphics[width=0.49\linewidth]{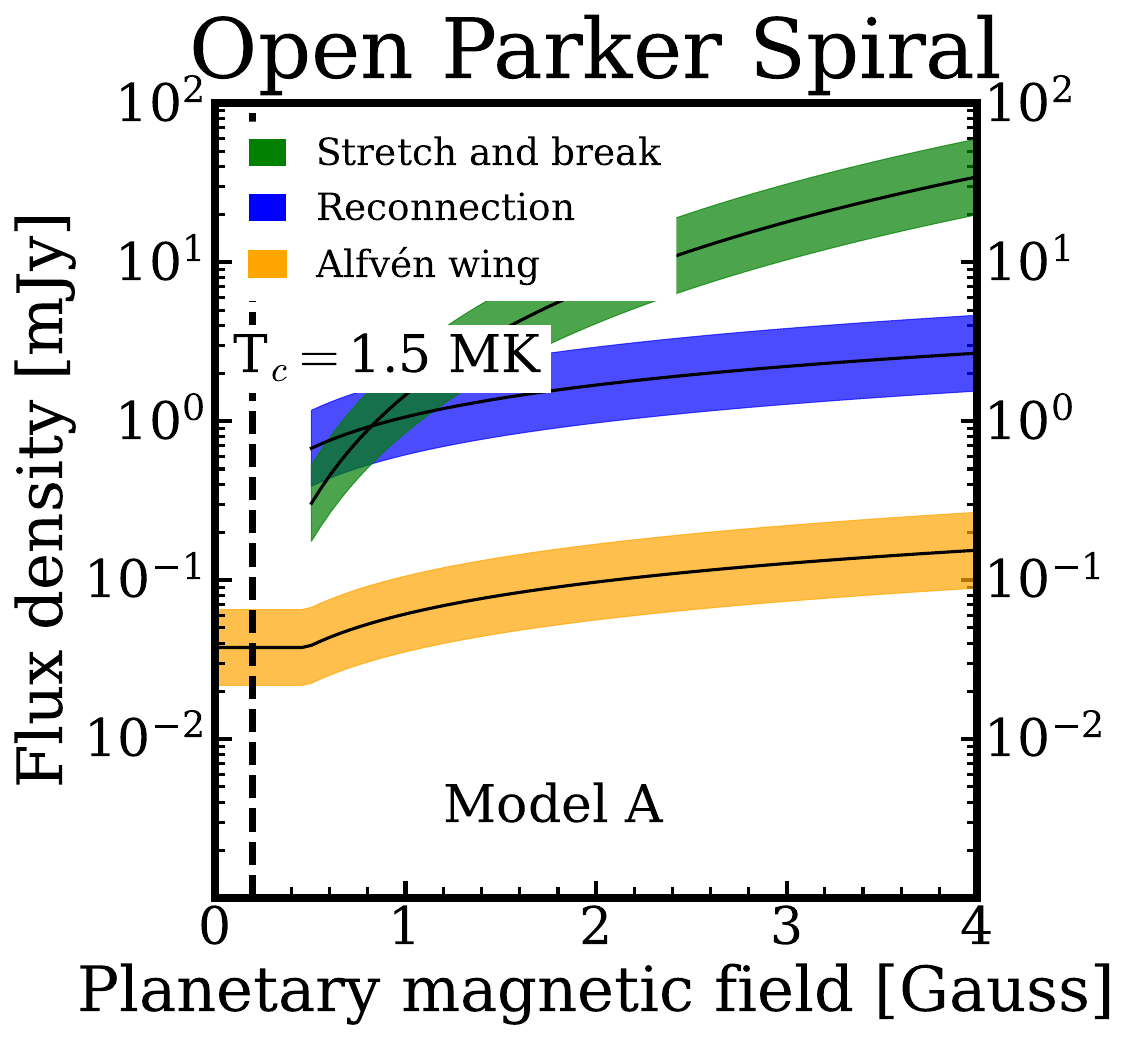}
\includegraphics[width=0.49\linewidth]{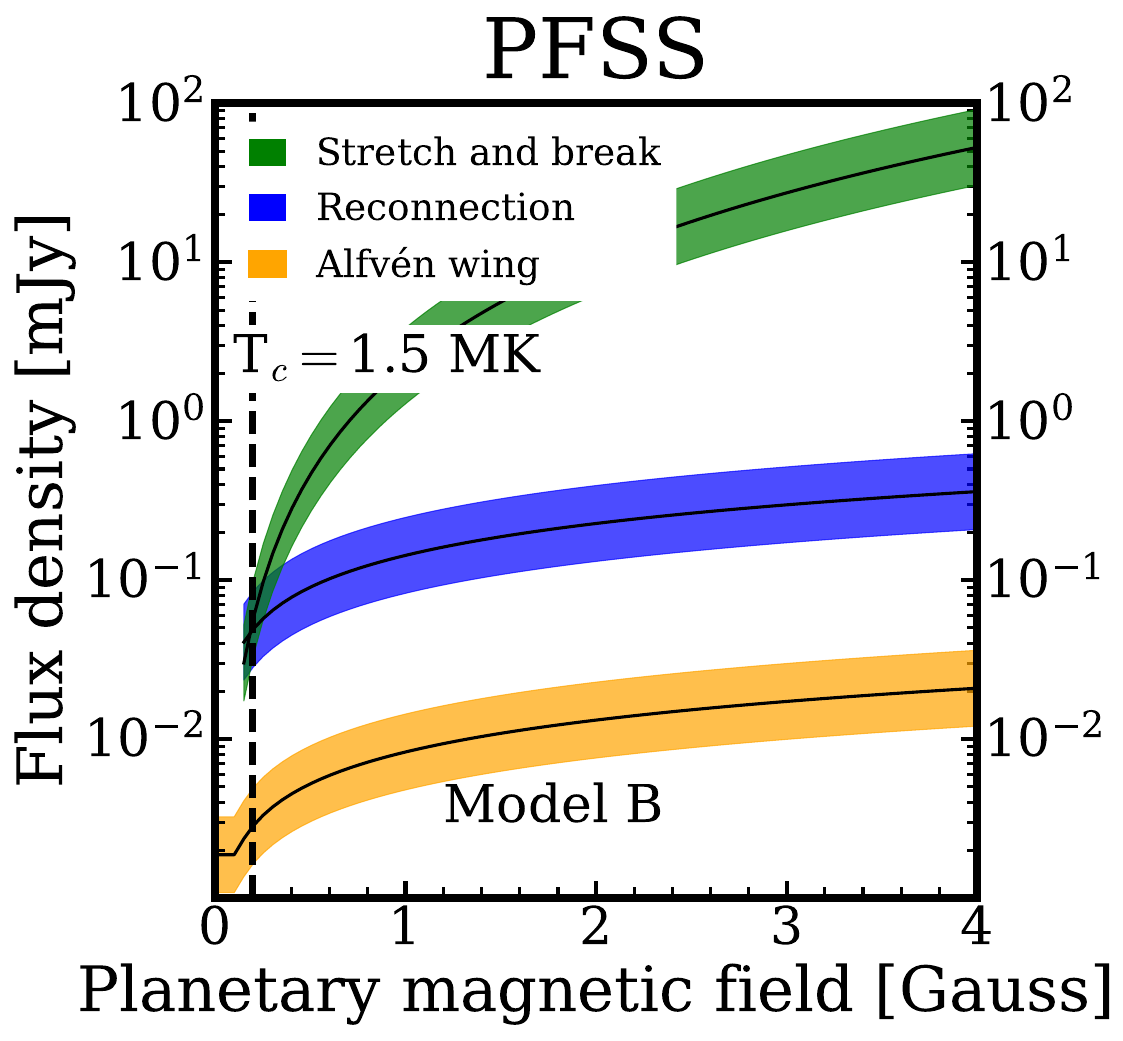}
\\
\includegraphics[width=0.49\linewidth]{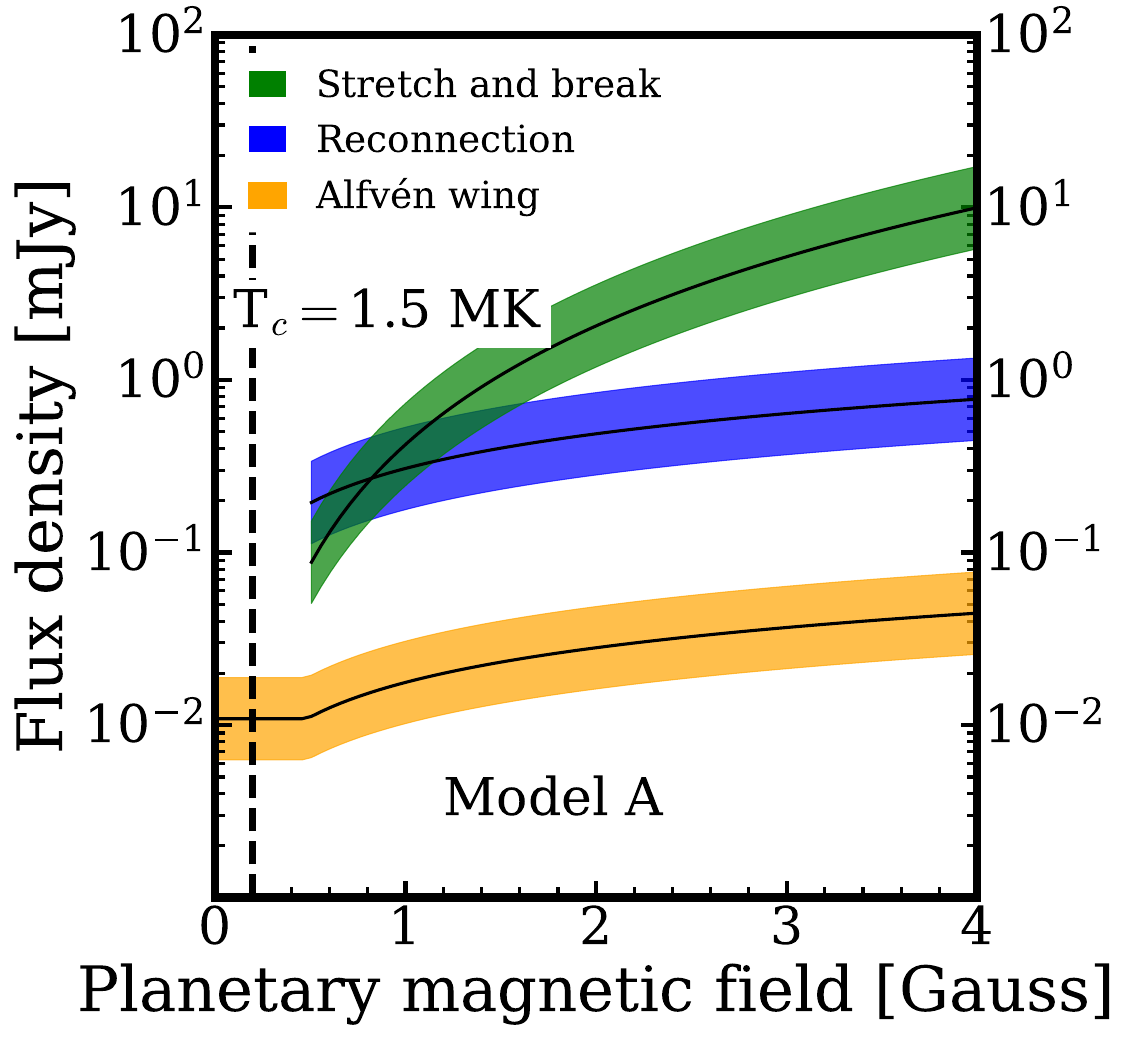}
\hspace{3pt}
\includegraphics[width=0.48\linewidth]{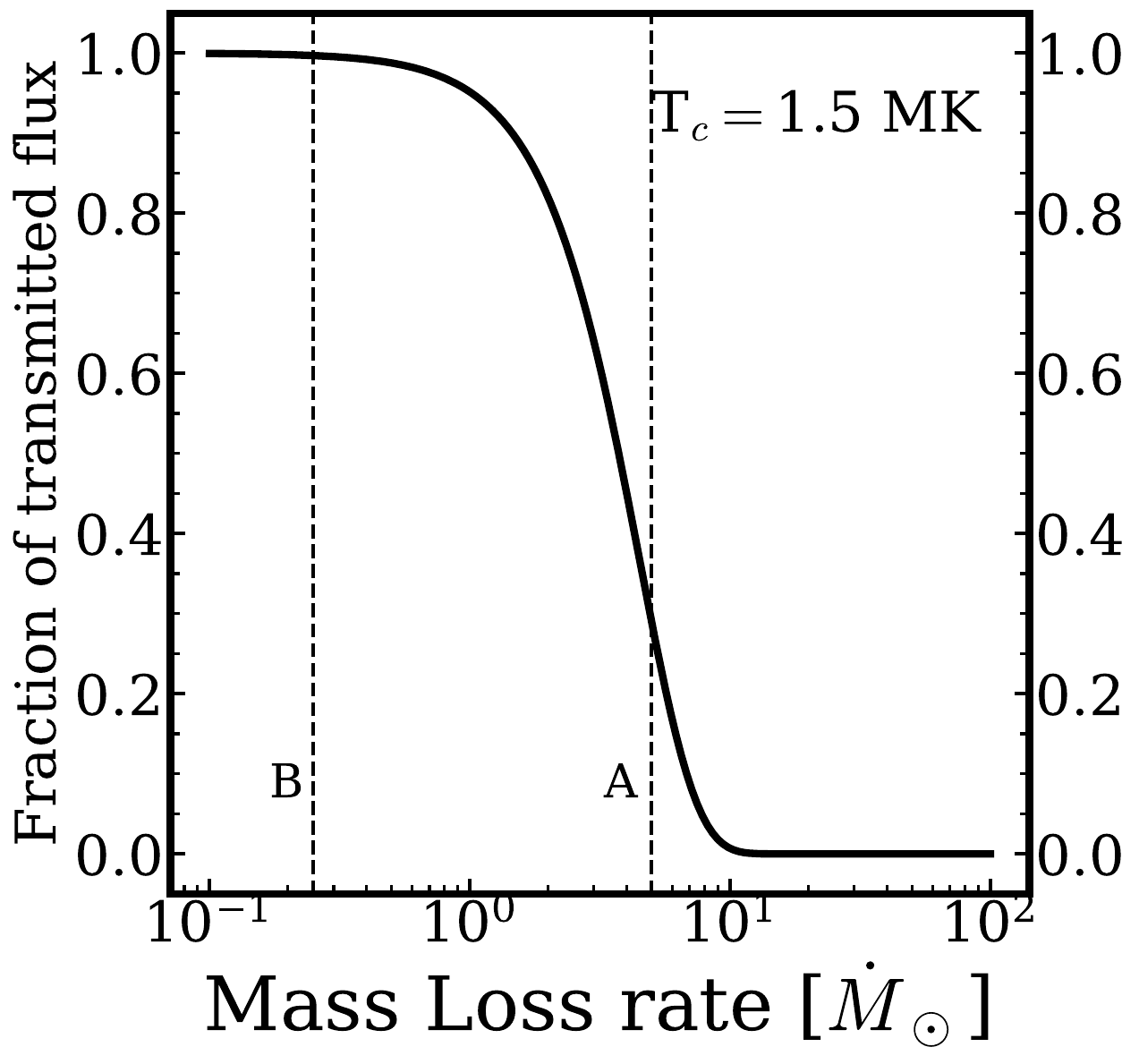}
\caption{
 {\small Predicted radio emission from star-planet interaction in YZ Cet, as function of the magnetic field of the planet, for models A and B. The vertical dashed lines correspond to the reference value of the exoplanetary magnetic field of 0.20 G (see Table~\ref{tab:sample}).
Top panels show the predicted unabsorbed radio emission,  while bottom panels show the attenuated radio emission in the same scenarios, taking into account free-free absorption from thermal electrons in the stellar wind plasma.
The bottom right panel shows the fraction of transmitted flux density, as a function of $\dot{M_\star}$. While the effect is completely negligible for the low mass-loss rate model (B), it has a strong effect for the high mass-loss model (A).
}}
\label{fig:YZCet_FLUX_Bfield}
\end{figure*}%

\begin{figure*}
\centering
\includegraphics[width=0.49\linewidth]{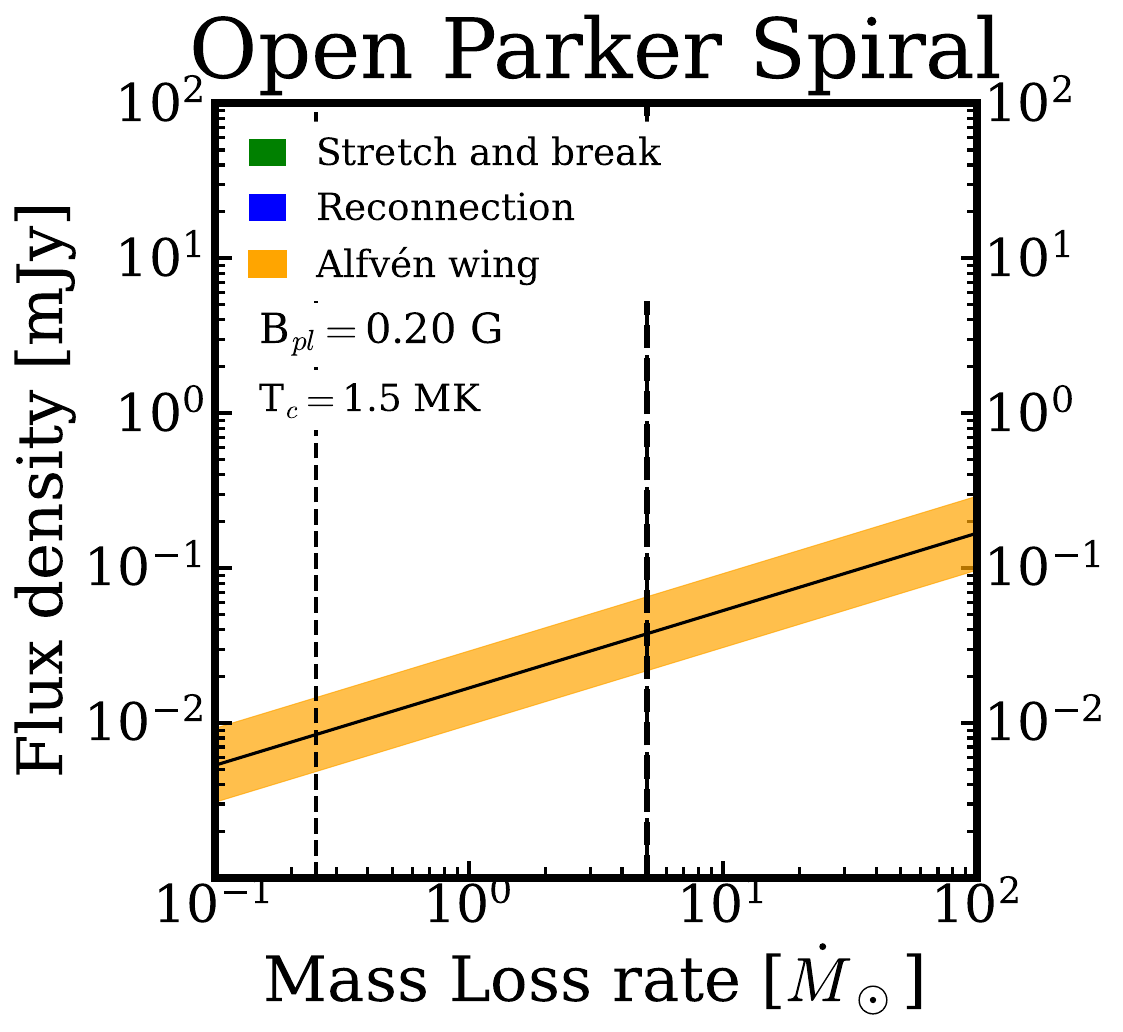}
\includegraphics[width=0.49\linewidth]{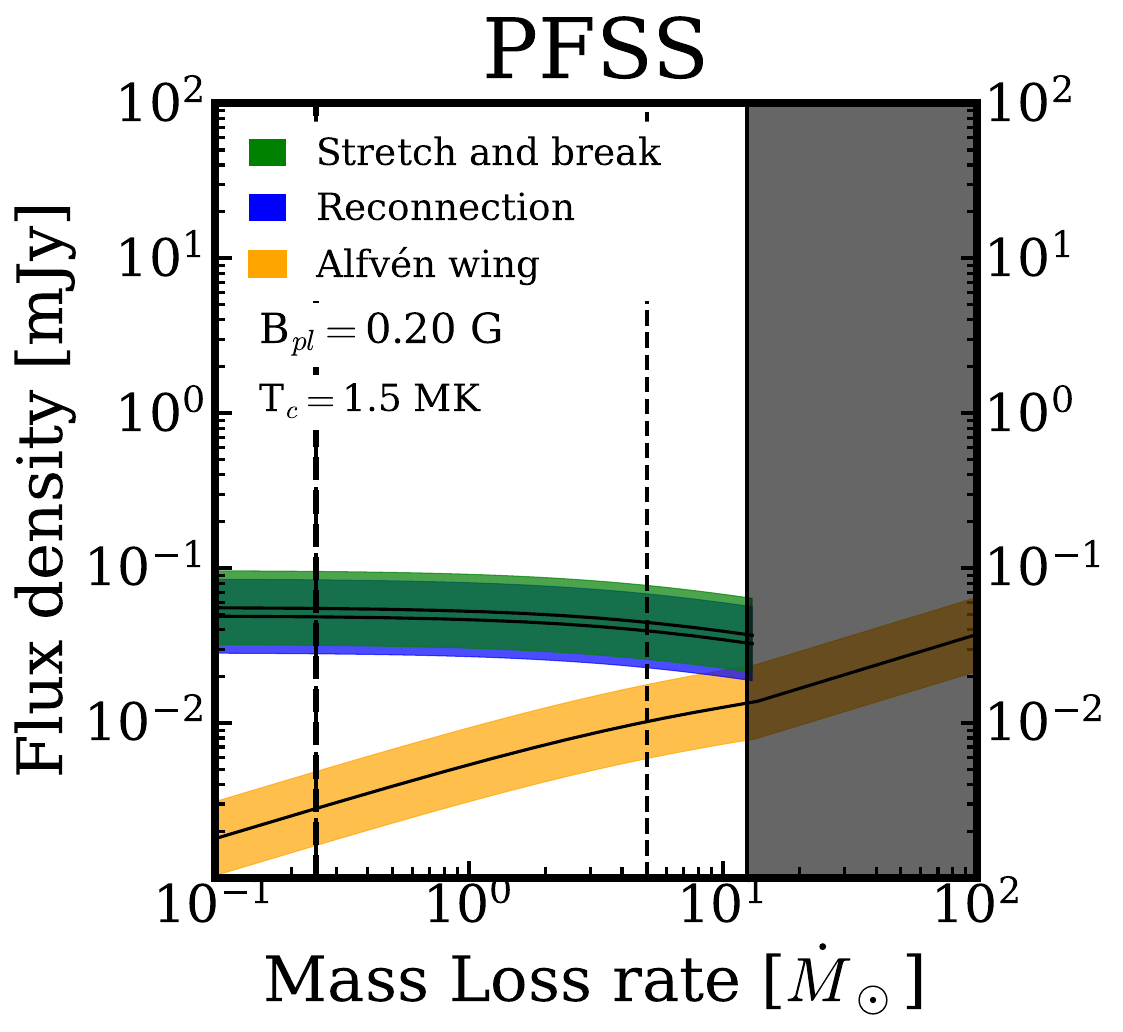}
\\
\includegraphics[width=0.49\linewidth]{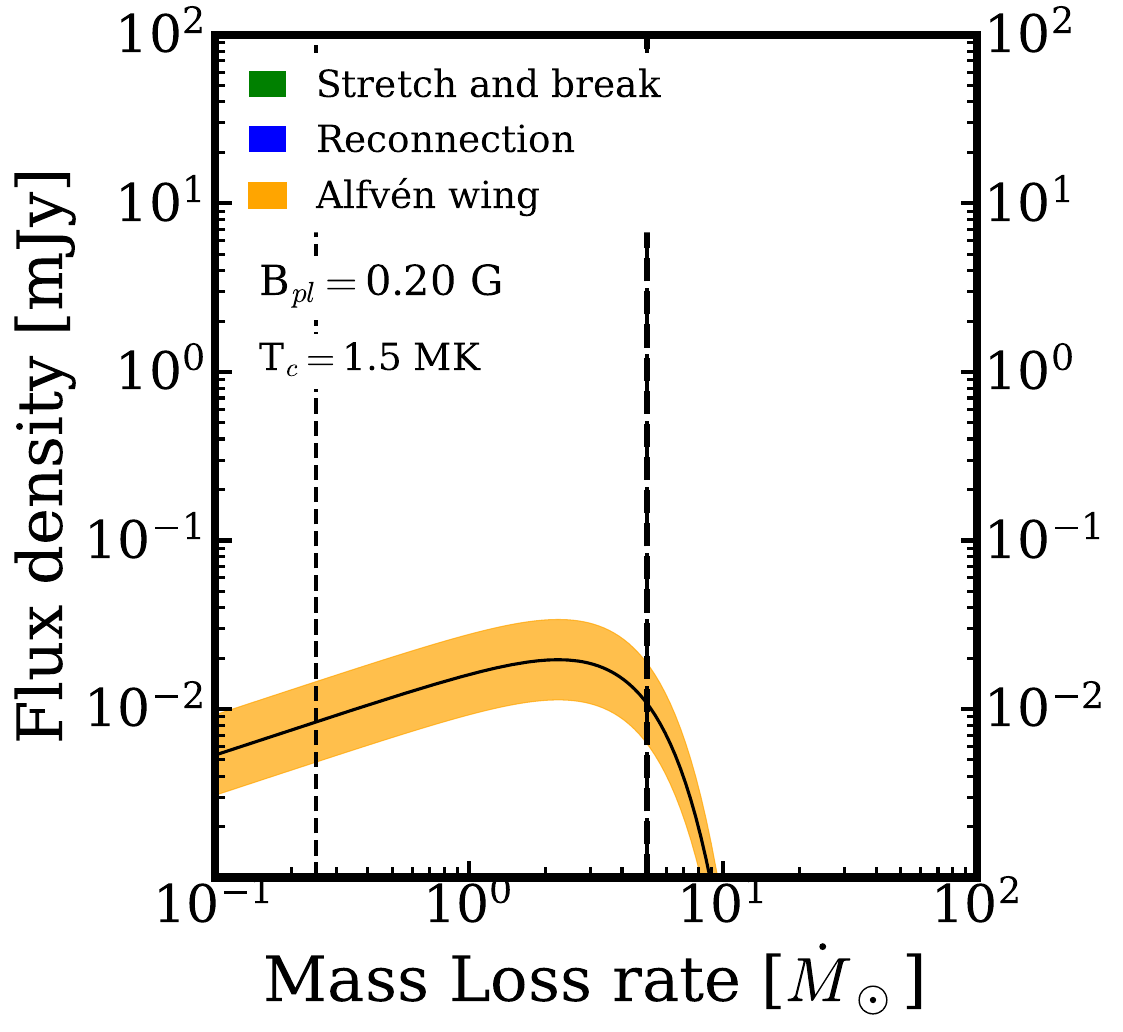}
\includegraphics[width=0.49\linewidth]{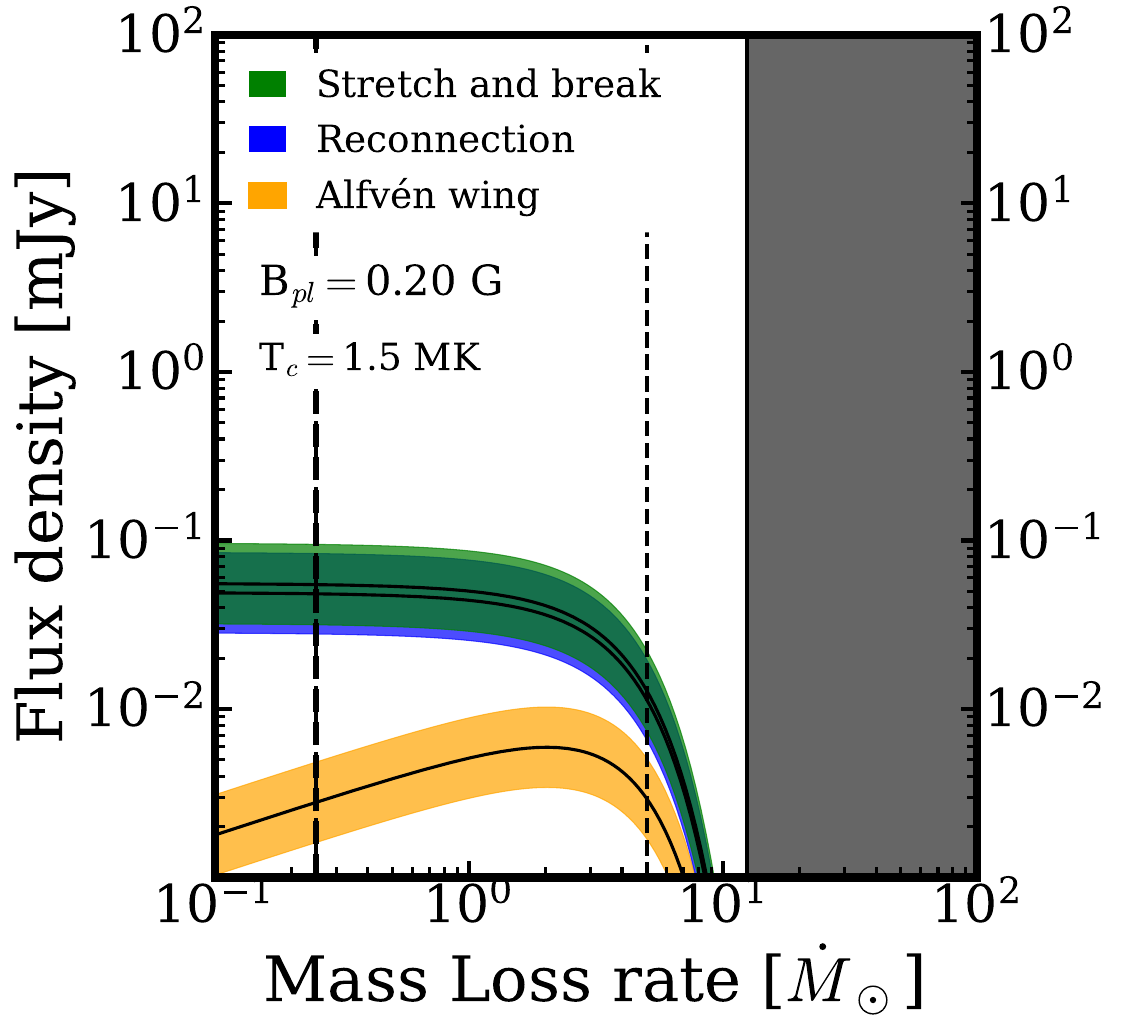}
\caption{
 {\small Predicted radio emission from star-planet interaction in YZ Cet, as function of the stellar mass-loss rate, for a Parker spiral geometry (left) and PFSS (right). 
The vertical dashed lines correspond to the reference values of \Mdotstar\ = 0.25 \Mdotsun\ (model B) and \Mdotstar\ = 5.0 \Mdotsun\ (model A). 
The upper row shows the predicted unabsorbed radio emission, neglecting free-free absorption, while the lower row show the attenuated radio emission in the same scenarios, taking into account free-free absorption from thermal electrons in the stellar wind plasma.}}
\label{fig:YZCet_FLUX_mdot}
\end{figure*}%

%%%%%%%%%%%%%%%%%%%%%%%%%%%%%%%%%%%%%%%%%%%%%%%%%%%%%%%%%%%%
%%%%%%%%%%%%%%%%%%% YZ Cet       %%%%%%%%%%%%%%%%%%%%%%%%%%%
%%%%%%%%%%%%%%%%%%%%%%%%%%%%%%%%%%%%%%%%%%%%%%%%%%%%%%%%%%%%
 \subsection{YZ Cet}
 \label{sec:YZCet}
 
YZ Ceti is a variable M-dwarf at at distance of 7 pc that hosts three close-in (semi-major axis of less than 0.03 au) Earth-like exoplanets, which provides also an excellent test case for \texttt{SIRIO}'s ability to model star-planet interactions. 
\citet{Pineda2023} and \citet{Trigilio2023} reported tentative detections of SPI emission in this system. We compare our modeling results with those presented in \citet{Pineda2023}, focusing on the impact of different stellar wind mass-loss rates and magnetic field geometries. 
We show in Fig.~\ref{fig:Bsw_MA_pineda} the intensity of the stellar wind magnetic field and the Alfvén Mach number as a function of the orbital separation, which allows a direct comparison with Fig.~11 in \citet{Pineda2023}. To this end, we used the same parameters and naming as in their paper: 
Model A corresponds to a pure Parker spiral magnetic field geometry, where the radial field declines from the surface of the star,  and has a high stellar wind mass-loss rate (\Mdotstar\ = 5 \Mdotsun; see Table \ref{tab:sample}); model B corresponds to a magnetically weaker potential field source surface (PFSS) extrapolation with a much smaller mass-loss rate (\Mdotstar\ = 0.25 \Mdotsun) . We also included, for comparison purposes, the ``PFSS base'' model of \citet{Pineda2023}, which is intermediate between models A and B,  with the PFSS magnetic field geometry of model B, but with the high mass-loss rate of model A. 

The stellar wind magnetic field of YZ Ceti as a function of the orbital separation looks exactly as in Fig.~11 of \citet{Pineda2023}. In particular, \Bsw\ decays more slowly in model A (blue solid line) than in model B (magenta dotted line) because the azimuthal component of the magnetic field in model A is negligible, so the field is mostly radial. 
\Bsw\ decays much faster in model B  up to the orbital separation \Rss\, because the magnetic wind geometry is that of a dipole. Beyond \Rss, the behavior is the same as in model A, as the geometry is open in both scenarios.
Note also that,since \Bsw\ does not depend on $\dot{M}$, the values of \Bsw\ for the PFSS base model (dash-dotted line) and model B coincide.   
The bottom panel of Fig.~\ref{fig:Bsw_MA_pineda} shows $\MA$ as a function of the orbital separation for the three models discussed in \citet{Pineda2023}. 
The values of \MA\ for the PFSS base model essentially coincide with those in their paper. A similar behaviour is shown  by model A, with our values being slightly smaller. Note also that the values of $\MA$ for model A and model B in Fig.~11 of \citet{Pineda2023} start to depart beyond \Rss, while in our case these values are the same (Fig.~\ref{fig:Bsw_MA_pineda}, bottom).

To explain this discrepancy, we first note that the value of $\MA$ at $\Rss$ is the same, both for model A (Parker spiral) and model B (PFSS), in agreement with the results in \citet{Pineda2023}. This is a consequence of their chosen values of $\dot{M}$. Indeed, the ratio of the Alfvén Mach numbers for both models is $\zeta = M_{\rm A} \rm{(A)} / M_{\rm A} \rm{(B)}$, so $\zeta = (R_\star / \Rss) \cdot [\dot{M} {\rm (A)} / \dot{M}{\rm (B)}]^{1/2}$. Since $\Rss = 4.5\, \Rstar$ and $\dot{M}{\rm (A)} / \dot{M}{\rm (B)} = 20$, then $\zeta = 1.0$. Further, model B assumes a PFSS geometry, which beyond $\Rss$ has the same slope as a Parker spiral. Since the  values of  \MA\ in both models were selected so that they would coincide at \Rss, and both models have the same magnetic field geometry beyond that distance, the coincidence of \MA\ for both models as a function of the orbital separation is expected.
Hence, the relatively small discrepancy with \citet{Pineda2023} is most likely due to our use of the Parker model to determine the stellar wind velocity profile, which assumes a purely radial wind,  rather than the use of the Weber-Davis model, as done in \citet{Pineda2023}, which considers also an azimuthal component of the wind velocity profile.  We note nevertheless that  despite using a simpler wind velocity profile, our model yields very similar results.

Fig.~\ref{fig:yzcet-vw-pressure} shows the stellar wind velocity and pressure radial profiles  of YZ Cet for Model B, to facilitate comparison with the results obtained by \citet{Pineda2023}. As shown in Fig.~\ref{fig:Bsw_MA_pineda}, the Alfvén speed is larger than the relative speed up to a distance beyond the orbit of YZ Cet d. Note that the main contribution to the relative velocity is that of the stellar wind, except if there was a planet within a few stellar radii, in which case the contribution of the Keplerian speed could be higher. The bottom panel shows in a slightly different way the fact that the planets b, c, and d are in the sub-Alfvénic regime: the pressure of the stellar wind magnetic field dominates over the ram (dynamic) pressure at least up to an orbital separation of more than 60 \Rstar.

%%%%%%%%%%%%%%%%%%%%%%%%%%%%%%%%%%%%%%%%%%%%%%%%%%%%%%%%%%%%
%%%%%%%%% free-free absorption effect in the YZ Cet system  
%%%%%%%%%%%%%%%%%%%%%%%%%%%%%%%%%%%%%%%%%%%%%%%%%%%%%%%%%%%%
\subsubsection{\rm{Free-free absorption effect in the radio emission from the YZ Cet - YZ Cet b system}}

We show in Figs.~\ref{fig:YZCet_FLUX_Bfield} and~\ref{fig:YZCet_FLUX_mdot} the expected radio emission from star-planet interaction in the YZ Cet - YZ Cet b system, for models A and B, as a function of  the magnetic field of the planet and of the stellar mass-loss rate, respectively. In each Figure, top panels correspond to simulations where free-free absorption is ignored, while bottom panels correspond to simulations where free-free absorption is taken into account. 

Fig.~\ref{fig:YZCet_FLUX_Bfield} shows the dependence of the radio emission as a function of the planetary magnetic field, both for the high mass-loss rate model A (left) and the low mass-loss rate model B (right).  In both cases, the planets are in the sub-Alfvénic regime. We also ran simulations for the pure dipolar geometry, which show that for the low mass-loss rate model, the planets would already be very close to the super-Alfvénic regime, and clearly in it for the high mass-loss rate scenario, so we do not discuss this scenario. As the bottom right panel illustrates, the high mass-loss rate of model A implies a significant free-free absorption of the outgoing radio emission. The shape of the plot is the same as in the panel above, but with the flux density reduced by a factor of four. On the contrary, the low mass-loss rate of model B implies that free-free absorption is negligible, so there is no difference between the unabsorbed and free-free absorbed predicted flux density. Therefore, we only show the unabsorbed case for model B (top right panel).

Note that the reconnection scenario does predict a flux density that is about one order of magnitude higher (even more for the stretch-and-break model)  than the Alfvén wing scenario both for models B and A, as long as the planet is able to carve out its own magnetosphere.  Given the reference planetary magnetic field of YZ Cet b ($\Bp \sim 0.20$ G; vertical dashed line in the graphs), the planet is able to carve out its own magnetosphere in model B, where the mass-loss rate is small, and so is the ram pressure of the stellar wind.
On the contrary, the higher mass-loss rate in model A leads to an increased ram pressure of the stellar wind at the orbit of YZ Cet b, which makes our reference value of the exoplanetary magnetic field too weak to sustain a magnetosphere. As a result, magnetic reconnection does not take place for planetary magnetic field strengths below about 0.4 G.

Fig.~\ref{fig:YZCet_FLUX_mdot} shows the predicted flux density as a function of the stellar wind mass-loss rate for 
a pure Parker spiral (left panels) and a PFSS geometry  (right panels), for the reference value of the planetary magnetic field of YZ Cet b.
As discussed earlier, the radio emission in the Alfvén wing scenario steadily increases with the stellar wind mass loss rate as $\Mdotstar^{1/2}$, as can be seen in the top panels, where free-free absorption is not taken into account.
On the contrary, 
in the reconnection and stretch-and-break models, the radio emission scales with the magnetopause radius squared 
($S_{\rm rec/SB} \propto {\rm R}^2_{\rm mp}$), which decreases as the ram pressure increases (see Eq.~\ref{eq:Rmp})

However, this picture changes when absorption effects are take into account, as shown in the bottom panels. While absorption effects can be neglected for values of \Mdotstar $\lesssim \Mdotsun$, for higher mass-loss rates the effect becomes increasingly noticeable, and for values \Mdotstar $\gtrsim 10 \Mdotsun$, essentially all radio emission is totally suppressed. 
We also note that, for our reference magnetic field value of the planet, neither the reconnection nor the stretch-and-break model are able to yield radio emission, and the predictions from the Alfvén wing model are much less than the values reported in \citet{Pineda2023} or \citet{Trigilio2023}. If a PFSS geometry is at place, the  flux density predicted by the Alfvén win model is rather small. The reconnection and stretch-and-break models predict  significantly larger radio fluxes, but still smaller than reported in the above papers. We note, however, that a slightly higher efficiency factor than our nominal value, say $\eta = (3-4)\times10^{-3}$, would be enough to be in agreement with observations.

In summary, our analysis reveals that the isothermal wind is a very good approximation for cases like YZ Cet, and highlights the role of free-free absorption, which can play a crucial role in determining the detectability of radio signals.

%%%%%%%%%%%%%%%%%%%%%%%%%%%%%%%%%%%%%%%%%%%%%%%%%%%%%%%%%%%%
%%%%%%%%%%%%%%%%%%% GJ 1151      %%%%%%%%%%%%%%%%%%%%%%%%%%%
%%%%%%%%%%%%%%%%%%%%%%%%%%%%%%%%%%%%%%%%%%%%%%%%%%%%%%%%%%%%
\subsection{GJ 1151}
\label{sec:gj1151}

GJ1151 is an M dwarf at a distance of 8.04 pc. \citet{Vedantham2020} reported a coherent and highly circularly polarized radio emission at a frequency range of 120 - 167 MHz for the entirety of one of the LoTSS pointings, which may have been induced by the interaction with a close-orbiting planet. However, at the time of the publication of that work there was no known planet orbiting GJ 1151. They concluded that the signal is induced by an undetected planet with a period ranging from 1 to 5 days.
Using data from HARPS-N and The Habitable-zone Planet Finder (HPF), \citet{Mahadevan2021} claimed the existence of close-orbiting planet, GJ 1551 b, with a period of 2.02 days. However, this claim was soon challenged by further CARMENES RV observations \citep{Perger2021}. Later, \citep{Blanco-Pozo2023} reported another planet, GJ 1551 c, with a semi-major axis of 0.57 au, but too far from the star to be the cause of the stellar radio emission. Nonetheless, these RV campaigns presented by \citep{Perger2021} and \citep{Blanco-Pozo2023} still placed upper limits on the minimum mass of the unconfirmed GJ 1151 b, constraining it to a range between 0.73$M_{\oplus}$ (with a 1-day period) and 1.25 $M_{\oplus}$ (with a 5-day period).

\begin{figure}
\centering

\includegraphics[width=\columnwidth]{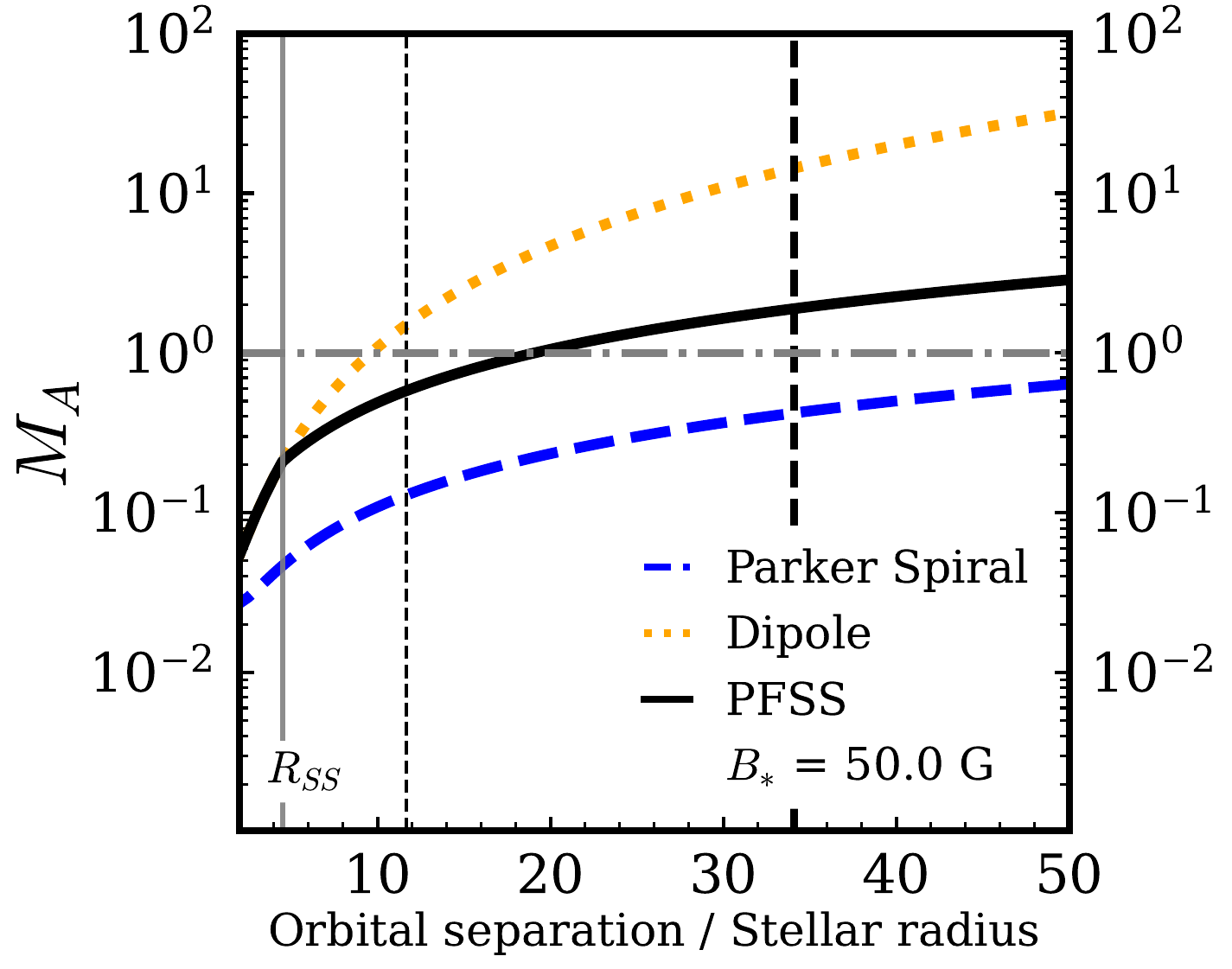}

\caption{\small 
Alfvén Mach number, $\MA$, as a function of orbital separation for the GJ 1151 system. We show $M_A$ for the three stellar wind geometries: dipole (orange dotted line), Parker spiral (blue dashed line), and hybrid PFSS model (solid black line). The vertical lines indicate the expected orbital separation of the planets corresponding to the 1-day (thin dashed line) and 5-day (thick dashed line) period estimates of GJ 1151 b. 
}
\label{fig:GJ1151_d_orb_diag}
\end{figure}%

Here we use \texttt{SIRIO} to study detection prospects for GJ 1151 b on the lower and higher end of the reported mass range. The stellar parameters considered here (see Table \ref{tab:sample}) are those from \citet{Vedantham2020} (except for the stellar mass-loss rate, which we assume to be solar-like), while the planetary constraints are from \citet{Blanco-Pozo2023}. We assume a stellar magnetic field of 50 G, which would result in cyclotron emission with a frequency consistent to the radio detections by \citet{Vedantham2020}.
In Fig. \ref{fig:GJ1151_d_orb_diag} we show $\MA$ as a function of the orbital separation for the three geometries. The left and right vertical dashed lines respectively show the orbital separation for the 1-day period, lower mass limit of GJ 1151 b (a=13 $R_*$) and the 5-day period, higher mass limit (a=38 $R_*$). For the dipolar geometry we can see that in both scenarios the planet falls in the super-Alfvénic regime, so there would be no star planet interaction.  In contrast, for the Parker spiral the planet is inside the Alfvén surface in both cases. Interestingly, for the hybrid PFSS geometry what we see is that for the short period estimate the planet is in the sub-Alfvénic regime, while the long period case planet lies beyond the Alfvén surface. The transition between both regimes takes place at a separation or around 20 $R_*$. This implies that, for a PFSS hybrid geometry, there would only be radio emission from star-planet interaction if the orbital period of the planet is smaller than 2 days.

\begin{figure*}
\centering

   \includegraphics[width=0.49\linewidth]{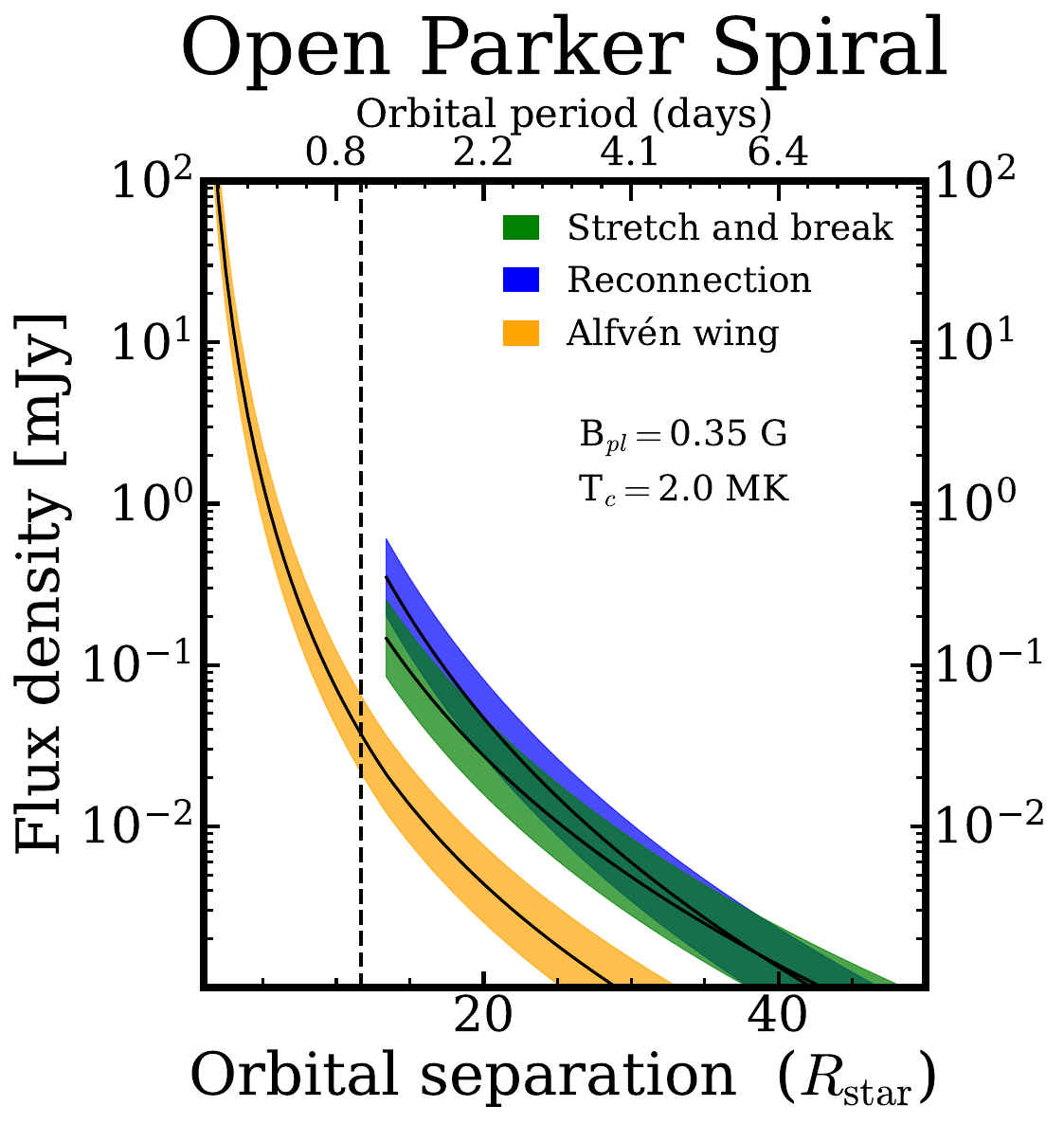}
   \includegraphics[width=0.49\linewidth]{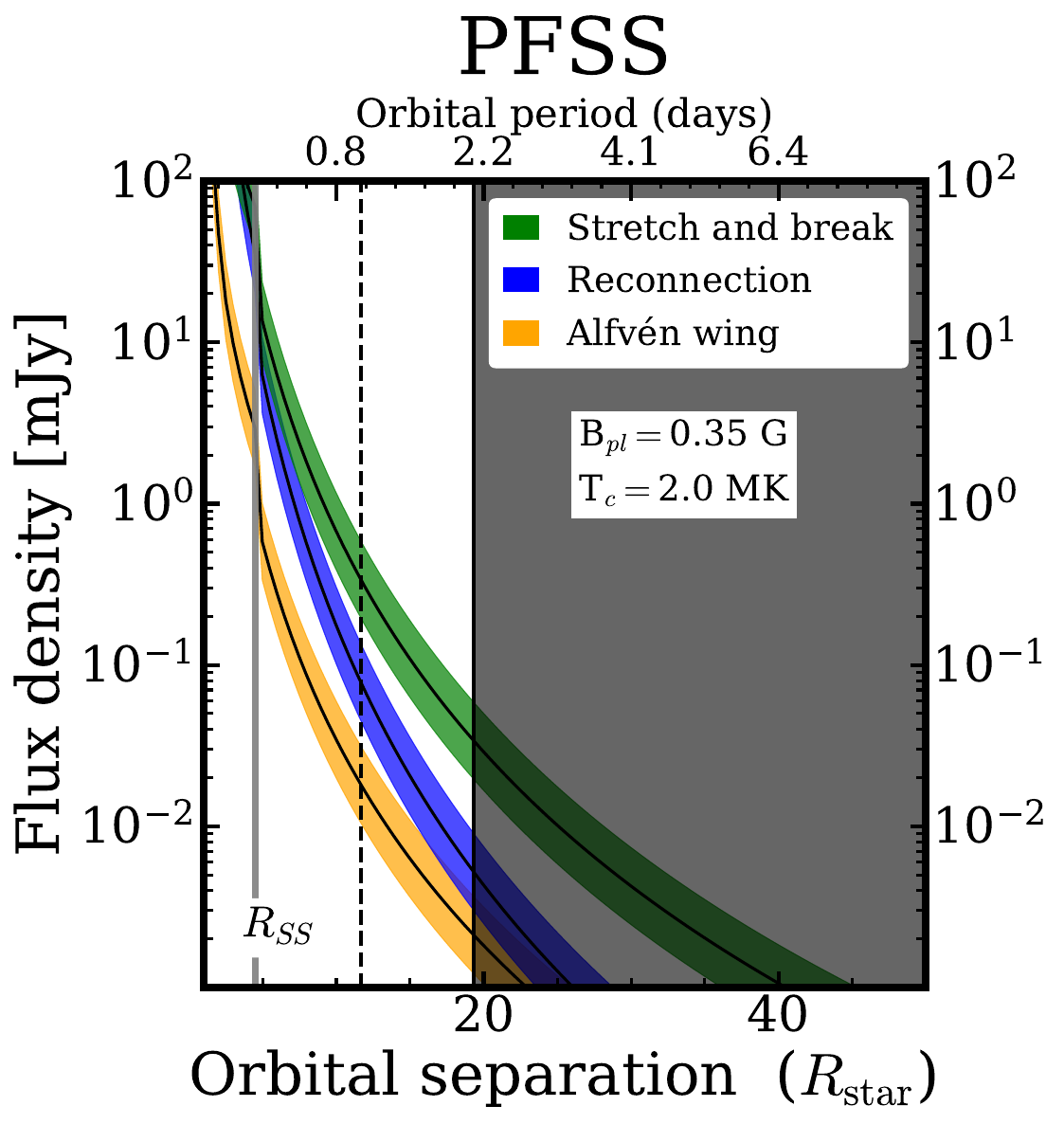}
   \\
  
   \includegraphics[width=0.49\linewidth]{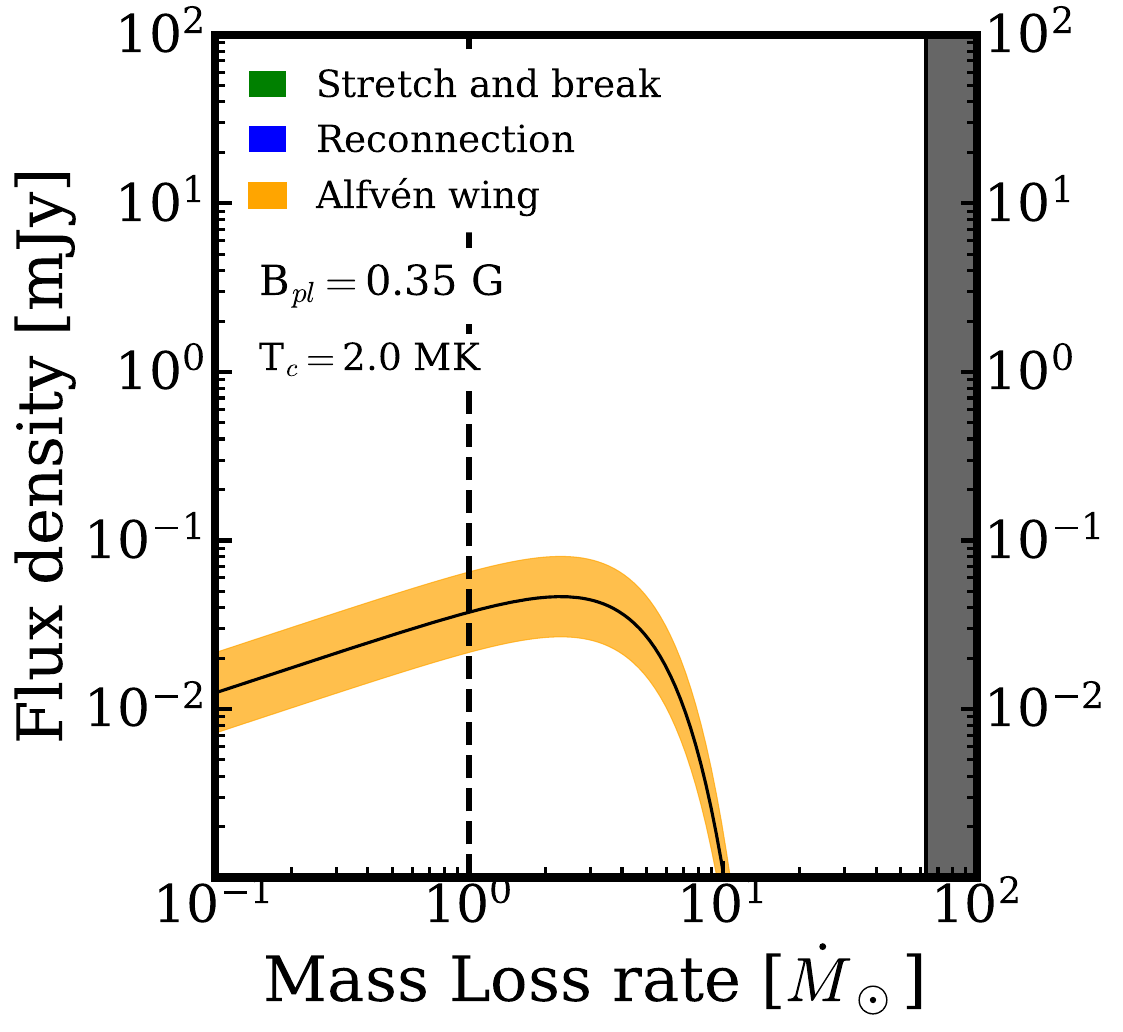}
   \includegraphics[width=0.49\linewidth]{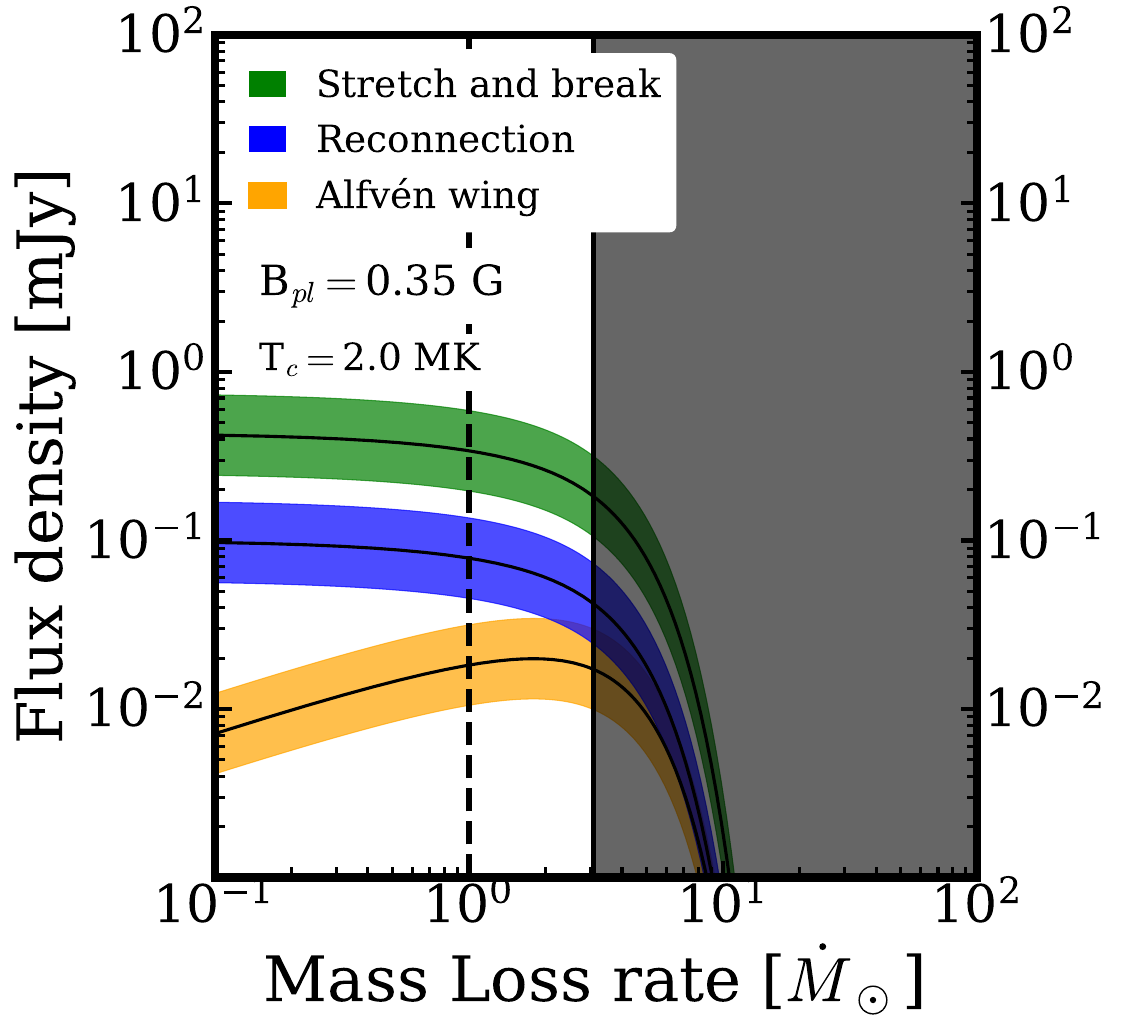}
\caption{
{\small Predicted SPI radio flux density for the interaction between GJ~1151 and a hypothetic 0.73 $M_{\oplus}$ planet, for a Parker spiral (left) and a hybrid PFSS (right) stellar wind geometry. 
Top panels show the predicted flux density as a function of the orbital separation, while bottom 
panels show it as a function of the stellar wind mass loss rate.} 
}
\label{fig:GJ1151_flux_close_planet}
\end{figure*}%

\begin{figure*}%[htbp!]e
\centering

    \includegraphics[width=0.49\linewidth]{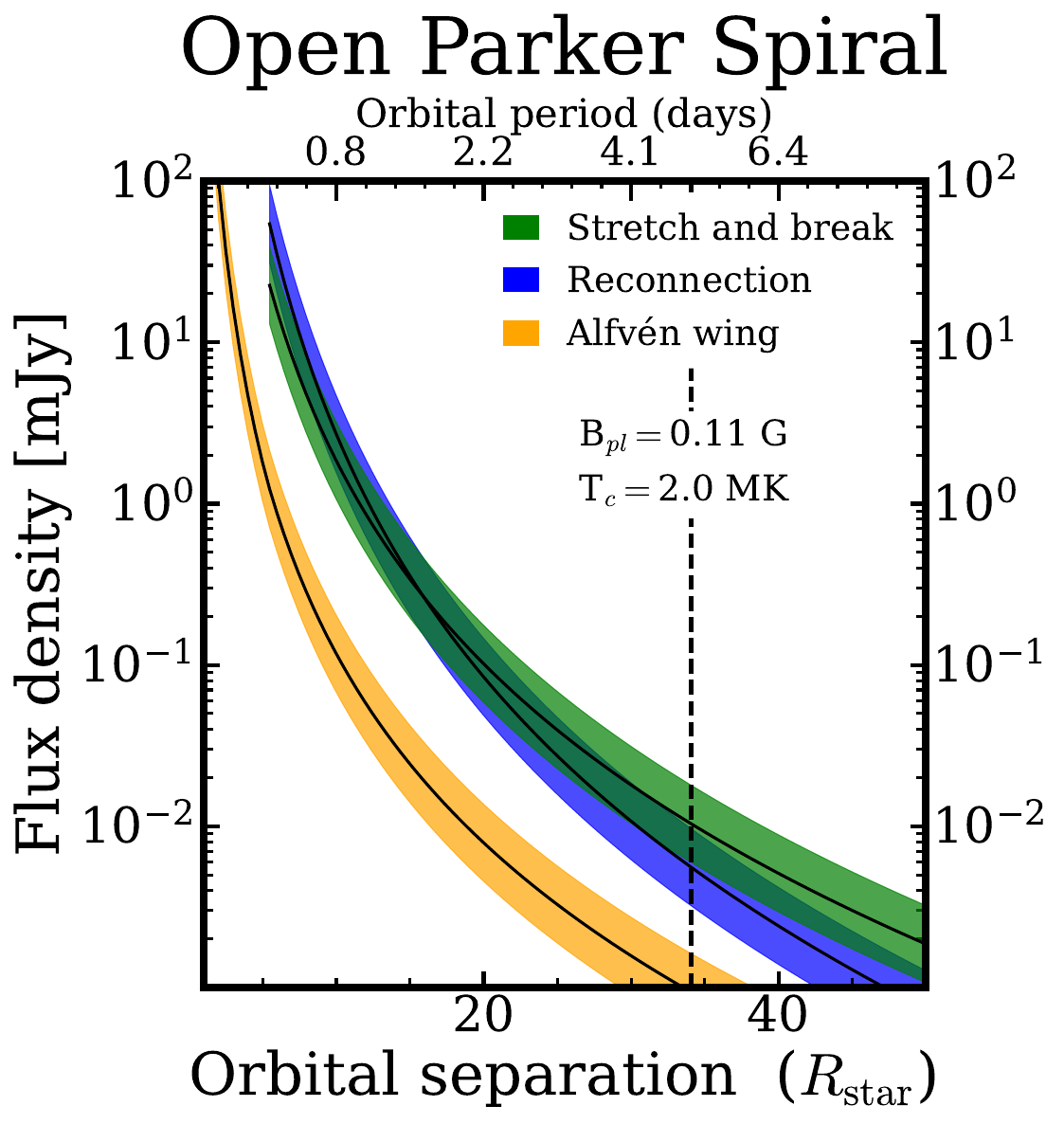}
   \includegraphics[width=0.49\linewidth]{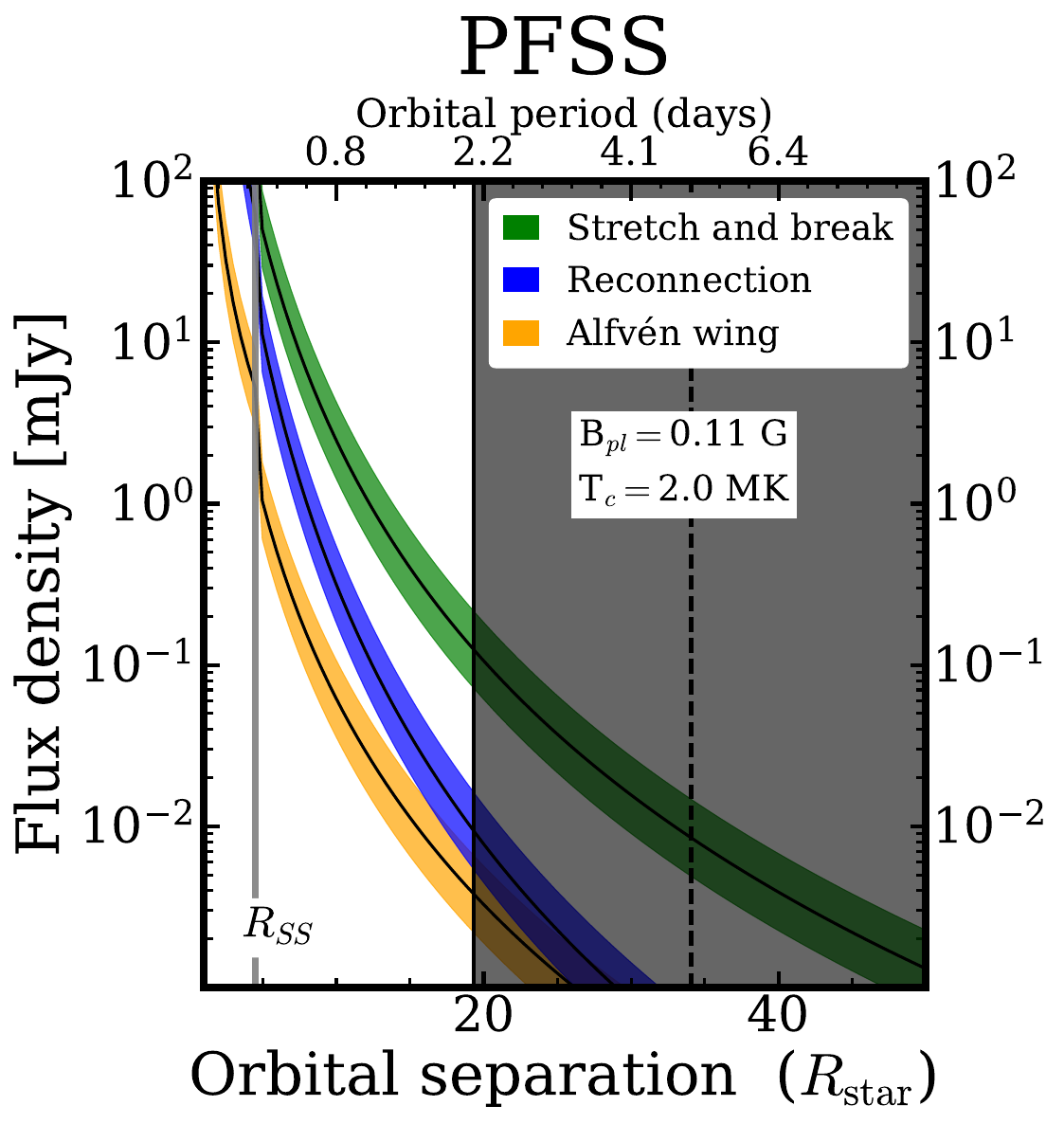}
   \\
   \includegraphics[width=0.49\linewidth]{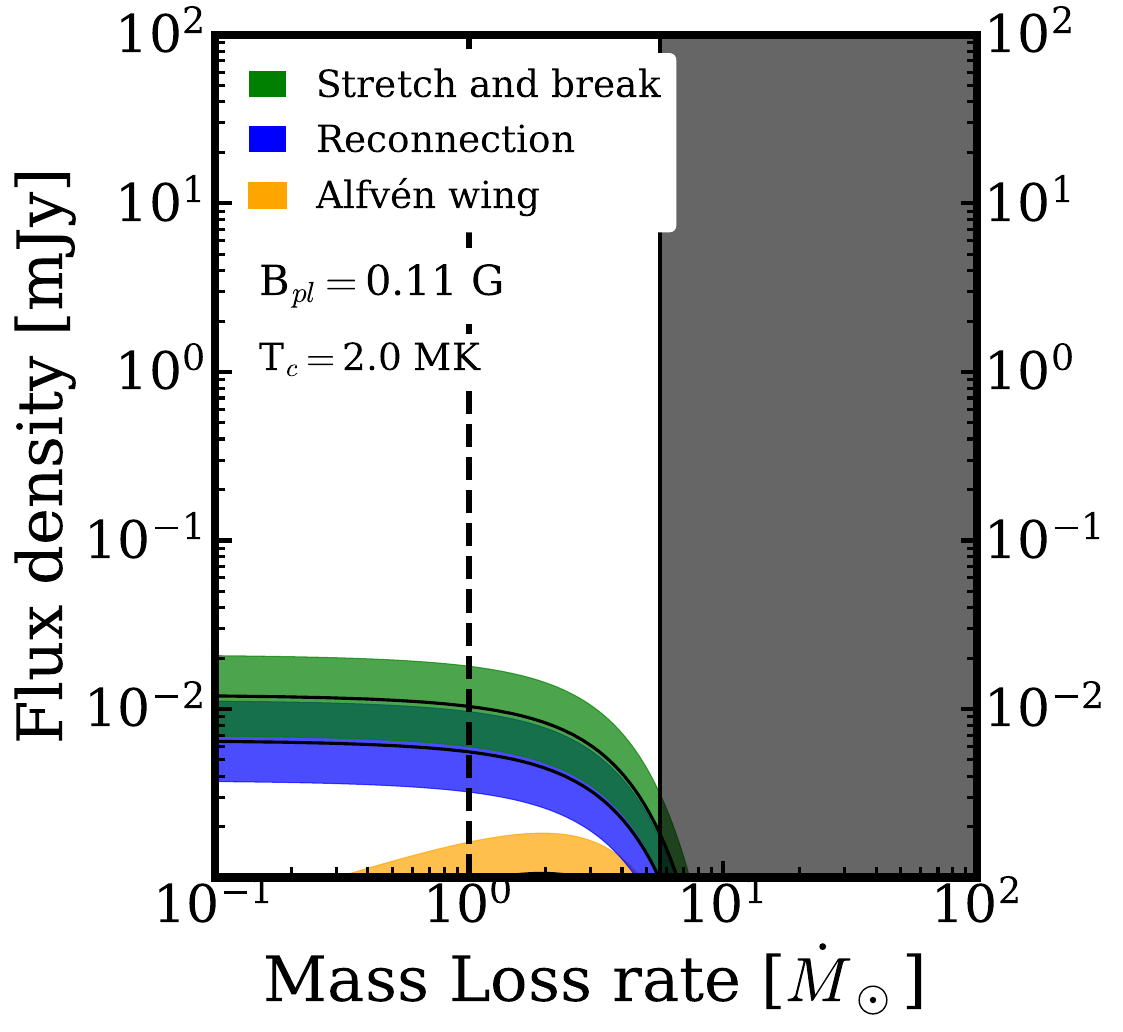}
   \includegraphics[width=0.49\linewidth]{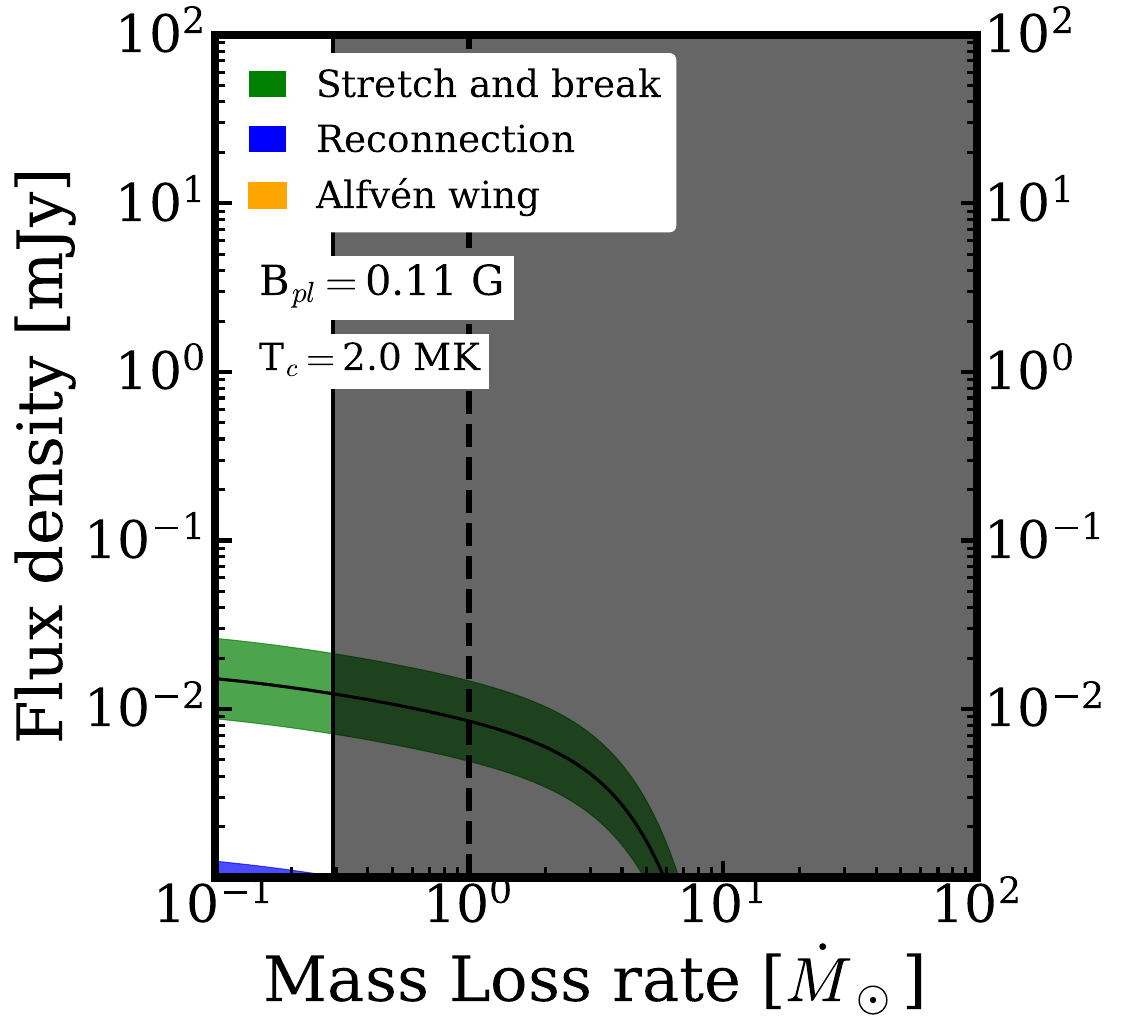}
\caption{
{\small  Same as in Fig.~\ref{fig:GJ1151_flux_close_planet}, but for a hypothetic high-mass (1.25 $M_{\oplus}$) planet.
}
}
\label{fig:GJ1151_flux_far_planet}
\end{figure*}%

We show in Figs.~\ref{fig:GJ1151_flux_close_planet} and ~\ref{fig:GJ1151_flux_far_planet} the expected radio flux for two scenarios of GJ~1151 b: a low-mass planet  (Fig.~\ref{fig:GJ1151_flux_close_planet}) and a high-mass planet (Fig.~\ref{fig:GJ1151_flux_far_planet}). The top panels in both figures show the flux as a function of the orbital separation, while the bottom panels show the predicted flux density as a function of stellar mass-loss rate. We show the results for two stellar wind geometries: a Parker spiral (left) and a hybrid PFSS model (right). 

As mentioned earlier, there is no star-planet interaction for the dipolar geometry case.
Overall, the predicted emission from the Alfvén wing model for GJ 1151 is stronger than that for other systems such as Proxima Centauri or YZ Cet. For instance, flux densities can reach up to $\sim$ 30--40~$\mu$Jy for the lower-mass planet and a Parker spiral geometry. Notably, regardless of the planet’s mass or orbital distance, it is likely to be sufficiently magnetized to sustain a magnetosphere for a PFSS hybrid geometry. In that case, the reconnection and stretch-and-break models predict an even stronger radio flux. The emission is also stronger the closer the planet is to the star, reaching up  $\sim 400 \mu$Jy for the lower mass estimate in a PFSS geometry, according to the stretch-and-break model. 
The effect of free free absorption is small for the assumed solar mass-loss rate, suppressing around 3 \% of the emitted flux.  For 
$\Mdotstar \gtrsim 10 \Mdotsun$ the emission is almost completely absorbed. Nevertheless,  
the strong predicted radio flux makes GJ 1151 one of the most promising candidates for detecting star-planet interactions (provided that the potentially interacting planet, GJ 1151 b, is eventually confirmed).
\citet{Vedantham2020} reported a time-frequency-averaged  Stokes I flux density of $\sim 890\mu$Jy, which is about a factor of two larger than 
the predicted value from the stretch-and-break model for the low-mass planet scenario. 
Therefore, if the detected radio signal would be confirmed as due to SPI, this would favor the lower-mass, short-period (most likely $P_{\rm orb} < 2$ days) exoplanet scenario for GJ 1151 b.

To summarize, we used \texttt{SIRIO} to study whether the radio signal detected by \citet{Vedantham2020} could be explained by SPI between GJ1151 and the hypothetical close-orbiting planet GJ 1151 b. We considered two scenarios, one where the mass of the planet is 0.73\,$M_{\oplus}$ and an orbital period of 1 day and another with a mass of 1.25\,$M_{\oplus}$  and a period of 5 days, representing the lower and upper bounds on the planet’s minimum mass, respectively \citep{Perger2021,Blanco-Pozo2023}. Our results show that under a dipolar stellar wind geometry no SPI occurs and for the hybrid PFSS SPI only takes place if the orbital period is less than 2 days, effectively ruling out the higher-mass limit of GJ 1151 b. The computed flux densities for the low-mass planet using the reconnection model are of the same order of magnitude as the observed values reported by \citet{Vedantham2020}, further reinforcing the low-mass limit for the hypothetical GJ 1151 b.

\section{Summary}
\label{sec:summary}

In this paper, we revisited the radio emission arising from magnetic star-planet interaction, for the paradigmatic cases of the M dwarf systems Proxima Centauri, YZ~Ceti and GJ~1151, for which putative detections of radio emission from sub-Alfvénic interaction have been claimed.
We  modelled the predicted radio emission using 
SIRIO (\textbf{S}tar-planet \textbf{I}nteraction and  \textbf{R}adio \textbf{I}nduced \textbf{O}bservations), a public Python code that predicts the radio emission arising from sub-Alfvénic SPI in close-in exoplanetary systems.
\texttt{SIRIO} implements a one-dimensional isothermal stellar wind model, with options for three different stellar wind magnetic field geometries: a pure dipole, an open Parker spiral, and a hybrid Potential Field Source Surface (PFSS) model. While the dipole and Parker spiral assume that all magnetic field lines are closed and open, respectively, the hybrid PFSS geometry bridges the two other geometries and offers a more realistic description by considering the lines are closed up to a certain separation from the star, $\Rss$, and open beyond that point. 

\texttt{SIRIO} computes the emission as a function of the orbital separation, mass loss rate of the star or the magnetic field of the planet. It also accounts for the effects of free-free absorption. To estimate the Poynting flux, \texttt{SIRIO} implements three alternative models: the Alfvén wing \citep{Zarka2007, Saur2013}, the reconnection \citep{Lanza2009} and the stretch-and-break model \citep{Lanza2013,Strugarek2022}.
While the Alfvén-wing model does not require the exoplanet to be magnetized to produce SPI radio emission, the other two models require that the exoplanetary magnetic be large enough so as to create a magnetosphere above the exoplanet surface; otherwise, no radio emission can be generated.  
\texttt{SIRIO} also accounts for the effects of free-free absorption within the stellar wind, which can significantly reduce the observed radio flux. 
We benchmarked the results of our \texttt{SIRIO} code against three iconic M dwarf systems 
that have been the subject of intense observational campaigns aimed at detecting SPI radio signals: Proxima Centauri, YZ Ceti, and GJ 1151. Those systems have also been subject of various modeling efforts to predict their radio emission from SPI, and are hence the most suitable ones for comparison purposes.  

For Proxima Centauri, our simulations reproduce exactly the results of \citet{Turnpenney2018}, which discussed only a Parker spiral geometry for the stellar wind magnetic field. We extend their analysis by including two additional stellar wind geometries, as well as two alternative models for yielding radio emission from sub-Alfvénic star-planet interaction: the reconnection and stretch-and-break models. 

For the reference values of $\Mdotstar=1.5\Mdotsun$ and $B_*$=600 G, we find that both Proxima b and the more inner planet, Proxima d, fall in the super-Alfvénic regime for a  dipolar geometry, but are well within the sub-Alfvénic region in the Parker spiral or PFSS geometries, and hence can yield SPI radio emission. We find that the radio emission yielded by the Alfvén wing model would be too faint to be detectable. On the other hand,  Proxima b can have a stable magnetosphere for the hybrid PFSS geometry, thus enabling magnetic reconnection. This boosts the predicted radio emission for the reconnection and stretch-and-break model and, if the magnetic field of Proxima is Earth-like, the expected flux density predicted by the stretch-and-break model would be in agreement with the observations by \citet{PerezTorres2021}.
Free-free absorption effects become noticeable only at high stellar mass-loss rates, above $30 \Mdotsun$, which seems a too high value for Proxima. Therefore, it is unlikely that free-free absorption plays a role in this system. 
We extended our analysis by running \texttt{SIRIO} under two additional conditions: one adopting $\Mdotstar=0.20 \Mdotsun$ (following \citealt{Reville2024}), and another with $\Mdotstar=0.25 \Mdotsun$ combined with a weaker stellar magnetic field of $B_*$=200 G, as in \citet{Kavanagh2021}. The modeling by \citet{Kavanagh2021} suggests the planet is extremely close to the Alfvén surface, while for the more recent modeling by \citet{Reville2024} this surface is much further out, at around 300 $R_{*}$, compared to the orbital distance of Proxima b, which lies at about 70,$R_\star$.

We also compared the results of \texttt{SIRIO} for YZ Cet to those in \citet{Pineda2023}. As in that publication, we considered two different models: One where the mass loss rate of the star is high (5$\Mdotsun$) and the magnetic field of stellar wind has only open magnetic field lines (model B), and another one where the mass loss rate is much lower (0.25$\Mdotsun$) and the magnetic field geometry is a hybrid/PFSS one (model B). 
Our modelling confirms the findings of \citet{Pineda2023} that this system likely hosts planets in the sub-Alfvénic regime capable of generating detectable radio emission.
We notice, though, that unlike the case of Proxima Cen, free-free absorption within the stellar wind of YZ Cet can strongly suppress the emitted signal, potentially explaining the non-detection of expected SPI signatures in some observations. In fact, while free-free absorption is negligible for model B, which has a very low value of $\dot{M}_\star$, for model A, with a higher mass-loss rate, about 75\% of the emission is absorbed and, if $\dot{M}_\star$ was as high as $\sim 10\,\dot{M}_\odot$, the signal would be entirely suppressed. 
Those results emphasize the need for precise measurements of the stellar wind mass-loss rates to accurately predict SPI radio fluxes and assess the potential detectability of a system.

Finally, we used \texttt{SIRIO} to assess whether the radio emission reported by \citet{Vedantham2020} could plausibly originate from SPI involving GJ 1151 and a hypothetical close-in planet. Based on the constraints from \citet{Blanco-Pozo2023}, we modeled two exoplanetary scenarios: a lower-mass planet (0.73 $M_{\oplus}$) with a 1-day orbital period, and a higher-mass planet (1.25 $M_{\oplus}$) with a 5-day period. Our simulations show if the stellar wind geometry is an open Parker spiral, then SPI may be at work in both cases, while for the hybrid PFSS geometry only the close-in, lower-mass planet would be able to interact with the star. If, instead, a dipolar geometry is at place, then the planet would be in the super-Alfvénic regime in both scenarios. The flux densities computed by \texttt{SIRIO} from the stretch-and-break model for the 0.73 $M_{\oplus}$ planet are of the same order of magnitude than the radio detections by \citet{Vedantham2020}, which also favors the low-mass planet scenario.

These case studies have proven that \texttt{SIRIO} 
is a flexible and computationally efficient tool for assessing the feasibility of radio observations targeting SPI, as well as for constraining stellar wind mass-loss rates and exoplanetary magnetic fields.
It is a reliable and versatile tool to model SPI, offering a very comprehensive framework that incorporates multiple stellar wind geometries and the effects of free-free absorption, and estimates the radio emission as a function of the orbital separation, stellar mass-loss rate and planetary magnetic field. As radio observations of exoplanetary systems continue to advance, tools like \texttt{SIRIO} will be essential for refining the target selection and interpreting the results of observing campaigns.

\section*{Acknowledgements}

We thank the anonymous referee for his/her constructive and detailed review, which has sinificantly improved our paper. 
We acknowledge financial support through the Severo Ochoa grant CEX2021-001131-S and through the Spanish National grant PID2023-147883NB-C21, funded by MCIU/AEI/ 10.13039/501100011033, as well as support through ERDF/EU. LPM also acknowledges funding through grant PRE2020-095421, funded by MCIU/AEI/10.13039/501100011033 and by FSE Investing in your future. 
% The Acknowledgements section is not numbered. Here you can thank helpful
% colleagues, acknowledge funding agencies, telescopes and facilities used etc.

%%%%%%%%%%%%%%%%%%%%%%%%%%%%%%%%%%%%%%%%%%%%%%%%%%
\section*{Data Availability}

All the data and code used to produce the plots in this  publication can be found in the SIRIO repository: \href{https://github.com/mapereztorres/sirio}{https://github.com/mapereztorres/sirio}, under the branch \texttt{sirio-paper}.
 
% The inclusion of a Data Availability Statement is a requirement for articles published in MNRAS. Data Availability Statements provide a standardised format for readers to understand the availability of data underlying the research results described in the article. The statement may refer to original data generated in the course of the study or to third-party data analysed in the article. The statement should describe and provide means of access, where possible, by linking to the data or providing the required accession numbers for the relevant databases or DOIs.



\bibliographystyle{mnras}
\bibliography{bibfile}

%\begin{thebibliography}{99}
%\end{thebibliography}

%%%%%%%%%%%%%%%%%%%%%%%%%%%%%%%%%%%%%%%%%%%%%%%%%%

%%%%%%%%%%%%%%%%% APPENDICES %%%%%%%%%%%%%%%%%%%%%

\appendix

\section{Input parameters for \texttt{SIRIO}}

The code has two alternative input modes, a table called "table.csv" or a text file "target.py". In both cases, the input parameters are distance of the star (in parsec), the radius and mass of the star (in solar units), the rotation period of the star, the magnetic field at the stellar equator (in Gauss), the mass and radius of the planet (in Earth units), the orbital period of the planet (in days), the temperature of the stellar corona (in MK), the mass-loss rate of the star (in solar units).
The code also needs the semi-major axis of the planet(in au). Since we consider a circular orbit, we equate a to the orbital radius. However, there is also an additional eccentricity parameter, to test the emission differences when the planet is in the periastron or the apastron.

\section{Stellar wind model}
\label{app:sw_model}
If the stellar wind is stationary, the equation of motion is (e.g., \citealt{Lamers-Cassinelli1999})

\begin{equation}
\vw \frac{d \vw}{dr} + \frac{1}{\rho} \frac{dp}{dr} + \frac{G M_\star}{r^2} = 0 
\label{eq:eq-of-motion}
\end{equation}

where \vw\ is the wind velocity,  $r$ is the radial distance, $\rho$ is the stellar wind mass density, $G$ is the gravitational constant, $M_\star$ is the stellar mass, and $p$ is the thermal pressure. The above equation describes the motion or the momentum of the gas in a stationary stellar wind. The acceleration of the wind plasma (first term) is produced by the pressure gradient (second term) and the gravity (third term).  The energy equation reduces to $ \Tw = $ constant, since we assume that the stellar wind of each star has a fixed temperature. 
If the flow behaves like an ideal gas, then the thermal pressure can be written as follows:

\begin{equation}
p = \nw\, \kB \Tw 
\label{eq:pressure}
\end{equation}

where $\nw = \rhow / \mu\, m_{\rm p}$ is the stellar wind number density ($\mu$ is the mean weight per particle, and $m_{\rm p}$ is the proton mass), \kB\ is the Boltzmann constant, and \Tw is the stellar wind temperature. 
Substituting Eq.~\ref{eq:pressure} into Eq.~\ref{eq:eq-of-motion} yields:

\begin{equation}
\frac{1}{\vw}\frac{d \vw}{dr} = \left( \frac{2a^2}{r} - \frac{G M_\star}{r^2} \right)/ (\vw^2 - a^2)  
\label{eq:momentum}
\end{equation}

where 
$a =\sqrt{\kB \Tw / \mu m_{\rm p}}$ is the sound speed, which is constant for an isothermal stellar wind. The critical point, where
 $\vw (r) = a$, occurs at:

\begin{equation}
r_{\rm c} = \frac{G M_\star}{2 a^2} 
 \label{eq:critical-distance}
\end{equation}

The only solution of the momentum equation that can have a positive velocity gradient at all distances is the one that goes through the critical point, where $\vw (r) = a$, the sonic point. In an isothermal stellar wind, the critical point coincides with the sonic point, and the stellar wind velocity profile, $\vw (r)$, depends only on the stellar wind temperature, \Tw, and the stellar mass, $M_\star$.
We obtain the stellar wind velocity profile, $\vw (r)$, by solving the following equation:

\begin{equation}
\rm{ln} (x) - x = \rm{ln} [D(r)],
\label{eq:Lambert}
\end{equation}

where $x=(\vw/a)^2$, and
$D(r)=\left(r/r_c\right)^{-4}\, \exp\left[4\, (1- r_c/r) - 1\right]$.
This equation is solved using the Lambert 
$W$ function \citep{Cranmer2004}:

\[
\vw^2 = 
\begin{cases} 
-a^2 W_0[-D(r)], & \text{if } r \leq r_c \\ 
-a^2 W_{-1}[-D(r)],  & \text{if } r \geq r_c 

\end{cases}
\]

\section{Stellar wind magnetic field geometries}
\label{app:b_field_geometries}

\texttt{SIRIO} computes the magnetic field intensity for three different geometries currently considered in the code: a pure dipole, an open Parker spiral, and a potential field source surface (PFSS), which is a hybrid geometry.

In the geometry of a classical dipole, all of its field lines are closed. The general expression for the magnetic field of a dipole is 
$\vec{B} = \left[ 3 (\vec{m} \cdot \hat{r}) \hat{r} - \vec{m} \right] / r^3$

\begin{figure}
   \centering
    \includegraphics[width=\linewidth]{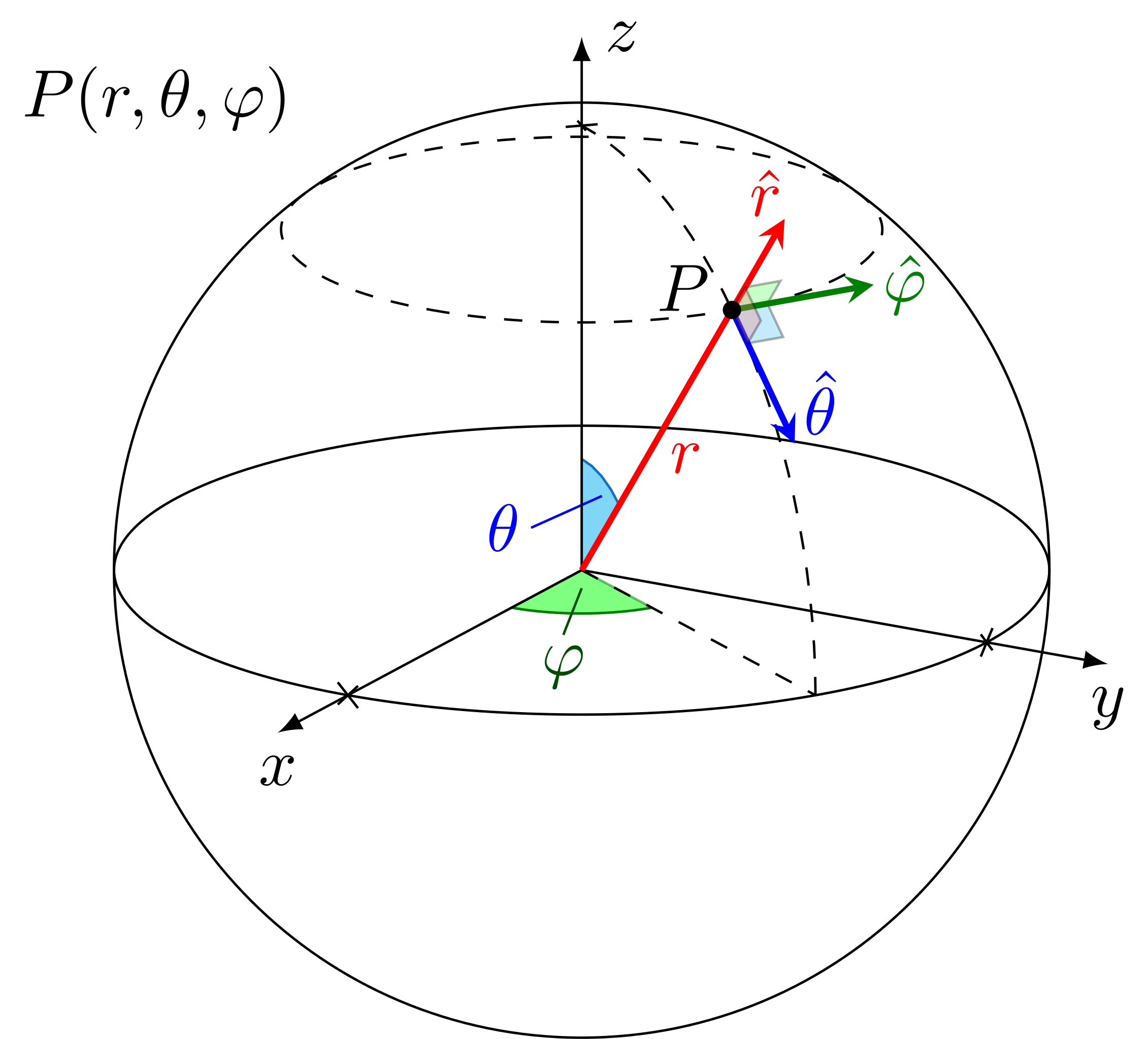}
\caption{
Diagram showing the spherical coordinates convention adopted here, superimposed on a Cartesian reference frame. $r$ denotes the radial distance, $\theta$ is the polar angle (0 < $\theta$ < $\pi$) and $\varphi$ is the azimuthal angle (0 < $\varphi$ < 2$\pi$). 
Image by Cristian Quinzacara, licensed under \href{https://creativecommons.org/licenses/by-sa/4.0/}{CC BY-SA 4.0}, via Wikimedia Commons.
}  
\label{fig:spherical-coordinates}
\end{figure}

In our chosen basis (see fig. \ref{fig:spherical-coordinates}), the magnetic moment of the dipole is on the Z axis: $\vec{m} = m \hat{z} =m (\cos \theta \cdot \hat{r} - \sin\theta \cdot \hat{\theta})
$. Substituting now $\vec{m} $ in the previous expresion, one obtains

\begin{equation}
     \vec{B} = \frac{m}{r^3} (2 \cos \theta \cdot \hat{r}  - \sin \theta \cdot \hat{\theta}).
\end{equation}

Hence, there is no $\phi$ component of the magnetic field in the dipolar geometry. To determine the modulus of the magnetic moment, we consider the case where the point of reference is in the XY axis ($\theta = \frac{\pi}{2}$) and located at the stellar surface ($r=R_*$), where the modulus of the local magnetic field is the measured magnetic field intensity of the star ($\|\vec{B}\|= B_*$). Since 
$\vec{B} =  (m/R_\star^3)\,\hat{\theta}$, so that $ m = R_\star^3\, \|\vec{B}\| =  R_\star^3\,  B_\star$. Therefore,

\begin{equation}
    \label{eq:B_dipole}
    B_{\rm r}=2 \frac{B_{*}\,R_\star^3}{r^3} \cos\theta; 
    B_{\theta}= \frac{B_{*}\,R_\star^3}{r^3} \sin\theta 
\end{equation}

\noindent

In the current version of the code, we assume that the orbital plane of the planet coincides with the equatorial plane of the star, so  that $\theta = \frac{\pi}{2}$. Therefore, $B_{\rm r}=0$ and the contribution to the magnetic field comes solely from the polar component, $\hat{\theta}$: $B_{\rm dipole}=B_{\theta}$.

We also consider a Parker spiral geometry for the stellar wind magnetic field, which consists of purely open field lines starting at the surface of the star. The stellar wind magnetic field has two components: a radial one, $B_{\rm r}$, and an azimuthal one, $B_{\phi}$,
        
\begin{equation}
    \label{eq:B_Parker}
    B_{\rm r} = B_\star \left( \frac{R_\star}{r} \right)^2; 
    B_{\phi} = B_{\rm r} \frac{\Omega_\star r}{\vsw} \rm sin\,\theta
\end{equation}

\noindent
where $\Omega_{*}$ is rotation speed of the star and $\vsw$ is the stellar wind speed. Note that, unlike in a dipolar geometry, in the Parker spiral geometry there is no polar component. We also emphasize that the azimuthal component of the stellar wind magnetic field at the orbital separation of all exoplanets discussed in this paper is negligible in the Parker spiral model, so the magnetic field is almost purely radial.

Finally, \texttt{SIRIO} considers a third geometry, which is a hybrid one between a purely closed dipolar geometry and an open Parker spiral one.
It is identical to a dipole from the star until a potential field source surface, where all the field lines are open and it becomes a Parker spiral.
Therefore, we consider a dipolar geometry if $r<\Rss$ and a Parker spiral if $r>\Rss$.  This is a simplified way of considering that most M dwarf stars can be well approximated by a main dipolar component, but whose dominance extend up to a given distance from the stellar surface, after which field lines are open.

\begin{figure}
    \centering
    \includegraphics[width=1\linewidth]{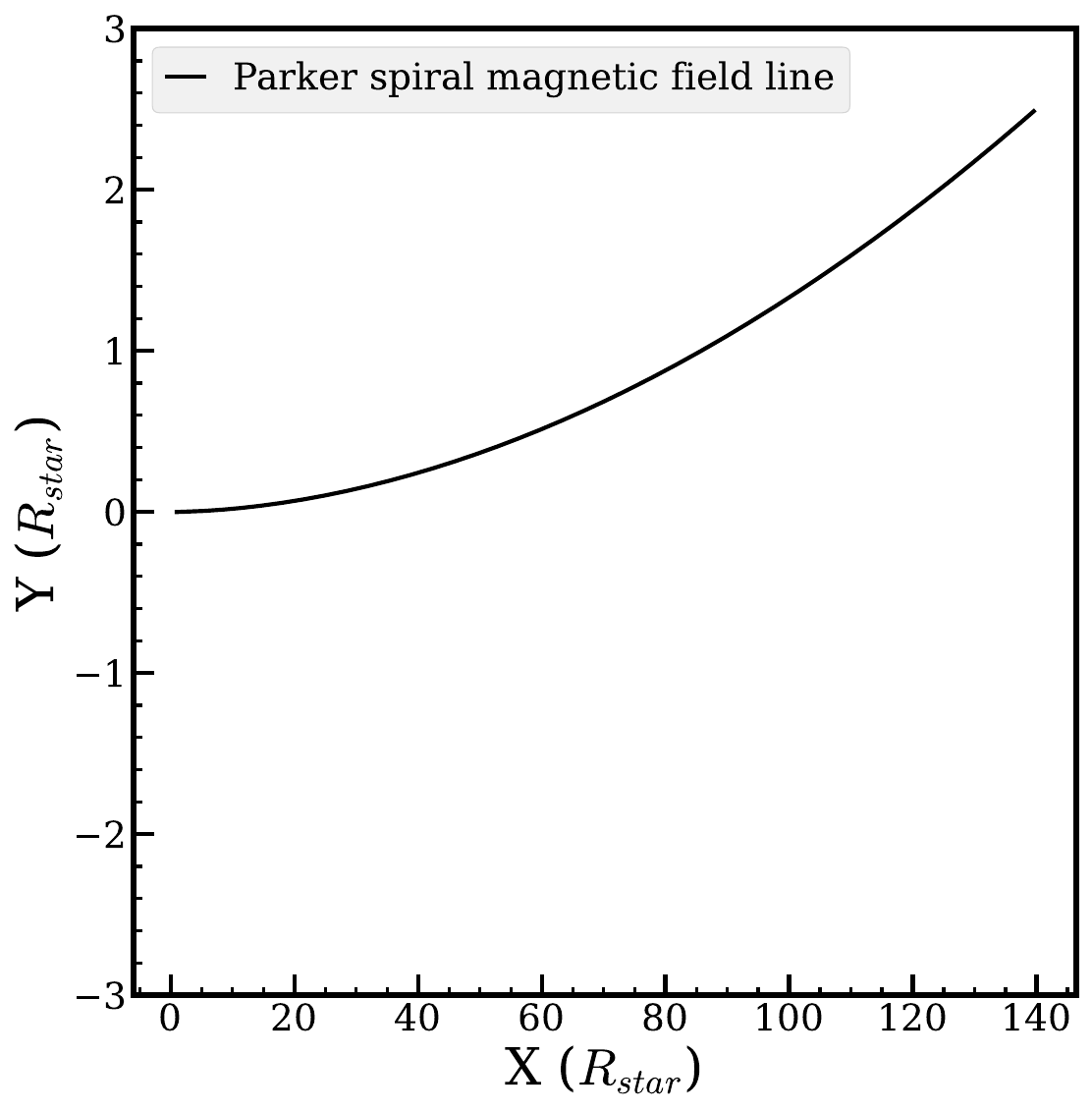}
    \caption{Stellar wind magnetic field for a Parker spiral geometry. Since the azimuthal component of the magnetic field is negligible in comparison to the radial one, the geometry can be approximated to a magnetic monopole or spherically symmetric field.%\com{change y axis to -5 to 5}
    }
    \label{fig:Parker_spiral_magnetic_field}
\end{figure}

\section{Interaction strength \texorpdfstring{$\overline\alpha$}{alpha}}
\label{app:alpha_interaction}

The $\overline\alpha$ parameter characterizes the strength of the sub-Alfvénic interaction. For $\overline\alpha$ = 0, the obstacle (i.e., the planet) does not perturb the surrounding plasma and no interaction takes place, while for $\overline\alpha$ = 1 the obstacle completely blocks the plasma flow and the interaction reaches its maximum strength \citep{Saur2013}. The interaction strength, $\overline\alpha$, can be approximated as

\begin{equation}
    \overline\alpha=\frac{\Sigma_{\rm P}}{\Sigma_{\rm P}+2\Sigma_{\rm A}}
\end{equation}

where $\Sigma_{\rm P}$ is the Pedersen conductance and $\Sigma_{\rm A}$ is the Alfvén conductance. $\Sigma_{\rm P}$ is estimated as follows \citep{Nichols2016}:

\begin{equation}
    \Sigma_{\rm P}=\kappa\, a^\lambda \left( \frac{B_{\rm Jup}}{B_{\rm pl}}\right) \left( \frac{L_{\rm XUV}}{L_{\rm XUV , \sun}} \right)^\mu
\end{equation}

where $a$ is the orbital separation of the planet, in au, $B_{\rm Jup}$ is the magnetic field of Jupiter, in Gauss, and $L_{\rm XUV}$ and $L_{\rm XUV , \sun}$ are the XUV luminosity for the star and the Sun, respectively. The constant factors in the equation are defined as $\kappa$= 15.475, $\lambda$ = -2.082, and $\mu$ = 1/2. The Alfvén conductance, in cgs, is given by \citep{Neubauer1980}:

\begin{equation}
\Sigma_A = \frac{1}{4 \pi V_{\rm A}\sqrt{1 + M_A^2 - 2 M_A \rm {cos^2\Theta},
}}
\end{equation}

where $\Theta$ is the angle in Fig.~\ref{fig:sketch}, which corresponds to the complementary angle in \citet{Saur2013} in \citet{Neubauer1980}. We see that for the three stellar systems discussed in this paper, $\Sigma_{\rm P} \gg \Sigma_{\rm A}$. For example, for Proxima Centauri using the parameters from \citet{Turnpenney2018}, $\Sigma_{\rm P}$ = 6.94$\times10^5$ mho and $\Sigma_A$ = 10.22 mho close to the stellar surface, where it reaches its maximum value. Therefore, $1 - \overline\alpha_{\rm min}$ = 1.16$\times10^{-5}$. We obtain similar values of $\overline\alpha$ for YZ Cet and GJ~1151, so in practice $\overline\alpha = 1$ for all three systems.

\end{document}